\documentclass[a4paper,11pt]{article}

\usepackage{hyperref} \hypersetup{colorlinks=true, linkcolor=blue,
  citecolor=magenta, urlcolor=black, bookmarksnumbered=true,
  bookmarkstype=toc, bookmarksopen=false, pdftitle={Intertwining
    Operator in Thermal CFT}, pdfauthor={Satoshi Ohya}}

\usepackage[margin=1in]{geometry}

\usepackage[bitstream-charter,cal=cmcal]{mathdesign}
\usepackage[T1]{fontenc}

\usepackage{xcolor}

\usepackage{amsmath,amsbsy}

\usepackage{epic,eepic}

\usepackage[hang,font=small,labelfont=bf]{caption}
\usepackage{subfigure}

\makeatletter

\@addtoreset{equation}{section} \makeatother

\usepackage{titlesec}
\titleformat{\section}[block]{\filright\bfseries\boldmath}{\thesection.}{0.5em}{}[]
\titleformat{\subsection}[block]{\filright\bfseries\boldmath}{\thesubsection.}{0.5em}{}

\usepackage{titletoc} \contentsmargin{0cm}
\titlecontents{section}[.8cm]{\filright\bfseries\boldmath}{\contentslabel{.8cm}}{}{\hfill\thecontentspage}[]
\titlecontents{subsection}[1.6cm]{\filright\small}{\contentslabel{.8cm}}{}{\;\titlerule*[.5pc]{.}\;\thecontentspage}[]
\titlecontents{subsubsection}[2.4cm]{\filright\small}{\contentslabel{.8cm}}{}{\;\titlerule*[.5pc]{.}\;\thecontentspage}[]

\usepackage{cite}

\title{\large\bfseries\boldmath Intertwining Operator in Thermal
  CFT$_{d}$}
\author{{\normalsize Satoshi Ohya\footnote{E-mail: \texttt{ohya@phys.cst.nihon-u.ac.jp}}}\\[1em]
  {\small\itshape Institute of Quantum Science, Nihon University}\\
  {\small\itshape Kanda-Surugadai 1-8-14, Chiyoda, Tokyo 101-8308,
    Japan}}
\date{\small (Dated: \today)}

\begin{document}
\maketitle
\flushbottom

\begin{abstract}
  It has long been known that two-point functions of conformal field
  theory (CFT) are nothing but the integral kernels of intertwining
  operators for two equivalent representations of conformal
  algebra. Such intertwining operators are known to fulfill some
  operator identities---the intertwining relations---in the
  representation space of conformal algebra. Meanwhile, it has been
  known that the S-matrix operator in scattering theory is nothing but
  the intertwining operator between the Hilbert spaces of in- and
  out-particles. Inspired by this algebraic resemblance, in this paper
  we develop a simple Lie-algebraic approach to momentum-space
  two-point functions of thermal CFT living on the hyperbolic
  spacetime $\mathbb{H}^{1}\times\mathbb{H}^{d-1}$ by exploiting the
  idea of Kerimov's intertwining operator approach to exact
  S-matrix. We show that in thermal CFT on
  $\mathbb{H}^{1}\times\mathbb{H}^{d-1}$ the intertwining relations
  reduce to certain linear recurrence relations for two-point
  functions in the complex momentum space. By solving these recurrence
  relations, we obtain the momentum-space representations of advanced
  and retarded two-point functions as well as positive- and
  negative-frequency two-point Wightman functions for a scalar primary
  operator in arbitrary spacetime dimension $d\geq3$.
\end{abstract}

\begingroup
\hypersetup{linkcolor=black}
\tableofcontents
\endgroup

\newpage
\section{Introduction and Summary}
\label{section:1}
It is widely recognized that, up to a few numerical factors, two-point
functions for primary operators of conformal field theory (CFT) are
completely determined through the $SO(2,d)$ conformal symmetry in any
spacetime dimension $d\geq1$. However, it is less recognized---or
generally unappreciated except to experts
\cite{Koller:1974ut,Dobrev:1977,Fradkin:1978pp,Todorov:1978rf}---that
two-point functions in CFT have special representation-theoretic
meanings: they are the integral kernels of intertwining operators for
two equivalent representations of conformal algebra
$\mathfrak{so}(2,d)$. Namely, once given a two-point function for a
primary operator of scaling dimension $\Delta$, we can construct an
operator $G_{\Delta}$ which maps a primary state of scaling dimension
$d-\Delta$ to another primary state of scaling dimension
$\Delta$.\footnote{This is essentially equivalent to the so-called
  ``shadow operator formalism'' \cite{Ferrara:1972uq}. For more
  details, see section \ref{section:3}.} Such operator
$G_{\Delta}$---the intertwining operator---satisfies the following
commutative diagram and operator identities (intertwining relations):
\begin{center}
\xdefinecolor{rgb_000000}{rgb}{0,0,0}%
\xdefinecolor{rgb_ff0000}{rgb}{1,0,0}%
\setlength{\unitlength}{1cm}%
\begin{picture}(10,3.2)(0,0)%
\allinethickness{0.8pt}%
\color{rgb_ff0000}%
\path(1,2.3)(1,1)
\allinethickness{0.0140584cm}%
\path(0.995607,1.01318)(1.00439,1.01318)
\path(0.991214,1.02636)(1.00879,1.02636)
\path(0.98682,1.03954)(1.01318,1.03954)
\path(0.982427,1.05272)(1.01757,1.05272)
\path(0.978034,1.0659)(1.02197,1.0659)
\path(0.973641,1.07908)(1.02636,1.07908)
\path(0.969247,1.09226)(1.03075,1.09226)
\path(0.964854,1.10544)(1.03515,1.10544)
\path(0.960461,1.11862)(1.03954,1.11862)
\path(0.956068,1.1318)(1.04393,1.1318)
\path(0.951674,1.14498)(1.04833,1.14498)
\path(0.947281,1.15816)(1.05272,1.15816)
\path(1.03954,1.11862)(1.03954,1.15816)
\path(1.02636,1.07908)(1.02636,1.15816)
\path(1.01318,1.03954)(1.01318,1.15816)
\path(1,1)(1,1.15816)
\path(0.98682,1.03954)(0.98682,1.15816)
\path(0.973641,1.07908)(0.973641,1.15816)
\path(0.960461,1.11862)(0.960461,1.15816)
\allinethickness{0.8pt}%
\path(1,1.15816)(0.947281,1.15816)(1,1)(1.05272,1.15816)(1,1.15816)
\path(1.6,2.6)(3.4,2.6)
\allinethickness{0.0140584cm}%
\path(3.38682,2.59561)(3.3829,2.6057)
\path(3.37364,2.59121)(3.36579,2.6114)
\path(3.36046,2.58682)(3.34869,2.6171)
\path(3.34728,2.58243)(3.33158,2.62281)
\path(3.3341,2.57803)(3.31448,2.62851)
\path(3.32092,2.57364)(3.29737,2.63421)
\path(3.30774,2.56925)(3.28027,2.63991)
\path(3.29456,2.56485)(3.26316,2.64561)
\path(3.28138,2.56046)(3.24606,2.65131)
\path(3.2682,2.55607)(3.24184,2.62386)
\path(3.25502,2.55167)(3.24184,2.58557)
\path(3.26161,2.64613)(3.24184,2.63844)
\path(3.28138,2.63954)(3.24184,2.62417)
\path(3.30115,2.63295)(3.24184,2.60989)
\path(3.32092,2.62636)(3.24184,2.59561)
\path(3.34069,2.61977)(3.24184,2.58134)
\path(3.36046,2.61318)(3.24184,2.56706)
\path(3.38023,2.60659)(3.24184,2.55278)
\allinethickness{0.8pt}%
\path(3.24184,2.6)(3.24184,2.54728)(3.4,2.6)(3.24184,2.65272)(3.24184,2.6)
\path(4,2.3)(4,1)
\allinethickness{0.0140584cm}%
\path(3.96777,1.15816)(3.95207,1.14378)
\path(3.98826,1.15816)(3.95687,1.1294)
\path(4.00874,1.15816)(3.96166,1.11502)
\path(4.02923,1.15816)(3.96645,1.10065)
\path(4.04972,1.15816)(3.97124,1.08627)
\path(4.04503,1.13509)(3.97604,1.07189)
\path(4.03603,1.10808)(3.98083,1.05751)
\path(4.02702,1.08106)(3.98562,1.04313)
\path(4.01801,1.05404)(3.99041,1.02876)
\path(4.00901,1.02702)(3.99521,1.01438)
\path(3.98972,1.03083)(4.00479,1.01438)
\path(3.97945,1.06165)(4.00959,1.02876)
\path(3.96917,1.09248)(4.01438,1.04313)
\path(3.9589,1.12331)(4.01917,1.05751)
\path(3.94862,1.15413)(4.02396,1.07189)
\path(3.9629,1.15816)(4.02876,1.08627)
\path(3.98086,1.15816)(4.03355,1.10065)
\path(3.99883,1.15816)(4.03834,1.11502)
\path(4.01679,1.15816)(4.04313,1.1294)
\path(4.03475,1.15816)(4.04793,1.14378)
\allinethickness{0.8pt}%
\path(4,1.15816)(3.94728,1.15816)(4,1)(4.05272,1.15816)(4,1.15816)
\path(1.4,0.6)(3.6,0.6)
\allinethickness{0.0140584cm}%
\path(3.44184,0.561809)(3.45942,0.553139)
\path(3.44184,0.576337)(3.47699,0.558996)
\path(3.44184,0.590865)(3.49456,0.564854)
\path(3.44184,0.605394)(3.51214,0.570712)
\path(3.44184,0.619922)(3.52971,0.576569)
\path(3.44184,0.63445)(3.54728,0.582427)
\path(3.44184,0.648978)(3.56485,0.588285)
\path(3.50924,0.630255)(3.58243,0.594142)
\path(3.58163,0.593877)(3.58682,0.604393)
\path(3.56326,0.587754)(3.57364,0.608786)
\path(3.5449,0.581632)(3.56046,0.61318)
\path(3.52653,0.575509)(3.54728,0.617573)
\path(3.50816,0.569386)(3.5341,0.621966)
\path(3.48979,0.563263)(3.52092,0.626359)
\path(3.47142,0.557141)(3.50774,0.630753)
\path(3.45305,0.551018)(3.49456,0.635146)
\path(3.44184,0.559403)(3.48138,0.639539)
\path(3.44184,0.590508)(3.4682,0.643932)
\path(3.44184,0.621614)(3.45502,0.648326)
\allinethickness{0.8pt}%
\path(3.44184,0.6)(3.44184,0.547281)(3.6,0.6)(3.44184,0.652719)(3.44184,0.6)
\put(1,2.6){\makebox(0,0)[c]{\hbox{\color{rgb_000000}$\mathcal{V}_{d-\Delta}$}}}
\put(4,2.6){\makebox(0,0)[c]{\hbox{\color{rgb_000000}$\mathcal{V}_{d-\Delta}$}}}
\put(1,0.6){\makebox(0,0)[c]{\hbox{\color{rgb_000000}$\mathcal{V}_{\Delta}$}}}
\put(4,0.6){\makebox(0,0)[c]{\hbox{\color{rgb_000000}$\mathcal{V}_{\Delta}$}}}
\put(0.859416,1.6){\makebox(0,0)[r]{\hbox{\color{rgb_000000}$G_{\Delta}$}}}
\put(4.14058,1.6){\makebox(0,0)[l]{\hbox{\color{rgb_000000}$G_{\Delta}$}}}
\put(2.5,2.70544){\makebox(0,0)[b]{\hbox{\color{rgb_000000}$J_{d-\Delta}^{ab}$}}}
\put(2.5,0.494562){\makebox(0,0)[t]{\hbox{\color{rgb_000000}$J_{\Delta}^{ab}$}}}
\put(8,1.6){\makebox(0,0)[c]{\hbox{\color{rgb_000000}$J_{\Delta}^{ab}G_{\Delta}=G_{\Delta}J_{d-\Delta}^{ab}$}}}
\end{picture}%
\end{center}
where $\mathcal{V}_{\alpha}$ is a representation space of scaling
dimension $\alpha\in\{\Delta,d-\Delta\}$ in which the quadratic
Casimir operator $C_{2}[\mathfrak{so}(2,d)]$ takes a definite value
and $J_{\alpha}^{ab}=-J_{\alpha}^{ba}$ are the $SO(2,d)$ generators
that act on the representation space $\mathcal{V}_{\alpha}$. An
important observation here is that the intertwining operator
$G_{\Delta}$ is quite analogous to an S-matrix operator $S$ in
scattering theory which satisfies the following commutative diagram
and operator identity:
\begin{center}
\xdefinecolor{rgb_000000}{rgb}{0,0,0}%
\xdefinecolor{rgb_ff0000}{rgb}{1,0,0}%
\setlength{\unitlength}{1cm}%
\begin{picture}(10,3.2)(0,0)%
\allinethickness{0.8pt}%
\color{rgb_ff0000}%
\path(1,2.3)(1,1)
\allinethickness{0.0140584cm}%
\path(0.995607,1.01318)(1.00439,1.01318)
\path(0.991214,1.02636)(1.00879,1.02636)
\path(0.98682,1.03954)(1.01318,1.03954)
\path(0.982427,1.05272)(1.01757,1.05272)
\path(0.978034,1.0659)(1.02197,1.0659)
\path(0.973641,1.07908)(1.02636,1.07908)
\path(0.969247,1.09226)(1.03075,1.09226)
\path(0.964854,1.10544)(1.03515,1.10544)
\path(0.960461,1.11862)(1.03954,1.11862)
\path(0.956068,1.1318)(1.04393,1.1318)
\path(0.951674,1.14498)(1.04833,1.14498)
\path(0.947281,1.15816)(1.05272,1.15816)
\path(1.03954,1.11862)(1.03954,1.15816)
\path(1.02636,1.07908)(1.02636,1.15816)
\path(1.01318,1.03954)(1.01318,1.15816)
\path(1,1)(1,1.15816)
\path(0.98682,1.03954)(0.98682,1.15816)
\path(0.973641,1.07908)(0.973641,1.15816)
\path(0.960461,1.11862)(0.960461,1.15816)
\allinethickness{0.8pt}%
\path(1,1.15816)(0.947281,1.15816)(1,1)(1.05272,1.15816)(1,1.15816)
\path(1.5,2.6)(3.5,2.6)
\allinethickness{0.0140584cm}%
\path(3.48682,2.59561)(3.4829,2.6057)
\path(3.47364,2.59121)(3.46579,2.6114)
\path(3.46046,2.58682)(3.44869,2.6171)
\path(3.44728,2.58243)(3.43158,2.62281)
\path(3.4341,2.57803)(3.41448,2.62851)
\path(3.42092,2.57364)(3.39737,2.63421)
\path(3.40774,2.56925)(3.38027,2.63991)
\path(3.39456,2.56485)(3.36316,2.64561)
\path(3.38138,2.56046)(3.34606,2.65131)
\path(3.3682,2.55607)(3.34184,2.62386)
\path(3.35502,2.55167)(3.34184,2.58557)
\path(3.36161,2.64613)(3.34184,2.63844)
\path(3.38138,2.63954)(3.34184,2.62417)
\path(3.40115,2.63295)(3.34184,2.60989)
\path(3.42092,2.62636)(3.34184,2.59561)
\path(3.44069,2.61977)(3.34184,2.58134)
\path(3.46046,2.61318)(3.34184,2.56706)
\path(3.48023,2.60659)(3.34184,2.55278)
\allinethickness{0.8pt}%
\path(3.34184,2.6)(3.34184,2.54728)(3.5,2.6)(3.34184,2.65272)(3.34184,2.6)
\path(4,2.3)(4,1)
\allinethickness{0.0140584cm}%
\path(3.96777,1.15816)(3.95207,1.14378)
\path(3.98826,1.15816)(3.95687,1.1294)
\path(4.00874,1.15816)(3.96166,1.11502)
\path(4.02923,1.15816)(3.96645,1.10065)
\path(4.04972,1.15816)(3.97124,1.08627)
\path(4.04503,1.13509)(3.97604,1.07189)
\path(4.03603,1.10808)(3.98083,1.05751)
\path(4.02702,1.08106)(3.98562,1.04313)
\path(4.01801,1.05404)(3.99041,1.02876)
\path(4.00901,1.02702)(3.99521,1.01438)
\path(3.98972,1.03083)(4.00479,1.01438)
\path(3.97945,1.06165)(4.00959,1.02876)
\path(3.96917,1.09248)(4.01438,1.04313)
\path(3.9589,1.12331)(4.01917,1.05751)
\path(3.94862,1.15413)(4.02396,1.07189)
\path(3.9629,1.15816)(4.02876,1.08627)
\path(3.98086,1.15816)(4.03355,1.10065)
\path(3.99883,1.15816)(4.03834,1.11502)
\path(4.01679,1.15816)(4.04313,1.1294)
\path(4.03475,1.15816)(4.04793,1.14378)
\allinethickness{0.8pt}%
\path(4,1.15816)(3.94728,1.15816)(4,1)(4.05272,1.15816)(4,1.15816)
\path(1.4,0.6)(3.6,0.6)
\allinethickness{0.0140584cm}%
\path(3.44184,0.561809)(3.45942,0.553139)
\path(3.44184,0.576337)(3.47699,0.558996)
\path(3.44184,0.590865)(3.49456,0.564854)
\path(3.44184,0.605394)(3.51214,0.570712)
\path(3.44184,0.619922)(3.52971,0.576569)
\path(3.44184,0.63445)(3.54728,0.582427)
\path(3.44184,0.648978)(3.56485,0.588285)
\path(3.50924,0.630255)(3.58243,0.594142)
\path(3.58163,0.593877)(3.58682,0.604393)
\path(3.56326,0.587754)(3.57364,0.608786)
\path(3.5449,0.581632)(3.56046,0.61318)
\path(3.52653,0.575509)(3.54728,0.617573)
\path(3.50816,0.569386)(3.5341,0.621966)
\path(3.48979,0.563263)(3.52092,0.626359)
\path(3.47142,0.557141)(3.50774,0.630753)
\path(3.45305,0.551018)(3.49456,0.635146)
\path(3.44184,0.559403)(3.48138,0.639539)
\path(3.44184,0.590508)(3.4682,0.643932)
\path(3.44184,0.621614)(3.45502,0.648326)
\allinethickness{0.8pt}%
\path(3.44184,0.6)(3.44184,0.547281)(3.6,0.6)(3.44184,0.652719)(3.44184,0.6)
\put(1,2.6){\makebox(0,0)[c]{\hbox{\color{rgb_000000}$\mathcal{H}_{\text{out}}$}}}
\put(4,2.6){\makebox(0,0)[c]{\hbox{\color{rgb_000000}$\mathcal{H}_{\text{out}}$}}}
\put(1,0.6){\makebox(0,0)[c]{\hbox{\color{rgb_000000}$\mathcal{H}_{\text{in}}$}}}
\put(4,0.6){\makebox(0,0)[c]{\hbox{\color{rgb_000000}$\mathcal{H}_{\text{in}}$}}}
\put(0.859416,1.6){\makebox(0,0)[r]{\hbox{\color{rgb_000000}$S$}}}
\put(4.14058,1.6){\makebox(0,0)[l]{\hbox{\color{rgb_000000}$S$}}}
\put(2.5,2.70544){\makebox(0,0)[b]{\hbox{\color{rgb_000000}$A_{\text{out}}$}}}
\put(2.5,0.494562){\makebox(0,0)[t]{\hbox{\color{rgb_000000}$A_{\text{in}}$}}}
\put(8,1.6){\makebox(0,0)[c]{\hbox{\color{rgb_000000}$A_{\text{in}}S=SA_{\text{out}}$}}}
\end{picture}%
\end{center}
where $\mathcal{H}_{\text{in}}$ and $\mathcal{H}_{\text{out}}$ are the
Hilbert spaces of in- and out-particles and $A_{\text{in}}$ and
$A_{\text{out}}$ are some operators (typically creation and
annihilation operators) that act on $\mathcal{H}_{\text{in}}$ and
$\mathcal{H}_{\text{out}}$.

The purpose of this paper is to pursue this analogy between
$G_{\Delta}$ and $S$. To be more concrete, in this paper we shall
develop a Lie-algebraic method to compute momentum-space two-point
functions of $d(\geq3)$-dimensional CFT at finite temperature by
exploiting the idea of Kerimov's intertwining operator approach to
exact S-matrix \cite{Kerimov:1998zz} (see also
\cite{Kerimov:1998,Kerimov:2002a,Kerimov:2002b}). In the late 1990s
Kerimov developed an elegant Lie-algebraic method to compute S-matrix
elements of exactly-solvable quantum mechanical models whose dynamics
are governed by dynamical symmetry $SO(p,q)$, $p\geq q>1$. He found
that in such models S-matrix operators are nothing but the
intertwining operators for two equivalent representations of $SO(p,q)$
and showed that, by using the intertwining relations, S-matrix
elements must fulfill certain linear recurrence relations. By solving
those recurrence relations, he obtained a large class of
quantum-mechanical exact S-matrices in a purely algebraic
fashion.\footnote{It should be noted that, though the Kerimov's
  S-matrix theory is limited to quantum-mechanical scattering
  problems, it is also true in relativistic quantum field theory that
  the S-matrix operator is nothing but an intertwining operator; see,
  e.g., chapter 6 of Strocchi's book \cite{Strocchi:2013awa}. Note
  also that quantum-mechanical exact S-matrices (i.e., reflection and
  transmission amplitudes) are related to exact two-body S-matrix
  elements of nonrelativistic quantum field theory; see, e.g.,
  \cite{Sogo:1981mc}.} We shall apply his method to the problem of
computing momentum-space correlation functions of finite-temperature
CFT in spacetime dimension $d\geq3$.\footnote{There are many parallels
  between the Kerimov's S-matrix theory and our Lie-algebraic approach
  to thermal two-point functions. There is, however, a crucial
  difference. The difference is the basis for the representation of
  $\mathfrak{so}(2,d)$: the Kerimov's method \cite{Kerimov:1998zz} is
  based on the basis that diagonalizes the maximal compact subgroup
  $SO(2)\times SO(d)\subset SO(2,d)$, whereas our method is based on
  the basis that diagonalizes the noncompact subgroup
  $SO(1,1)\times SO(1,d-1)\subset SO(2,d)$.} In general, it is a very
hard task to calculate Fourier transforms of position-space conformal
correlators. This is true even at zero temperature and the situation
gets worse at finite temperature. In fact, to the best of our
knowledge, momentum-space correlation functions of thermal CFT in
generic dimension $d\geq3$ have not been explicitly derived even for
scalar two-point functions.\footnote{For the case of two-dimensional
  thermal CFT, the momentum-space two- and three-point Wightman
  functions for a scalar primary operator were calculated in
  \cite{Maldacena:1997ih,Gubser:1997cm,Bredberg:2009pv} and
  \cite{Becker:2014jla}. See also our previous works on holographic
  approach to momentum-space thermal two-point functions of one- and
  two-dimensional CFT \cite{Ohya:2013xva,Ohya:2013vba}.} We shall see
that, at least for the case of two-point functions, the intertwining
relations may well provide an alternative to Fourier transform.

\begin{figure}[t]
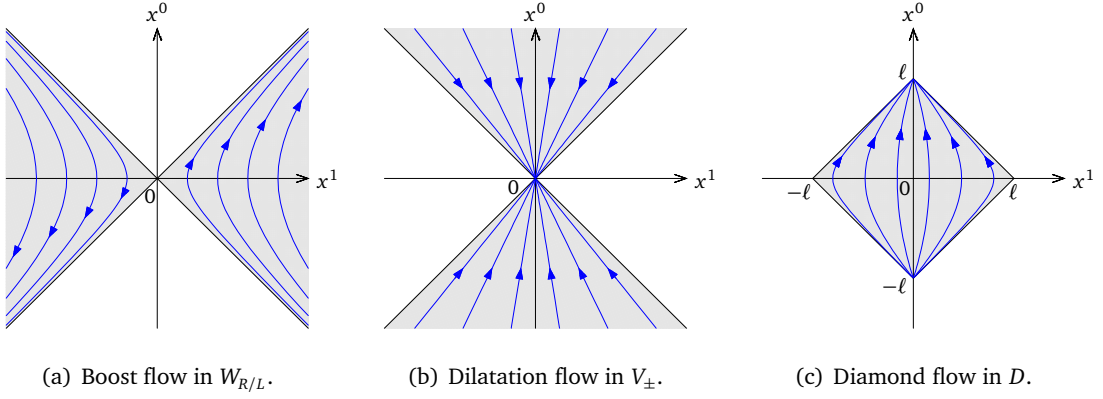

  \centering \subfigure[Boost flow in
  $W_{R/L}$.\label{figure:1a}]{\input{figures/figure1a.eepic}} \quad
  \subfigure[Dilatation flow in
  $V_{\pm}$.\label{figure:1b}]{\input{figures/figure1b.eepic}} \quad
  \subfigure[Diamond flow in
  $D$.\label{figure:1c}]{\input{figures/figure1c.eepic}}
  \caption{Flows of $SO(1,1)$ conformal Killing vectors projected onto
    the $(x^{0},x^{1})$-plane. Blue curves are the orbits of the
    action of the one-parameter subgroup $SO(1,1)\subset SO(2,d)$ on a
    spacetime point $x^{\mu}\in W_{R/L}, V_{\pm}, D$, where the
    subgroup $SO(1,1)$ is generated by (a) the Lorentz boost generator
    $M^{10}=J^{10}$, (b) the dilatation generator $D=J^{d,d+1}$, and
    (c) the combination
    $\tfrac{1}{2}(\ell P^{0}-\tfrac{1}{\ell}K^{0})=J^{d0}$; see
    sections \ref{section:2.2}--\ref{section:2.4} for more details.}
  \label{figure:1}
\end{figure}

In addition to the intertwining operator, there is another key
component for the study of thermal conformal correlators: the Unruh
effect. In order to thermalize $d$-dimensional CFT, we shall put CFT
on the Rindler wedge $W_{R/L}$---the causally connected region for an
eternally uniformly accelerating observer---and identify the temporal
coordinate with the $SO(1,1)\subset SO(2,d)$ Lorentz boost
parameter. According to Sewell's theorem \cite{Sewell:1982zz}, which
follows from the celebrated Bisognano--Wichmann theorem
\cite{Bisognano:1975ih,Bisognano:1976za}, any such quantum field
theory (not necessarily conformally-invariant) automatically satisfies
the Kubo--Martin--Schwinger (KMS) thermal equilibrium condition
\cite{Haag:1967sg} for Wightman functions and hence gets thermalized
geometrically.\footnote{A nice exposition of Sewell's theorem can be
  found in chapter 5 of \cite{Takagi:1986kn}.} It should be noted
that, though Sewell's theorem is proved axiomatically and hence can be
applied to any relativistic quantum field theory with or without
conformal symmetry, special things happen when the theory enjoys
conformal invariance. The point is that the conformal group $SO(2,d)$
admits other geometrical realizations of the action of the
one-parameter subgroup $SO(1,1)$. One is the subgroup $SO(1,1)$
generated by the dilatation and the other by the linear combination of
the temporal components of momentum and special conformal
transformation generators; see figure \ref{figure:1}. Orbits of the
actions of these two subgroups coincide with the worldlines for an
observer in semi-eternal uniform motion and a uniformly accelerating
observer with finite lifetime, each of whose causally connected
regions are the future (past) light-cone $V_{\pm}$ and diamond $D$,
respectively.\footnote{Note, however, that proper times of these
  observers do \emph{not} coincide with the $SO(1,1)$ group
  parameters: they are related by certain coordinate transformations
  \cite{Martinetti:2002sz} (see also section \ref{section:2}).} As
discussed first by Buchholz for $V_{\pm}$ \cite{Buchholz:1977ze} and
later by Hislop and Longo for $D$ and $V_{\pm}$ \cite{Hislop:1981uh}
in the context of modular theory in operator algebra, CFTs restricted
on these regions are shown to be thermal as well under the
identifications of temporal coordinates with these $SO(1,1)$ group
parameters. In this way, CFT in any spacetime dimension $d$ easily
gets thermalized by just putting it on the Rindler wedge $W_{R/L}$,
light-cone $V_{\pm}$, or diamond $D$ with suitably-chosen temporal
coordinates. The price to pay, however, is that thus obtained thermal
CFTs are \emph{not} conformal to those on the flat Minkowski
spacetime: they all describe finite-temperature quantum field theory
living on the hyperbolic spacetime
$\mathbb{H}^{1}\times\mathbb{H}^{d-1}$ rather than
$\mathbb{R}^{1,d-1}$.\footnote{$\mathbb{H}^{1}\times\mathbb{H}^{d-1}$
  is often written as $\mathbb{R}\times\mathbb{H}^{d-1}$ in the
  literature. (Note that $\mathbb{H}^{1}$ is isometric to
  $\mathbb{R}$.)} Though less obvious its physical applications
are,\footnote{Thermal CFT on $\mathbb{H}^{1}\times\mathbb{H}^{d-1}$
  (or its Euclidean version on $\mathbb{S}^{1}\times\mathbb{H}^{d-1}$)
  has recently been attracted much attention in the context of
  (holographic) entanglement entropy; see, e.g.,
  \cite{Casini:2011kv}. Unfortunately, however, immediate applications
  of our results are less obvious.} for the sake of simplicity this
paper studies finite-temperature CFT living on the hyperbolic
spacetime by restricting zero-temperature CFT on the Rindler wedge,
light-cone, or diamond.

The rest of the paper is organized as follows. The next two sections
are devoted to preliminary materials. Section \ref{section:2} is a
description of $d$-dimensional CFT on the Rindler wedge, light-cone,
and diamond in terms of the embedding space formalism. The embedding
space formalism is quite an old idea for describing conformal symmetry
\cite{Dirac:1936fq} but has recently been revived thanks to its
usefulness for studying correlation functions and conformal blocks;
see, e.g., \cite{Weinberg:2010fx,Costa:2011mg,Costa:2011dw}. After
reviewing everything we need in section \ref{section:2.1}, we
introduce in sections \ref{section:2.2}--\ref{section:2.4} the
$d$-dimensional Rindler wedge, light-cone, and diamond as particular
subspaces of $(d+1)$-dimensional cone and discuss thermal correlation
functions in the position space. We shall see that the two-point
function for a scalar primary operator of scaling dimension $\Delta$
living on the hyperbolic spacetime
$\mathbb{H}^{1}\times\mathbb{H}^{d-1}$ takes the following form:
\begin{align}
  \left[\frac{2\pi^{2}T^{2}}{-\cosh(2\pi T(t-t^{\prime}))-H\cdot H^{\prime}}\right]^{\Delta}, \label{eq:1.1}
\end{align}
where
$(t,H),(t^{\prime},H^{\prime})\in\mathbb{H}^{1}\times\mathbb{H}^{d-1}$
and $T$ is the Unruh temperature. We then introduce the intertwining
operator for a scalar primary operator in section \ref{section:3}. We
first review the basics of intertwining operator and then show that,
for the case of zero-temperature CFT, the intertwining relations just
reduce to the well-known conformal Ward--Takahashi identities for
two-point functions. Section \ref{section:4} is the main part of the
present paper and discusses implications of intertwining relations for
thermal two-point functions. In section \ref{section:4.1} we construct
the representation of conformal algebra $\mathfrak{so}(2,d)$ in which
the $SO(1,1)$ Lorentz boost generator becomes diagonal. This is quite
unconventional yet the most important part of this paper because, in
geometrically thermalized CFT, the time-translation generator
generates the noncompact subgroup $SO(1,1)$ rather than, say, the
compact subgroup $SO(2)$. By using this representation, we show in
section \ref{section:4.2} that, for the case of finite-temperature CFT
on the hyperbolic spacetime $\mathbb{H}^{1}\times\mathbb{H}^{d-1}$,
the intertwining relations reduce to the following linear recurrence
relations for two-point function $\Tilde{G}_{\Delta}(\omega,k)$ in the
complex momentum space:
\begin{subequations}
  \begin{align}
    \Tilde{G}_{\Delta}(\omega\pm i2\pi T,k\pm i2\pi T)
    &=\frac{\Delta-(d-2)/2\mp i(\omega+k)/2\pi T}{\Tilde{\Delta}-(d-2)/2\mp i(\omega+k)/2\pi T}
      \Tilde{G}_{\Delta}(\omega,k), \label{eq:1.2a}\\
    \Tilde{G}_{\Delta}(\omega\pm i2\pi T,k\mp i2\pi T)
    &=\frac{\Delta-(d-2)/2\mp i(\omega-k)/2\pi T}{\Tilde{\Delta}-(d-2)/2\mp i(\omega-k)/2\pi T}
      \Tilde{G}_{\Delta}(\omega,k), \label{eq:1.2b}
  \end{align}
\end{subequations}
where $\omega$ is the frequency conjugate to the $SO(1,1)$ boost
parameter---the proper-time for an eternally uniformly accelerating
observer---and $k(>0)$ is the modulus of spatial momentum which is
related to the eigenvalue of the quadratic Casimir operator of the
subalgebra $\mathfrak{so}(1,d-1)$. Here $\Tilde{\Delta}=d-\Delta$ is
the scaling dimension of the so-called ``shadow operator''
\cite{Ferrara:1972uq}. The goal of this paper is to derive and solve
these recurrence relations. We shall see that the retarded, advanced
and positive- as well as negative-frequency two-point Wightman
functions are all obtained by solving these recurrence relations. We
conclude in section \ref{section:5}. Appendix \ref{section:A} presents
computational details for the Fourier transforms of positive- and
negative-frequency two-point Wightman functions, which are given by
just replacing $t-t^{\prime}$ with $t-t^{\prime}\mp i\epsilon$ in
\eqref{eq:1.1}. Making use of the harmonic analysis on the hyperbolic
spacetime $\mathbb{H}^{1}\times\mathbb{H}^{d-1}$, we show that the
Fourier transforms of Wightman functions---very hard $d$-dimensional
integrals that involve the modified Bessel functions---indeed coincide
with two of the solutions to the recurrence relations \eqref{eq:1.2a}
and \eqref{eq:1.2b}.

\section{Embedding Space Formalism}
\label{section:2}
Conformal transformations are nonlinear transformations of spacetime
coordinates. However, they can be linearly realized if we embed the
$d$-dimensional Minkowski spacetime into the $(d+2)$-dimensional
ambient space $\mathbb{R}^{2,d}$. In this section we first review the
basics of embedding space formalism for zero-temperature CFT. We then
introduce the Rindler wedge, light-cone, and diamond as particular
subspaces of $(d+1)$-dimensional cone which can be embedded into
$\mathbb{R}^{2,d}$. We shall see that these subspaces are all
conformal to the hyperbolic spacetime
$\mathbb{H}^{1}\times\mathbb{H}^{d-1}$ and the Wightman functions on
$\mathbb{H}^{1}\times\mathbb{H}^{d-1}$ indeed satisfy the KMS thermal
equilibrium condition under the identification of temporal coordinates
with the $SO(1,1)$ group parameters.

We emphasize that this section is mostly a condensation of known
results, yet contains some small new results in the construction of
thermal CFTs and thermal correlation functions. We have collected all
the necessary ingredients in order to make the paper
self-contained. Notice that a similar construction of Euclidean
thermal CFT on $\mathbb{S}^{1}\times\mathbb{H}^{d-1}$ can be found in
\cite{Faulkner:2014jva,Agon:2015ftl}.

Throughout this section the spacetime dimension is $d\geq2$. It is
also briefly commented on the case $d=1$.

\subsection{Embedding the Minkowski spacetime into the cone}
\label{section:2.1}
To begin with, let us first consider the $(d+1)$-dimensional null
cone, $\text{Cone}_{d+1}$, which can be embedded into the
$(d+2)$-dimensional ambient space $\mathbb{R}^{2,d}$ as follows:
\begin{align}
  \text{Cone}_{d+1}=\left\{X^{a}=(X^{0},\cdots,X^{d+1}): X\cdot X=0~~\&~~X^{a}\neq0\right\}, \label{eq:2.1}
\end{align}
where $X\cdot X=\eta_{ab}X^{a}X^{b}$ with
$\eta_{ab}=\eta^{ab}=\mathrm{diag}(-1,+1,\cdots,+1,-1)$ being the
ambient space metric. Thus defined cone is obviously invariant under
the following transformations:
\begin{align}
  X^{a}\mapsto{g^{a}}_{b}X^{b}
  \quad\text{and}\quad
  X^{a}\mapsto\lambda X^{a}, \label{eq:2.2}
\end{align}
where $g$ is an element of $O(2,d)$ and $\lambda$ is a non-vanishing
real. As usual, in this paper we shall focus on the identity component
of indefinite orthogonal group $O(2,d)$ which, with a slight abuse of
notation, we just denote by $SO(2,d)$. The cone \eqref{eq:2.1} can be
parameterized as follows:
\begin{align}
  X^{\mu}=\ell\frac{x^{\mu}}{z}, \quad
  X^{d}=\frac{\ell^{2}-x\cdot x}{2z}, \quad
  X^{d+1}=\frac{\ell^{2}+x\cdot x}{2z}, \label{eq:2.3}
\end{align}
or, conversely,
\begin{align}
  z=\frac{\ell^{2}}{X^{d}+X^{d+1}}
  \quad\text{and}\quad
  x^{\mu}=\ell\frac{X^{\mu}}{X^{d}+X^{d+1}}, \label{eq:2.4}
\end{align}
where $z\in\mathbb{R}_{\times}=\{z\in\mathbb{R}: z\neq0\}$,
$x^{\mu}=(x^{0},\cdots,x^{d-1})\in\mathbb{R}^{1,d-1}$, and
$x\cdot x=\eta_{\mu\nu}x^{\mu}x^{\nu}$. Here $\ell>0$ is an arbitrary
reference length scale which is just introduced to adjust the length
dimension of $X^{a}$ and $(z,x^{\mu})$. We call the coordinate system
\eqref{eq:2.4} the Poincar\'{e} coordinate system. The $d$-dimensional
Minkowski spacetime can then be identified with the following section
of the cone:
\begin{align}
  \text{Poincar\'{e}}_{d}=\{X^{a}\in\text{Cone}_{d+1}: X^{d}+X^{d+1}=\ell\}. \label{eq:2.5}
\end{align}
This section---the Poincar\'{e} section---is the $z=\ell$ hypersurface
of the cone and hence can be parameterized as follows:
\begin{align}
  X^{\mu}=x^{\mu},\quad
  X^{d}=\frac{\ell^{2}-x\cdot x}{2\ell},\quad
  X^{d+1}=\frac{\ell^{2}+x\cdot x}{2\ell}. \label{eq:2.6}
\end{align}
It is easy to check that the induced metric on this section indeed
becomes the Minkowski metric in the Cartesian coordinate system:
\begin{align}
  ds_{\text{Poincar\'{e}}_{d}}^{2}
  &=\left.-(dX^{0})^{2}+(dX^{1})^{2}+\cdots+(dX^{d})^{2}-(dX^{d+1})^{2}\right|_{X\in\text{Poincar\'{e}}_{d}} \nonumber\\
  &=\eta_{\mu\nu}dx^{\mu}dx^{\nu}. \label{eq:2.7}
\end{align}
Hence the Poincar\'{e} section \eqref{eq:2.5} is isometric to the
Minkowski spacetime.

Let us next move on to the $SO(2,d)$ transformations on the cone
\eqref{eq:2.1}. In terms of the ambient space coordinates $X^{a}$, the
$SO(2,d)$ transformations are linearly realized as
$g: X^{a}\mapsto X_{g}^{a}={g^{a}}_{b}X^{b}$ with $g\in SO(2,d)$. In
terms of the Poincar\'{e} coordinates $(z,x^{\mu})$, on the other
hand, the $SO(2,d)$ transformations are nonlinearly realized as
$g:
(z,x^{\mu})\mapsto(z_{g},x_{g}^{\mu})=(\frac{\ell^{2}}{X_{g}^{d}+X_{g}^{d+1}},\ell\frac{X_{g}^{\mu}}{X_{g}^{d}+X_{g}^{d+1}})$.
There are four physically distinct subgroups in $SO(2,d)$:
\begin{itemize}
\item \emph{Dilatation.} The one-dimensional subgroup
  \begin{align}
    g
    =
    \begin{pmatrix}
      \boldsymbol{1}_{d} & & \\
      &\cosh\varphi &\sinh\varphi \\
      &\sinh\varphi &\cosh\varphi
    \end{pmatrix}
                      \in SO(1,1) \label{eq:2.8}
  \end{align}
  induces the dilatation $g: (z,x^{\mu})\mapsto(z_{g},x_{g}^{\mu})$,
  where
  \begin{align}
    z_{g}=\mathrm{e}^{-\varphi}z
    \quad\text{and}\quad
    x_{g}^{\mu}=\mathrm{e}^{-\varphi}x^{\mu}. \label{eq:2.9}
  \end{align}

\item \emph{Translation.} The $d$-dimensional subgroup\footnote{$E(1)$
    stands for the one-dimensional Euclidean group. (In general, the
    Euclidean group consists of translations and rotations. In one
    dimension, however, there is no rotation such that $E(1)$ is just
    the translation group.)}
  \begin{align}
    g
    =
    \begin{pmatrix}
      \boldsymbol{1}_{d} & \tfrac{a^{\mu}}{\ell} & \tfrac{a^{\mu}}{\ell} \\[1ex]
      -\frac{a_{\nu}}{\ell} & 1-\tfrac{1}{2}\tfrac{a\cdot a}{\ell^{2}} & -\tfrac{1}{2}\frac{a\cdot a}{\ell^{2}} \\[1ex]
      \tfrac{a_{\nu}}{\ell} & \tfrac{1}{2}\tfrac{a\cdot a}{\ell^{2}} &
      1+\tfrac{1}{2}\tfrac{a\cdot a}{\ell^{2}}
    \end{pmatrix}
                                                                       \in E(1)^{d} \label{eq:2.10}
  \end{align}
  induces the translation $g: (z,x^{\mu})\mapsto(z_{g},x_{g}^{\mu})$,
  where
  \begin{align}
    z_{g}=z
    \quad\text{and}\quad
    x_{g}^{\mu}=x^{\mu}+a^{\mu}. \label{eq:2.11}
  \end{align}
  
\item \emph{Special conformal transformation.} The $d$-dimensional
  subgroup
  \begin{align}
    g
    =
    \begin{pmatrix}
      \boldsymbol{1}_{d} & \ell b^{\mu} & -\ell b^{\mu} \\[1ex]
      -\ell b_{\nu} & 1-\tfrac{\ell^{2}}{2}b \cdot b & \tfrac{\ell^{2}}{2}b \cdot b \\[1ex]
      -\ell b_{\nu} & -\tfrac{\ell^{2}}{2}b \cdot b &
      1+\tfrac{\ell^{2}}{2}b \cdot b
    \end{pmatrix}
                                                      \in E(1)^{d} \label{eq:2.12}
  \end{align}
  induces the special conformal transformation
  $g: (z,x^{\mu})\mapsto(z_{g},x_{g}^{\mu})$, where
  \begin{align}
    z_{g}=\frac{z}{1-2(b\cdot x)+(b\cdot b)(x\cdot x)}
    \quad\text{and}\quad
    x_{g}^{\mu}=\frac{x^{\mu}-b^{\mu}(x\cdot x)}{1-2(b\cdot x)+(b\cdot b)(x\cdot x)}. \label{eq:2.13}
  \end{align}

\item \emph{Lorentz transformation.} The $d(d-1)/2$-dimensional
  subgroup
  \begin{align}
    g
    =
    \begin{pmatrix}
      \Lambda 	& \\
      & \boldsymbol{1}_{2}
    \end{pmatrix}
	\in SO(1,d-1) \label{eq:2.14}
  \end{align}
  induces the Lorentz transformation
  $g: (z,x^{\mu})\mapsto(z_{g},x_{g}^{\mu})$, where
  \begin{align}
    z_{g}=z
    \quad\text{and}\quad
    x_{g}^{\mu}={\Lambda^{\mu}}_{\nu}x^{\nu}. \label{eq:2.15}
  \end{align}
\end{itemize}
It should be noted that the translation \eqref{eq:2.11} and Lorentz
transformation \eqref{eq:2.15} are isometries of the Poincar\'{e}
section; that is, they leave the coordinate
$z=\ell^{2}/(X^{d}+X^{d+1})$ unchanged. The dilatation \eqref{eq:2.9}
and special conformal transformation \eqref{eq:2.13}, on the other
hand, change the coordinate $z$ and hence map a point on the section
to a point outside the section. But such a point outside the section
can be pulled back to a point on the section by (coordinate-dependent)
rescaling $X_{g}^{a}\mapsto\lambda(X_{g})X_{g}^{a}$ with
$\lambda(X_{g})=(X^{d}+X^{d+1})/(X_{g}^{d}+X_{g}^{d+1})$, without
making any change in the ratio
$\ell X_{g}^{\mu}/(X_{g}^{d}+X_{g}^{d+1})=x_{g}^{\mu}$. This
combination of $SO(2,d)$ rotations and rescaling furnishes the
conformal transformations on the Poincar\'{e} section which, as
mentioned before, can be identified with the $d$-dimensional Minkowski
spacetime.

Now, let us next move on to infinitesimal conformal transformations.
Let ${g^{a}}_{b}={\delta^{a}}_{b}+{\epsilon^{a}}_{b}$ be a generic
element of $SO(2,d)$ with $\epsilon_{ab}=-\epsilon_{ba}$ being
antisymmetric infinitesimal parameters. Under the identification
\begin{align}
  \begin{pmatrix}
    {\epsilon^{\mu}}_{\nu} &{\epsilon^{\mu}}_{d}
    &{\epsilon^{\mu}}_{d+1} \\
    {\epsilon^{d}}_{\nu} &0
    &{\epsilon^{d}}_{d+1} \\
    {\epsilon^{d+1}}_{\nu} &{\epsilon^{d+1}}_{d} &0
  \end{pmatrix}
                                                   =
                                                   \begin{pmatrix}
                                                     {\omega^{\mu}}_{\nu}
                                                     &\tfrac{a^{\mu}}{\ell}+\ell
                                                     b^{\mu}
                                                     &\tfrac{a^{\mu}}{\ell}-\ell b^{\mu} \\
                                                     -\tfrac{a_{\nu}}{\ell}-\ell
                                                     b_{\nu} &0
                                                     &\varphi \\
                                                     \tfrac{a_{\nu}}{\ell}-\ell
                                                     b_{\nu} &\varphi
                                                     &0
                                                   \end{pmatrix}, \label{eq:2.16}
\end{align}
where $\omega_{\mu\nu}=-\omega_{\nu\mu}$ is related to $\Lambda$ in
\eqref{eq:2.14} by
${\Lambda^{\mu}}_{\nu}={\delta^{\mu}}_{\nu}+{\omega^{\mu}}_{\nu}$, the
infinitesimal conformal transformation
$x^{\mu}\mapsto x_{1+\epsilon}^{\mu}$ associated with $g=1+\epsilon$
is just given by the sum of infinitesimal forms of \eqref{eq:2.9},
\eqref{eq:2.11}, \eqref{eq:2.13} and \eqref{eq:2.15}:
\begin{align}
  x_{1+\epsilon}^{\mu}
  =x^{\mu}
  -\varphi x^{\mu}
  +a^{\mu}
  +b_{\nu}(-\eta^{\mu\nu}x\cdot x+2x^{\mu}x^{\nu})
  +{\omega^{\mu}}_{\nu}x^{\nu}. \label{eq:2.17}
\end{align}
Substituting the parameterizations
\begin{align}
  \varphi={\epsilon^{d}}_{d+1}, \quad
  a^{\mu}=\frac{\ell}{2}({\epsilon^{\mu}}_{d}+{\epsilon^{\mu}}_{d+1}), \quad
  b^{\mu}=\frac{1}{2\ell}({\epsilon^{\mu}}_{d}-{\epsilon^{\mu}}_{d+1}), \quad
  {\omega^{\mu}}_{\nu}={\epsilon^{\mu}}_{\nu}, \label{eq:2.18}
\end{align}
one immediately sees that the infinitesimal conformal transformation
\eqref{eq:2.17} can be put into the following form:
\begin{align}
  x_{1+\epsilon}^{\mu}=x^{\mu}+\frac{1}{2}\epsilon_{ab}k^{\mu ab}(x), \label{eq:2.19}
\end{align}
where $k^{\mu ab}(x)=-k^{\mu ba}(x)$ are conformal Killing vectors
given by
\begin{subequations}
  \begin{align}
    k^{\mu \nu\lambda}
    &=\eta^{\mu\nu}x^{\lambda}-\eta^{\mu\lambda}x^{\nu}, \label{eq:2.20a}\\
    k^{\mu \nu d}
    &=\frac{\ell^{2}-x\cdot x}{2\ell}\eta^{\mu\nu}
      +\frac{x^{\mu}x^{\nu}}{\ell}, \label{eq:2.20b}\\
    k^{\mu \nu,d+1}
    &=\frac{\ell^{2}+x\cdot x}{2\ell}\eta^{\mu\nu}
      -\frac{x^{\mu}x^{\nu}}{\ell}, \label{eq:2.20c}\\
    k^{\mu d,d+1}
    &=-x^{\mu}. \label{eq:2.20d}
  \end{align}
\end{subequations}
It is easy to check that these vectors indeed satisfy the conformal
Killing equations in the flat spacetime background,
$\partial_{\mu}{k_{\nu}}^{ab}+\partial_{\nu}{k_{\mu}}^{ab}=\tfrac{2}{d}\eta_{\mu\nu}\partial_{\rho}k^{\rho
  ab}$. We shall repeatedly use these conformal Killing vectors in the
subsequent sections.

Before closing this section let us finally recall the construction of
conformal correlation functions in the embedding space formalism. A
basic ingredient for us is a $(d+2)$-dimensional homogeneous scalar
field $\mathcal{O}_{\Delta}(X)$ that satisfies the following
transformation laws:
\begin{align}
  \mathcal{O}_{\Delta}(gX)=\mathcal{O}_{\Delta}(X)
  \quad\text{and}\quad
  \mathcal{O}_{\Delta}(\lambda X)=\lambda^{-\Delta}\mathcal{O}_{\Delta}(X), \label{eq:2.21}
\end{align}
where $g\in SO(2,d)$ and $\lambda$ is a non-vanishing real. The
two-point function for $\mathcal{O}_{\Delta}(X)$ that satisfies
$G_{\Delta}(gX,gX^{\prime})=G_{\Delta}(X,X^{\prime})$ and
$G_{\Delta}(\lambda X,\lambda
X^{\prime})=\lambda^{-2\Delta}G_{\Delta}(X,X^{\prime})$ is uniquely
determined (up to the overall normalization and the possible
$i\epsilon$-prescription) and given by
\begin{align}
  G_{\Delta}(X,X^{\prime})
  =\frac{1}{(-2X\cdot X^{\prime})^{\Delta}}
  =\frac{1}{[(X-X^{\prime})^{2}]^{\Delta}}, \label{eq:2.22}
\end{align}
where the second equality follows from
$X\cdot X=X^{\prime}\cdot X^{\prime}=0$ and we have set the
normalization constant to be unity. It then follows from the
parameterization \eqref{eq:2.6} that the two-point function projected
onto the Poincar\'{e} section gives the well-known two-point function
for a scalar primary operator of zero-temperature CFT in $d$
dimensions:
\begin{align}
  \left.G_{\Delta}(X,X^{\prime})\right|_{X,X^{\prime}\in\text{Poincar\'{e}}_{d}}
  =\frac{1}{[(x-x^{\prime})^{2}]^{\Delta}}
  =\frac{1}{[-(x^{0}-x^{0\prime})^{2}+|\boldsymbol{x}-\boldsymbol{x}^{\prime}|^{2}]^{\Delta}}. \label{eq:2.23}
\end{align}
Now, if we worked in Euclidean signature, this would be the end of the
story for the construction of conformal two-point function. In
Lorentzian signature, however, this is not the end of the story
because we have to specify the $i\epsilon$-prescription in order to
avoid including the branch points at
$x^{0}=x^{0\prime}\pm|\boldsymbol{x}-\boldsymbol{x}^{\prime}|$ which
lie exactly on the real $x^{0}$-axis. There are several basic
two-point functions in Lorentzian quantum field theory. Among them are
the positive- and negative-frequency two-point Wightman functions
$G_{\Delta}^{\pm}(x,x^{\prime})$, which are the vacuum expectation
values of primary operators
$\langle\Omega|\mathcal{O}_{\Delta}(x)\mathcal{O}_{\Delta}(x^{\prime})|\Omega\rangle$
and
$\langle\Omega|\mathcal{O}_{\Delta}(x^{\prime})\mathcal{O}_{\Delta}(x)|\Omega\rangle$
and given by the following $i\epsilon$-prescriptions:
\begin{align}
  G_{\Delta}^{\pm}(x,x^{\prime})
  &=\frac{1}{[-(x^{0}-x^{0\prime}\mp i\epsilon)^{2}+|\boldsymbol{x}-\boldsymbol{x}^{\prime}|^{2}]^{\Delta}}\nonumber\\
  &=\frac{1}{[-(x^{0}-x^{0\prime})^{2}+|\boldsymbol{x}-\boldsymbol{x}^{\prime}|^{2}\pm i\epsilon\text{sgn}(x^{0}-x^{0\prime})]^{\Delta}},\label{eq:2.24}
\end{align}
where $\epsilon$ is a positive infinitesimal and $\text{sgn}(x)=x/|x|$
is the sign function. All the other two-point functions are given in
terms of these Wightman functions. For example, the time-ordered
two-point function
$G_{\Delta}^{F}(x,x^{\prime})=\langle\Omega|T\{\mathcal{O}_{\Delta}(x)\mathcal{O}_{\Delta}(x^{\prime})\}|\Omega\rangle$
is given by
\begin{align}
  G_{\Delta}^{F}(x,x^{\prime})
  &=\theta(x^{0}-x^{0\prime})G_{\Delta}^{+}(x,x^{\prime})+\theta(x^{0\prime}-x^{0})G_{\Delta}^{-}(x,x^{\prime})\nonumber\\
  &=\frac{1}{[-(x^{0}-x^{0\prime})^{2}+|\boldsymbol{x}-\boldsymbol{x}^{\prime}|^{2}+i\epsilon]^{\Delta}}.\label{eq:2.25}
\end{align}
Similarly, the retarded and advanced correlators
$G_{\Delta}^{R}(x,x^{\prime})=\theta(x^{0}-x^{0\prime})\langle\Omega|[\mathcal{O}_{\Delta}(x),\mathcal{O}_{\Delta}(x^{\prime})]|\Omega\rangle$
and
$G_{\Delta}^{A}(x,x^{\prime})=\theta(x^{0\prime}-x^{0})\langle\Omega|[\mathcal{O}_{\Delta}(x^{\prime}),\mathcal{O}_{\Delta}(x)]|\Omega\rangle$
are given by
\begin{subequations}
  \begin{align}
    G_{\Delta}^{R}(x,x^{\prime})
    &=\theta(x^{0}-x^{0\prime})\left(G_{\Delta}^{+}(x,x^{\prime})-G_{\Delta}^{-}(x,x^{\prime})\right)\nonumber\\
    &=G_{\Delta}^{F}(x,x^{\prime})-G_{\Delta}^{-}(x,x^{\prime}),\label{eq:2.26a}\\
    G_{\Delta}^{A}(x,x^{\prime})
    &=\theta(x^{0\prime}-x^{0})\left(G_{\Delta}^{-}(x,x^{\prime})-G_{\Delta}^{+}(x,x^{\prime})\right)\nonumber\\
    &=G_{\Delta}^{F}(x,x^{\prime})-G_{\Delta}^{+}(x,x^{\prime}).\label{eq:2.26b}
  \end{align}
\end{subequations}
In this way, one can as well study two-point functions for generic
primary tensors or higher-point functions by the embedding space
formalism; see, e.g.,
\cite{Weinberg:2010fx,Costa:2011mg,Costa:2011dw}. In this paper,
however, we will just focus on the scalar two-point function for
simplicity.

\subsection{Embedding the Rindler wedge into the cone}
\label{section:2.2}
Now let us turn to the problem of constructing finite-temperature CFT
in $d$ dimensions. To this end, let us first introduce the following
wedge regions of the cone:
\begin{subequations}
  \begin{align}
    \overline{W}_{R}&=\{X^{a}\in\text{Cone}_{d+1}: X^{1}\pm X^{0}>0~~\&~~X^{d+1}>0\}, \label{eq:2.27a}\\
    \overline{W}_{L}&=\{X^{a}\in\text{Cone}_{d+1}: X^{1}\pm X^{0}<0~~\&~~X^{d+1}>0\}. \label{eq:2.27b}
  \end{align}
\end{subequations}
We first wish to specify these wedge regions in terms of the
Poincar\'{e} coordinate system $(z,x^{\mu})$ given in
\eqref{eq:2.3}. To do this, let us first note that the conditions
$X^{1}\pm X^{0}>0$ together with $X\cdot X=0$ imply
$(X^{d+1})^{2}-(X^{d})^{2}=(X^{1}+X^{0})(X^{1}-X^{0})+\cdots+(X^{d-1})^{2}>0$,
which, together with $X^{d+1}>0$, implies the conditions
$X^{d+1}\pm X^{d}>0$. Hence $z=\ell^{2}/(X^{d}+X^{d+1})$ must be
positive-definite on $\overline{W}_{R/L}$. With this in mind, we
immediately see that the conditions $X^{1}\pm X^{0}>0$ and $X^{d+1}>0$
are translated into $x^{1}\pm x^{0}>0$ and
$\ell^{2}+(x^{1}+x^{0})(x^{1}-x^{0})+(x^{2})^{2}+\cdots+(x^{d-1})^{2}>0$
in the Poincar\'{e} coordinate system, the latter is automatically
satisfied if the former is fulfilled. Hence these wedge regions can
also be specified as follows:
\begin{subequations}
  \begin{align}
    \overline{W}_{R}&=\{(z,x^{\mu}): z>0~~\&~~x^{1}\pm x^{0}>0\}, \label{eq:2.28a}\\
    \overline{W}_{L}&=\{(z,x^{\mu}): z>0~~\&~~x^{1}\pm x^{0}<0\}. \label{eq:2.28b}
  \end{align}
\end{subequations}
Now it is obvious that the intersections of these wedges and the
Poincar\'{e} section give the following $d$-dimensional hypersurfaces
of the cone:
\begin{align}
  W_{R/L}
  &=\{X^{a}\in\overline{W}_{R/L}: X^{d}+X^{d+1}=\ell\}\nonumber\\
  &=\begin{cases}
    \displaystyle
    \{x^{\mu}: x^{1}\pm x^{0}>0\},\\
    \displaystyle
    \{x^{\mu}: x^{1}\pm x^{0}<0\},\label{eq:2.29}
  \end{cases}
\end{align}
which are nothing but the standard right and left Rindler wedges,
respectively.\footnote{The conditions $x^{1}\pm x^{0}>0$ and
  $x^{1}\pm x^{0}<0$ are often written as $x^{1}>|x^{0}|$ and
  $x^{1}<|x^{0}|$ in the literature.}  Hence, after fixing the
coordinate $z$ the wedge regions \eqref{eq:2.27a} and \eqref{eq:2.27b}
indeed describe the Rindler wedges of $d$-dimensional Minkowski
spacetime.

Let us next consider the isometries of the wedge
$\overline{W}_{R/L}$. The most important isometry of the wedge
$\overline{W}_{R/L}$ is the following hyperbolic rotation on the
$(X^{0},X^{1})$-plane:
\begin{align}
  g(\theta)=
  \begin{pmatrix}
    \cosh\theta & \sinh\theta & \\
    \sinh\theta & \cosh\theta & \\
    & & \boldsymbol{1}_{d}
  \end{pmatrix}
        \in SO(1,1), \label{eq:2.30}
\end{align}
which maps the wedge into itself; that is, the one-parameter subgroup
\eqref{eq:2.30} maps a point $X^{a}$ on the wedge to another point
$X^{a}(\theta)={g(\theta)^{a}}_{b}X^{b}$ on the wedge and hence gives
a globally well-defined transformation on
$\overline{W}_{R/L}$. Substituting
$X^{a}(\theta)=(X^{0}\cosh\theta+X^{1}\sinh\theta,X^{0}\sinh\theta+X^{1}\cosh\theta,X^{2},\cdots,X^{d+1})$
into \eqref{eq:2.4}, one immediately sees that the subgroup
\eqref{eq:2.30} induces the following one-parameter family of boost
flow on the wedge $\overline{W}_{R/L}$:
\begin{subequations}
  \begin{align}
    z(\theta)&=z, \label{eq:2.31a}\\
    x^{0}(\theta)&=x^{0}\cosh\theta+x^{1}\sinh\theta, \label{eq:2.31b}\\
    x^{1}(\theta)&=x^{0}\sinh\theta+x^{1}\cosh\theta, \label{eq:2.31c}\\
    \boldsymbol{x}_{\perp}(\theta)&=\boldsymbol{x}_{\perp}, \label{eq:2.31d}
  \end{align}
\end{subequations}
where $\boldsymbol{x}_{\perp}=(x^{2},\cdots,x^{d-1})$. Notice that
eqs.~\eqref{eq:2.31b}--\eqref{eq:2.31d} can also be obtained as
solutions to the following flow equation:
\begin{align}
  \frac{\partial x^{\mu}}{\partial\epsilon_{10}}=k^{\mu10}(x),\label{eq:2.32}
\end{align}
where $k^{\mu10}(x)$ is the conformal Killing vector given in
\eqref{eq:2.20a} and ${\epsilon^{1}}_{0}=\theta$. In other words,
eqs.~\eqref{eq:2.31b}--\eqref{eq:2.31d} are the flow of conformal
Killing vector $k^{\mu10}$ under the initial condition
$x^{\mu}(\theta=0)=(x^{0},x^{1},\boldsymbol{x}_{\perp})$. We also note
that eqs.~\eqref{eq:2.31b}--\eqref{eq:2.31d} coincide with the
worldline for an eternally uniformly accelerating observer in the
$d$-dimensional Minkowski spacetime. In fact, if we choose the initial
point as $x^{\mu}(0)=(0,\pm1/a,\boldsymbol{x}_{\perp})$,
eqs.~\eqref{eq:2.31b}--\eqref{eq:2.31d} reduce to the well-known
worldlines for uniformly accelerating observers of constant proper
acceleration $a>0$ moving on the right and left Rindler wedges:
\begin{align}
  x^{0}(\tau)=\pm\frac{1}{a}\sinh(a\tau),\quad
  x^{1}(\tau)=\pm\frac{1}{a}\cosh(a\tau),\quad
  \boldsymbol{x}_{\perp}(\tau)=\boldsymbol{x}_{\perp}. \label{eq:2.33}
\end{align}
Here the proper-time $\tau$ and group parameter $\theta$ are related
as $\theta=\pm a\tau$, where `$+$' for an observer in the right
Rindler wedge and `$-$' for an observer in the left Rindler
wedge. Notice that the proportional coefficient $a$ gives the Unruh
temperature $T=a/(2\pi)$. The conformal Killing vector flow associated
with the one-parameter subgroup \eqref{eq:2.30} is depicted in figure
\ref{figure:1a}.

Let us next move on to the parameterization of Rindler wedge
$W_{R/L}$. We wish to find a coordinate patch for $W_{R/L}$ where the
temporal coordinate is identified with the $SO(1,1)$ group parameter
(up to an overall dimensionful parameter that gives the Unruh
temperature $2\pi T$). Such coordinate system is constructed as
follows. First, let us pick up a reference point
$Y_{0}^{a}=(Y_{0}^{0},\cdots,Y_{0}^{d+1})$ on the Rindler wedge
$W_{R/L}$:
\begin{align}
  Y_{0}^{a}=(0,\pm\ell,0,\cdots,0,\ell)\in W_{R/L}. \label{eq:2.34}
\end{align}
Second, we map this point $Y_{0}^{a}$ on the Rindler wedge $W_{R/L}$
to a point $Y^{a}$ on the wedge $\overline{W}_{R/L}$ by applying the
following transformation matrix:
\begin{align}
  g=
  \begin{pmatrix}
    \cosh(t/\ell) &\sinh(t/\ell) &\\
    \sinh(t/\ell) &\cosh(t/\ell) &\\
    & &\Lambda
  \end{pmatrix}
        \in SO(1,1)\times SO(1,d-1), \label{eq:2.35}
\end{align}
where $\Lambda\in SO(1,d-1)$ is a $d\times d$ matrix that satisfies
the conditions $\eta_{ab}{\Lambda^{a}}_{c}{\Lambda^{b}}_{d}=\eta_{cd}$
($a,b,c,d\in\{2,\cdots,d+1\}$), $\det\Lambda=1$ and
${\Lambda^{d+1}}_{d+1}\geq1$. The resultant point
$Y^{a}={g^{a}}_{b}Y_{0}^{b}$ takes the following form:
\begin{align}
  Y^{a}=(\pm\ell\sinh(t/\ell),\pm\ell\cosh(t/\ell),\ell H^{a})\in\overline{W}_{R/L}, \label{eq:2.36}
\end{align}
where
$H^{a}=(H^{2},\cdots,H^{d+1})\equiv({\Lambda^{2}}_{d+1},\cdots,{\Lambda^{d+1}}_{d+1})$
is the rightmost column vector of $\Lambda\in SO(1,d-1)$ and
parameterizes the $(d-1)$-dimensional hyperbolic space
$\mathbb{H}^{d-1}$:
\begin{align}
  H\cdot H\equiv(H^{2})^{2}+\cdots+(H^{d})^{2}-(H^{d+1})^{2}=-1
  \quad\text{and}\quad
  H^{d+1}\geq1. \label{eq:2.37}
\end{align}
Notice that eq.~\eqref{eq:2.36} parameterizes the foliation
$\mathbb{H}^{1}\times\mathbb{H}^{d-1}\subset\overline{W}_{R/L}$ of the
cone. Finally, we bring back this point $Y^{a}$ on the wedge
$\overline{W}_{R/L}$ to a point $X^{a}$ on the Rindler wedge $W_{R/L}$
by multiplying the coordinate-dependent scaling factor:
\begin{align}
  X^{a}=\lambda(Y)Y^{a}
  \quad\text{with}\quad
  \lambda(Y)=\frac{\ell}{Y^{d}+Y^{d+1}}=\frac{1}{H^{d}+H^{d+1}}>0. \label{eq:2.38}
\end{align}
One can easily check that thus defined parameterization
\eqref{eq:2.38} indeed satisfies the conditions $X\cdot X=0$,
$X^{1}\pm X^{0}>0$, $X^{d+1}>0$ and $X^{d}+X^{d+1}=\ell$ and hence
covers the whole Rindler wedge $W_{R/L}$. By construction it is
obvious that in this coordinate system the time translation
$t\to t+\alpha$ is given by the $SO(1,1)$ Lorentz boost transformation
\eqref{eq:2.30} with $\theta=\alpha/\ell$. A straightforward
calculation shows that the induced metric on the Rindler wedge
$W_{R/L}$ takes the following form:
\begin{align}
  ds_{W_{R/L}}^{2}=\lambda^{2}(Y)dY\cdot dY=\frac{-dt^{2}+\ell^{2}dH\cdot dH}{(H^{d}+H^{d+1})^{2}}, \label{eq:2.39}
\end{align}
which is manifestly conformal to the hyperbolic spacetime
$\mathbb{H}^{1}\times\mathbb{H}^{d-1}$. As we will see shortly, the
metric \eqref{eq:2.39} indeed reduces to the well-known Rindler metric
if we choose an appropriate coordinate system for $\mathbb{H}^{d-1}$.

Now, let us finally move on to the study of thermal correlation
functions of $d$-dimensional CFT. As mentioned in section
\ref{section:1}, any quantum field theory restricted on the Rindler
wedge becomes thermal in the sense of the KMS condition for Wightman
functions. To see this, let us first consider the two-point function
on the Rindler wedge, which is obtained by just restricting
$G_{\Delta}(X,X^{\prime})=(-2X\cdot X^{\prime})^{-\Delta}$ on
$W_{R/L}$:
\begin{align}
  \left.G_{\Delta}(X,X^{\prime})\right|_{X,X^{\prime}\in W_{R/L}}
  =\lambda^{-\Delta}(Y)\lambda^{-\Delta}(Y^{\prime})
  \left[
  \frac{1/(2\ell^{2})}{-\cosh\left(\frac{t-t^{\prime}}{\ell}\right)-H\cdot H^{\prime}}
  \right]^{\Delta}. \label{eq:2.40}
\end{align}
The overall factors are irrelevant for us and can be removed by a
conformal transformation. Rescaling back to the hyperbolic spacetime,
$X^{a}\mapsto Y^{a}=\lambda^{-1}(Y)X^{a}$, we get the following
two-point function on $\mathbb{H}^{1}\times\mathbb{H}^{d-1}$:
\begin{align}
  \left.G_{\Delta}(Y,Y^{\prime})\right|_{Y,Y^{\prime}\in\mathbb{H}^{1}\times\mathbb{H}^{d-1}}
  =\frac{1}{(-2Y\cdot Y^{\prime})^{\Delta}}
  =\left[
  \frac{2\pi^{2}T^{2}}{-\cosh\left(2\pi T(t-t^{\prime})\right)-H\cdot H^{\prime}}
  \right]^{\Delta}, \label{eq:2.41}
\end{align}
where $T=1/(2\pi\ell)$ is the Unruh temperature. This is the two-point
function for a scalar primary operator of scaling dimension $\Delta$
in finite-temperature CFT on the hyperbolic spacetime
$\mathbb{H}^{1}\times\mathbb{H}^{d-1}$.

Several comments are in order at this stage:
\begin{itemize}
\item \emph{KMS condition.} Thermal equilibrium of quantum field
  theory is characterized by the KMS condition, which is the boundary
  condition for positive- and negative-frequency two-point Wightman
  functions in the complex time plane; see, e.g., chapter V of Haag's
  book \cite{Haag:1992hx}. In general, in finite-temperature quantum
  field theory in Lorentzian signature, the positive-frequency
  two-point Wightman function
  $G_{\Delta}^{+}(t)=\langle\mathcal{O}_{\Delta}(t)\mathcal{O}_{\Delta}(0)\rangle$
  is analytic in the strip $-\beta<\text{Im}\,t<0$ as a complex
  function of time $t$, whereas the negative-frequency two-point
  Wightman function
  $G_{\Delta}^{-}(t)=\langle\mathcal{O}_{\Delta}(0)\mathcal{O}_{\Delta}(t)\rangle$
  is analytic in the strip $0<\text{Im}\,t<\beta$,\footnote{These
    analyticity properties stem from statistical mechanics. The
    positive-frequency two-point Wightman function
    $G_{\Delta}^{+}(t)=\langle\mathcal{O}_{\Delta}(t)\mathcal{O}_{\Delta}(0)\rangle$
    corresponds to the statistical average
    $G_{\Delta}^{+}(t)=\frac{1}{Z}\text{Tr}\,(\mathrm{e}^{-\beta
      H}\mathcal{O}_{\Delta}(t)\mathcal{O}(0))=\frac{1}{Z}\sum_{E,E^{\prime}>0}\langle
    E|\mathrm{e}^{-\beta
      H}\mathrm{e}^{iHt}\mathcal{O}_{\Delta}(0)\mathrm{e}^{-iHt}|E^{\prime}\rangle\langle
    E^{\prime}|\mathcal{O}_{\Delta}(0)|E\rangle=\frac{1}{Z}\sum_{E,E^{\prime}>0}\mathrm{e}^{iE(t+i\beta)-iE^{\prime}t}|\langle
    E|\mathcal{O}_{\Delta}(0)|E^{\prime}\rangle|^{2}$, where
    $Z=\text{Tr}\,\mathrm{e}^{-\beta H}$ and $|E\rangle$ is an
    eigenstate of Hamiltonian $H$ which is assumed to be
    positive-definite. This infinite series expansion does not
    converge unless $t$ is in the strip $-\beta<\text{Im}\,t<0$, which
    is the analytic domain of $G_{\Delta}^{+}(t)$. Likewise,
    $G_{\Delta}^{-}(t)=\langle\mathcal{O}_{\Delta}(0)\mathcal{O}_{\Delta}(t)\rangle$
    is analytic in the domain $0<\text{Im}\,t<\beta$.} where
  $\beta=1/T$ is the inverse temperature and we have suppressed the
  spatial coordinates for simplicity. The KMS condition is the
  relation between the boundary values of $G_{\Delta}^{+}(t)$ and
  $G_{\Delta}^{-}(t)$ in these strips and expressed as the following
  equality:\footnote{The KMS condition stems from the cyclic property
    of the trace in statistical average. Noting the identity
    $\mathrm{e}^{-\beta H}\mathcal{O}_{\Delta}(t)\mathrm{e}^{\beta
      H}=\mathcal{O}_{\Delta}(t+i\beta)$, one easily gets
    $G_{\Delta}^{+}(t)=\frac{1}{Z}\text{Tr}\,(\mathrm{e}^{-\beta
      H}\mathcal{O}_{\Delta}(t)\mathcal{O}_{\Delta}(0))=\frac{1}{Z}\text{Tr}\,(\mathcal{O}_{\Delta}(t+i\beta)\mathrm{e}^{-\beta
      H}\mathcal{O}_{\Delta}(0))=\frac{1}{Z}\text{Tr}\,(\mathrm{e}^{-\beta
      H}\mathcal{O}_{\Delta}(0)\mathcal{O}_{\Delta}(t+i\beta))=G_{\Delta}^{-}(t+i\beta)$. Similarly,
    it follows from the trace cyclicity that
    $G_{\Delta}^{-}(t)=G_{\Delta}^{+}(t-i\beta)$.}
  \begin{align}
    G_{\Delta}^{\pm}(t)=G_{\Delta}^{\mp}(t\pm i\beta),\quad t\in\mathbb{R}. \label{eq:2.42}
  \end{align}
  In our problem, the positive- and negative-frequency two-point
  Wightman functions $G_{\Delta}^{+}(t)$ and $G_{\Delta}^{-}(t)$
  correspond to $G_{\Delta}(Y,Y^{\prime})$ and
  $G_{\Delta}(Y^{\prime},Y)$. Naively, these are the same functions
  because $G_{\Delta}(Y,Y^{\prime})$ is a symmetric function of $Y$
  and $Y^{\prime}$. One might therefore think that the KMS condition
  \eqref{eq:2.42} is trivially satisfied for \eqref{eq:2.41} since the
  hyperbolic cosine $\cosh(2\pi Tt)$ is periodic in the imaginary time
  direction with period $\beta=1/T$. There is, however, a subtle point
  because there are branch points lying exactly on the boundaries of
  the strips $-\beta<\text{Im}\,t<0$ and $0<\text{Im}\,t<\beta$. As we
  will see shortly in \eqref{eq:2.46}, $G_{\Delta}(Y,Y^{\prime})$
  possesses infinitely many branch points at $t=\pm r+in\beta$
  ($n\in\mathbb{Z}$) in the complex time plane. One way to avoid
  including such branch points is to deform the boundaries of the
  strips as depicted in figure \ref{figure:2}. An alternative (yet
  equivalent) way to avoid including the branch points is the
  following $i\epsilon$-prescriptions:
  \begin{align}
    G_{\Delta}^{\pm}(t)
    \equiv
    \begin{cases}
      \displaystyle \left[\frac{2\pi^{2}T^{2}}{-\cosh(2\pi T(t\mp
          i\epsilon))-H\cdot H^{\prime}}\right]^{\Delta}
      &\text{for $t\in\mathbb{R}$},\\[1em]
      \displaystyle \left[\frac{2\pi^{2}T^{2}}{-\cosh(2\pi T(t\pm
          i\epsilon))-H\cdot H^{\prime}}\right]^{\Delta} &\text{for
        $t\in\mathbb{R}\mp i\beta$},\label{eq:2.43}
    \end{cases}
  \end{align}
  where $\epsilon$ is a positive infinitesimal. It is now almost
  trivial to show that the Wightman functions satisfy the KMS
  condition \eqref{eq:2.42} under the $i\epsilon$-prescriptions
  \eqref{eq:2.43}. It should be pointed out here that, in terms of the
  Fourier transform
  $\Tilde{G}_{\Delta}^{\pm}(\omega)=\int_{-\infty}^{\infty}\!dt\,\mathrm{e}^{i\omega
    t}G_{\Delta}^{\pm}(t)$, the KMS condition is simply expressed as
  follows:\footnote{Here is the proof:
    $\Tilde{G}_{\Delta}^{+}(\omega)=\int_{-\infty}^{\infty}\!dt\,\mathrm{e}^{i\omega
      t}G_{\Delta}^{+}(t)=\int_{-\infty-i\beta}^{\infty-i\beta}\!dt\,\mathrm{e}^{i\omega
      t}G_{\Delta}^{+}(t)=\mathrm{e}^{\beta\omega}\int_{-\infty}^{\infty}\!dt\,\mathrm{e}^{i\omega
      t}G_{\Delta}^{+}(t-i\beta)=\mathrm{e}^{\beta\omega}\int_{-\infty}^{\infty}\!dt\,\mathrm{e}^{i\omega
      t}G_{\Delta}^{-}(t)=\mathrm{e}^{\beta\omega}\Tilde{G}_{\Delta}^{-}(\omega)$,
    where the second equality follows from the deformation of
    integration contour to the lower boundary of the strip
    $-\beta<\text{Im}\,t<0$. In the third equality we have changed the
    integration variable as $t\to t-i\beta$ and in the fourth equality
    we have used the KMS condition \eqref{eq:2.42}.}
  \begin{align}
    \Tilde{G}_{\Delta}^{+}(\omega)=\mathrm{e}^{\beta\omega}\Tilde{G}_{\Delta}^{-}(\omega).\label{eq:2.44}
  \end{align}
  In section \ref{section:4.2} and appendix \ref{section:A.2} we shall
  see that the momentum-space representations of \eqref{eq:2.43}
  indeed satisfy \eqref{eq:2.44}.

  \begin{figure}[t]
    \centering \input{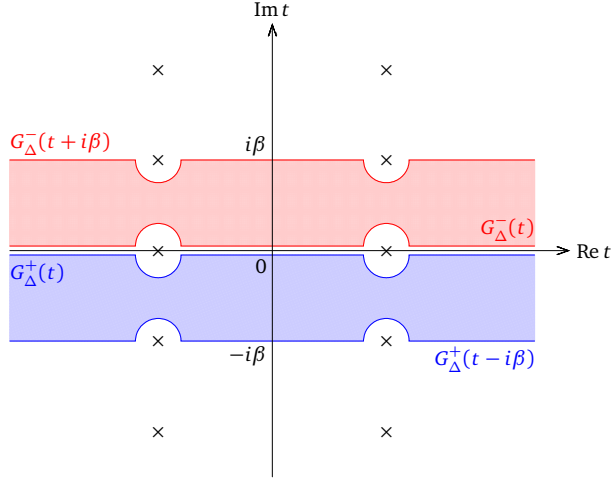}
    \caption{Analytic domains of the positive- and negative-frequency
      two-point Wightman functions. Black crosses represent the branch
      points at $t=\pm r+in\beta$, $n\in\mathbb{Z}$. The KMS
      conditions $G_{\Delta}^{\pm}(t)=G_{\Delta}^{\mp}(t\pm i\beta)$
      are the equalities between the boundary values of
      $G_{\Delta}^{+}$ and $G_{\Delta}^{-}$ in the blue and red
      strips.}
    \label{figure:2}
  \end{figure}

\item \emph{Zero-temperature limit.} The thermal two-point function
  \eqref{eq:2.41} correctly reduces to the zero-temperature correlator
  \eqref{eq:2.23} in the zero-temperature limit $T\to0$. To see this,
  let us first note that $H\cdot H^{\prime}$ just depends on the
  hyperbolic angle between the points $H^{a}$ and
  ${H^{a}}^{\prime}$. In fact, without any loss of generality, we can
  choose the point
  $Y^{a\prime}\in\mathbb{H}^{1}\times\mathbb{H}^{d-1}$ as the
  reference point $Y^{a\prime}=(0,\pm\ell,0,\cdots,0,\ell)$, which
  corresponds to $t^{\prime}=0$, $H^{2\prime}=\cdots=H^{d\prime}=0$
  and $H^{d+1\prime}=\ell$ in \eqref{eq:2.41}. Then, by parameterizing
  $H^{a}=(H^{2},\cdots,H^{d+1})$ into the hyperbolic spherical
  coordinates
  \begin{align}
    H^{i}=\sinh(r/\ell)\Omega^{i},\quad
    H^{d+1}=\cosh(r/\ell),\label{eq:2.45}
  \end{align}
  where $i\in\{2,\cdots,d\}$ and
  $(\Omega^{2})^{2}+\cdots+(\Omega^{d})^{2}=1$, we get the following
  expression for the thermal two-point function:
  \begin{align}
    \left.G_{\Delta}(Y,Y^{\prime})\right|_{Y,Y^{\prime}\in\mathbb{H}^{1}\times\mathbb{H}^{d-1}}
    &=\left[
      \frac{2\pi^{2}T^{2}}{-\cosh(2\pi Tt)+\cosh(2\pi Tr)}
      \right]^{\Delta} \nonumber\\
    &=\left[
      \frac{\pi^{2}T^{2}}{-\sinh(\pi T(t-r))\sinh(\pi T(t+r))}
      \right]^{\Delta},\label{eq:2.46}
  \end{align}
  which is quite similar to the well-known thermal two-point function
  of two-dimensional CFT that can be factorized into the left- and
  right-moving sectors. (Note, however, that $r$ ranges from $0$ to
  $\infty$ for $d\geq3$. For $d=2$, on the other hand, $r$ runs
  through $(-\infty,\infty)$ in \eqref{eq:2.45}.) Now it is obvious
  that \eqref{eq:2.46} reduces to the zero-temperature correlator in
  the zero-temperature limit,
  $G_{\Delta}(Y,Y^{\prime})\to1/(-t^{2}+r^{2})^{\Delta}$ as $T\to0$,
  where $r$ plays the role of spatial radial coordinate for $d\geq3$.

\item \emph{Rindler coordinates.} We have seen that the two-point
  function is remarkably simplified in the hyperbolic spherical
  coordinates \eqref{eq:2.45}. For practical calculations, however, it
  is more convenient to work in the following coordinate patch for
  $\mathbb{H}^{d-1}$ ($d\geq3$):
  \begin{align}
    H^{i}=\frac{x_{\perp}^{i}}{x},\quad
    H^{d}=\frac{1}{\ell}\frac{\ell^{2}-x^{2}-\boldsymbol{x}_{\perp}^{2}}{2x},\quad
    H^{d+1}=\frac{1}{\ell}\frac{\ell^{2}+x^{2}+\boldsymbol{x}_{\perp}^{2}}{2x},\label{eq:2.47}
  \end{align}
  where $i\in\{2,\cdots,d-1\}$, $x\in(0,\infty)$ and
  $\boldsymbol{x}_{\perp}\in\mathbb{R}^{d-2}$. In this
  parameterization, $X^{\mu}(=x^{\mu})$ in \eqref{eq:2.38} just
  becomes the standard Rindler coordinates
  \begin{align}
    x^{\mu}=(\pm x\sinh(t/\ell),\pm x\cosh(t/\ell),\boldsymbol{x}_{\perp}),\label{eq:2.48}
  \end{align}
  and the induced metric on the Rindler wedge $W_{R/L}$ becomes the
  well-known Rindler metric
  $ds_{W_{R/L}}^{2}=-(x/\ell)^{2}dt^{2}+dx^{2}+d\boldsymbol{x}_{\perp}^{2}$.
  A straightforward calculation shows that in this coordinate system
  the two-point function takes the following form:
  \begin{align}
    \left.G_{\Delta}(Y,Y^{\prime})\right|_{Y,Y^{\prime}\in\mathbb{H}^{1}\times\mathbb{H}^{d-1}}
    =\left[
    \frac{2\pi^{2}T^{2}}{-\cosh(2\pi T(t-t^{\prime}))+\frac{x^{2}+x^{\prime2}+|\boldsymbol{x}_{\perp}-\boldsymbol{x}_{\perp}^{\prime}|^{2}}{2xx^{\prime}}}
    \right]^{\Delta}.\label{eq:2.49}
  \end{align}
  Though it looks more complicated than \eqref{eq:2.46},
  eq.~\eqref{eq:2.49} is the best for calculating its Fourier
  transform (see appendix \ref{section:A}). The reason for this is
  that the complete orthonormal basis on $\mathbb{H}^{d-1}$ is most
  simply expressed in the coordinate system \eqref{eq:2.47}. We will
  also use the Rindler coordinates in section \ref{section:4}.
\end{itemize}

To summarize, CFT restricted on the Rindler wedge describes thermal
CFT on the hyperbolic spacetime $\mathbb{H}^{1}\times\mathbb{H}^{d-1}$
if we identify the temporal coordinate with the $SO(1,1)$ Lorentz
boost parameter (up to the dimensionful parameter $2\pi T$). It should
be noted that, though we have just focused on the scalar two-point
function for simplicity, one can also study thermal two-point
functions for generic primary tensors and/or thermal $n$-point
correlation functions for $n\geq3$. For example, the thermal two-point
function for energy-momentum tensor in Euclidean CFT on
$\mathbb{S}^{1}\times\mathbb{H}^{d-1}$ was given in
\cite{Faulkner:2015csl}. Instead of studying those thermal correlation
functions, however, let us proceed to construct thermal CFTs in which
the temporal flows are identified with other conformal Killing vector
flows associated with the one-parameter subgroup
$SO(1,1)\subset SO(2,d)$.

\subsection{Embedding the light-cone into the cone}
\label{section:2.3}
Let us next introduce the following wedge regions of the cone:
\begin{subequations}
  \begin{align}
    \overline{V}_{+}&=\{X^{a}\in\text{Cone}_{d+1}: X^{d}\pm X^{d+1}>0~~\&~~X^{0}>0\},\label{eq:2.50a}\\
    \overline{V}_{-}&=\{X^{a}\in\text{Cone}_{d+1}: X^{d}\pm X^{d+1}>0~~\&~~X^{0}<0\}.\label{eq:2.50b}
  \end{align}
\end{subequations}
Plugging the Poincar\'{e} coordinates \eqref{eq:2.3} into these, we
see that the conditions $X^{d}\pm X^{d+1}>0$ and $X^{0}>0$ are
translated into the conditions $z>0$, $-x\cdot x>0$ and
$x^{0}>0$. Hence, in terms of the Poincar\'{e} coordinate system these
wedge regions are specified as follows:
\begin{subequations}
  \begin{align}
    \overline{V}_{+}&=\{(z,x^{\mu}): z>0~~\&~~(x^{0})^{2}>\boldsymbol{x}^{2}~~\&~~x^{0}>0\},\label{eq:2.51a}\\
    \overline{V}_{-}&=\{(z,x^{\mu}): z>0~~\&~~(x^{0})^{2}>\boldsymbol{x}^{2}~~\&~~x^{0}<0\}.\label{eq:2.51b}
  \end{align}
\end{subequations}
The intersections of $\overline{V}_{\pm}$ and the Poincar\'{e} section
therefore give the standard future and past light-cones of
$d$-dimensional Minkowski spacetime:
\begin{align}
  V_{\pm}
  &=\{X^{a}\in\overline{V}_{\pm}: X^{d}+X^{d+1}=\ell\} \nonumber\\
  &=\{x^{\mu}: (x^{0})^{2}>\boldsymbol{x}^{2}~~\&~~\pm x^{0}>0\}.\label{eq:2.52}
\end{align}
Just as in the case of previous section, the wedge regions
$\overline{V}_{\pm}$ are mapped to themselves under the following
hyperbolic rotation on the $(X^{d},X^{d+1})$-plane:
\begin{align}
  g(\theta)=
  \begin{pmatrix}
    \boldsymbol{1}_{d}
    & & \\
    &\cosh\theta
    &\sinh\theta \\
    &\sinh\theta &\cosh\theta
  \end{pmatrix}
                   \in SO(1,1),\label{eq:2.53}
\end{align}
which, as we saw in eq.~\eqref{eq:2.9}, induces the following
one-parameter family of dilatation flow on the wedge
$\overline{V}_{\pm}$:
\begin{subequations}
  \begin{align}
    z(\theta)&=z\mathrm{e}^{-\theta},\label{eq:2.54a}\\
    x^{\mu}(\theta)&=x^{\mu}\mathrm{e}^{-\theta}.\label{eq:2.54b}
  \end{align}
\end{subequations}
Notice that \eqref{eq:2.54b} can also be obtained as a solution to the
following flow equation:
\begin{align}
  \frac{\partial x^{\mu}}{\partial\epsilon_{d,d+1}}=k^{\mu d,d+1}(x),\label{eq:2.55}
\end{align}
where $k^{\mu d,d+1}(x)$ is the conformal Killing vector given in
\eqref{eq:2.20d} and ${\epsilon^{d}}_{d+1}=\theta$. Note also that,
though \eqref{eq:2.53} is the isometry of the wedges
$\overline{V}_{\pm}$, it is not the isometry of the light-cones
$V_{\pm}$ because the coordinate $z$ does not remain intact under the
transformation \eqref{eq:2.53}. As mentioned in section
\ref{section:2.1}, however, any change in $z$ can be removed by
simultaneous rescaling $X^{a}\mapsto \lambda(X)X^{a}$ of the ambient
space coordinates without making any change in the ratio
$\ell X^{\mu}/(X^{d}+X^{d+1})=x^{\mu}$.

Before proceeding further, it is wise to point out here a physical
interpretation of the dilatation flow \eqref{eq:2.54b}. As discussed
in \cite{Martinetti:2002sz}, this dilatation flow coincides with the
worldline for an observer in semi-eternal uniform motion:
\begin{align}
  x^{\mu}(\tau)=v^{\mu}\tau
  \quad\text{with}\quad
  v^{\mu}=(\tfrac{1}{\sqrt{1-\boldsymbol{v}^{2}}},\tfrac{\boldsymbol{v}}{\sqrt{1-\boldsymbol{v}^{2}}}),\label{eq:2.56}
\end{align}
where $\tau$ is the proper-time for the observer and $\boldsymbol{v}$
is the spatial velocity whose modulus is less than the speed of light,
$|\boldsymbol{v}|<c=1$. The proper-time $\tau$ ranges from $0$ to
$\infty$ for an observer who is born at $x^{\mu}=0$ and then goes away
to the future infinity, whereas it lies between $-\infty$ and $0$ for
an observer who comes from the past infinity and then dies at
$x^{\mu}=0$. The causally connected regions for such observers are
nothing but the future and past light-cones $V_{\pm}$. Notice that the
proper-time $\tau$ does not exactly coincide with the dilatation flow
parameter $\theta$: they are related by the scale transformation
$\tau=\pm\ell\mathrm{e}^{-\theta}$, where `$+$' for the observer in
$V_{+}$ and `$-$' for the observer in $V_{-}$. Here at this stage
$\ell>0$ is just an arbitrary reference length scale that must be
introduced on the dimensional ground. The conformal Killing vector
flow associated with the one-parameter subgroup \eqref{eq:2.53} is
depicted in figure \ref{figure:1b}.

Now, just as we did in the previous section, we wish to find a
coordinate patch for $V_{\pm}$ where the temporal coordinate coincides
with the $SO(1,1)$ group parameter $\theta$ (but not with the proper
time $\tau$ for an observer in semi-eternal uniform
motion).\footnote{If we work in a coordinate system in which the
  temporal coordinate is identified with the proper time, we have to
  deal with time-dependent temperature in the would-be thermal
  equilibrium system \cite{Martinetti:2002sz}.} Since the construction
of such coordinate system is parallel to the case of Rindler wedge, we
here just present the result. The desired coordinate system is turned
out to be of the following form:
\begin{align}
  X^{a}=\lambda(Y)Y^{a}\in V_{\pm},\label{eq:2.57}
\end{align}
where $Y^{a}=(Y^{0},\cdots,Y^{d+1})$ parameterizes the foliation
$\mathbb{H}^{1}\times\mathbb{H}^{d-1}\subset\overline{V}_{\pm}$ of the
cone
\begin{align}
  Y^{a}=(\pm\ell H^{\mu},\ell\cosh(t/\ell),\ell\sinh(t/\ell))\in\overline{V}_{\pm},\label{eq:2.58}
\end{align}
and $\lambda(Y)$ is the coordinate-dependent scaling factor that pulls
back a point $Y^{a}$ on $\overline{V}_{\pm}$ to a point $X^{a}$ on
$V_{\pm}$:
\begin{align}
  \lambda(Y)=\frac{\ell}{Y^{d}+Y^{d+1}}=\mathrm{e}^{-t/\ell}>0.\label{eq:2.59}
\end{align}
Here $H^{\mu}=(H^{0},\cdots,H^{d-1})$ parameterizes the
$(d-1)$-dimensional unit hyperbolic space $\mathbb{H}^{d-1}$ and
satisfies the following conditions:
\begin{align}
  H\cdot H\equiv-(H^{0})^{2}+(H^{1})^{2}+\cdots+(H^{d-1})^{2}=-1
  \quad\text{and}\quad
  H^{0}\geq1.\label{eq:2.60}
\end{align}
It is easy to check that thus defined parameterization indeed
satisfies the conditions $X\cdot X=0$, $X^{d}\pm X^{d+1}>0$,
$X^{0}>0$, and $X^{d}+X^{d+1}=\ell$ and covers the whole future (past)
light-cones $V_{\pm}$. The induced metric on the light-cone $V_{\pm}$
takes the following form:
\begin{align}
  ds_{V_{\pm}}^{2}=\mathrm{e}^{-2t/\ell}\left(-dt^{2}+\ell^{2}dH\cdot dH\right),\label{eq:2.61}
\end{align}
which is again manifestly conformal to the hyperbolic spacetime
$\mathbb{H}^{1}\times\mathbb{H}^{d-1}$. The two-point function
projected out onto the light-cone $V_{\pm}$ is then given by
\begin{align}
  \left.G_{\Delta}(X,X^{\prime})\right|_{X,X^{\prime}\in V_{\pm}}
  =\lambda^{-\Delta}(Y)\lambda^{-\Delta}(Y^{\prime})
  \left[
  \frac{1/(2\ell^{2})}{-\cosh\left(\frac{t-t^{\prime}}{\ell}\right)-H\cdot H^{\prime}}
  \right]^{\Delta}.\label{eq:2.62}
\end{align}
The overall scale factors can be removed by a conformal transformation
and the resulting thermal two-point function on
$\mathbb{H}^{1}\times\mathbb{H}^{d-1}$ takes exactly the same form as
\eqref{eq:2.41}. Remaining discussions are exactly the same as those
in the previous section such that we shall not repeat here anymore.

\subsection{Embedding the diamond into the cone}
\label{section:2.4}
Let us complete our construction of thermal CFT by considering the
following wedge region of the cone:
\begin{align}
  \overline{D}=\{X^{a}\in\text{Cone}_{d+1}: X^{d}\pm X^{0}>0~~\&~~X^{d+1}>0\}.\label{eq:2.63}
\end{align}
Let us first rewrite \eqref{eq:2.63} in terms of the Poincar\'{e}
coordinate system $(z,x^{\mu})$ given in \eqref{eq:2.3}. To this end,
we first note that the condition $X^{d}\pm X^{0}>0$ (i.e.,
$X^{d}>|X^{0}|\geq0$) together with $X^{d+1}>0$ implies
$X^{d}+X^{d+1}>0$, which implies $z=\ell^{2}/(X^{d}+X^{d+1})$ must be
positive-definite on $\overline{D}$. Using this constraint on $z$, one
immediately sees that the conditions $X^{d}\pm X^{0}>0$ and
$X^{d+1}>0$ are translated into the conditions
$(x^{0}\pm\ell)^{2}-\boldsymbol{x}^{2}>0$ and
$\ell^{2}-(x^{0})^{2}+\boldsymbol{x}^{2}>0$, which can be put into a
single expression $|x^{0}|+|\boldsymbol{x}|<\ell$. Hence in the
Poincar\'{e} coordinate system the wedge region $\overline{D}$ is
described as follows:
\begin{align}
  \overline{D}=\{(z,x^{\mu}): z>0~~\&~~|x^{0}|+|\boldsymbol{x}|<\ell\}.\label{eq:2.64}
\end{align}
The intersection of the wedge $\overline{D}$ and the Poincar\'{e}
section then gives the well-known diamond (also known as the
double-cone) of $d$-dimensional Minkowski spacetime:
\begin{align}
  D
  &=\{X^{a}\in\overline{D}: X^{d}+X^{d+1}=\ell\}\nonumber\\
  &=\{x^{\mu}: |x^{0}|+|\boldsymbol{x}|<\ell\}.\label{eq:2.65}
\end{align}

Let us next consider the isometries of the wedge $\overline{D}$. Just
as in the previous sections, the wedge $\overline{D}$ is mapped to
itself under the following hyperbolic rotation on the
$(X^{0},X^{d})$-plane:
\begin{align}
  g(\theta)
  =
  \begin{pmatrix}
    \cosh\theta & & \sinh\theta	&\\
    & \boldsymbol{1}_{d-1} & &\\
    \sinh\theta & & \cosh\theta	&\\
    & & & 1
  \end{pmatrix}
          \in SO(1,1).\label{eq:2.66}
\end{align}
One can easily show that, if the initial point $X^{a}$ satisfies the
conditions $X^{d}\pm X^{0}>0$ and $X^{d+1}>0$, then the transformed
point $X^{a}(\theta)={g^{a}}_{b}X^{b}$ also satisfies the conditions
$X^{d}(\theta)\pm X^{0}(\theta)>0$ and $X^{d+1}(\theta)>0$ and hence
\eqref{eq:2.66} gives a globally well-defined transformation on the
wedge $\overline{D}$. Substituting $X^{a}(\theta)$ into
$z(\theta)=\ell^{2}/(X^{d}(\theta)+X^{d+1}(\theta))$ and
$x^{\mu}(\theta)=\ell
X^{\mu}(\theta)/(X^{d}(\theta)+X^{d+1}(\theta))$, one can also easily
show that the one-parameter subgroup \eqref{eq:2.66} induces the
following flow on the wedge $\overline{D}$:
\begin{subequations}
  \begin{align}
    z(\theta)
    &=\mathrm{e}^{-\varphi}\frac{z}{1-2(b\cdot x)+(b\cdot b)(x\cdot x)},\label{eq:2.67a}\\
    x^{\mu}(\theta)
    &=\mathrm{e}^{-\varphi}\frac{x^{\mu}-b^{\mu}(x\cdot x)}{1-2(b\cdot x)+(b\cdot b)(x\cdot x)}+a^{\mu},\label{eq:2.67b}
  \end{align}
\end{subequations}
where
\begin{align}
  a^{\mu}=(\ell\tanh(\tfrac{\theta}{2}),0,\cdots,0), \quad
  b^{\mu}=(\tfrac{1}{\ell}\tanh(\tfrac{\theta}{2}),0,\cdots,0), \quad
  \varphi=2\log\cosh(\tfrac{\theta}{2}).\label{eq:2.68}
\end{align}
As is evident from these expressions, this flow---the diamond
flow---can also be obtained by the following three successive
transformations: i) special conformal transformation \eqref{eq:2.13};
ii) dilatation \eqref{eq:2.9}; and iii) translation
\eqref{eq:2.11}. In fact, the $SO(1,1)$ matrix \eqref{eq:2.66} admits
the following decomposition:\footnote{The decomposition
  \eqref{eq:2.69} can also be written as follows:
  \begin{align}
    \exp\left[i\frac{\theta}{2}\left(\ell P^{0}-\frac{1}{\ell}K^{0}\right)\right]
    =\exp\left[-i\ell\tanh\left(\frac{\theta}{2}\right)P^{0}\right]
    \exp\left[i2\log\cosh\left(\frac{\theta}{2}\right)D\right]
    \exp\left[i\frac{1}{\ell}\tanh\left(\frac{\theta}{2}\right)K^{0}\right], \nonumber
  \end{align}
  where $D$, $P^{0}$, and $K^{0}$ are the $(d+2)\times(d+2)$-matrix
  representations given in \eqref{eq:3.26}. Notice that the set of
  generators $\{D,P^{0},K^{0}\}$ forms the one-dimensional conformal
  algebra $\mathfrak{so}(2,1)\subset\mathfrak{so}(2,d)$.}
\begin{align}
  \begin{pmatrix}
    \boldsymbol{1}_{d} &\tfrac{a^{\mu}}{\ell}
    &\tfrac{a^{\mu}}{\ell} \\[1ex]
    -\tfrac{a_{\nu}}{\ell} &1-\tfrac{1}{2}\tfrac{a\cdot a}{\ell^{2}}
    &-\tfrac{1}{2}\tfrac{a\cdot a}{\ell^{2}} \\[1ex]
    \tfrac{a_{\nu}}{\ell} &\tfrac{1}{2}\tfrac{a\cdot a}{\ell^{2}}
    &1+\tfrac{1}{2}\tfrac{a\cdot a}{\ell^{2}}
  \end{pmatrix}
      \begin{pmatrix}
        \boldsymbol{1}_{d}
        & & \\[1ex]
        &\cosh\varphi
        &\sinh\varphi \\[1ex]
        &\sinh\varphi &\cosh\varphi
      \end{pmatrix}
                        \begin{pmatrix}
                          \boldsymbol{1}_{d} &\ell b^{\mu}
                          &-\ell b^{\mu} \\[1ex]
                          -\ell b_{\nu} &1-\tfrac{\ell^{2}}{2}b\cdot b
                          &\tfrac{\ell^{2}}{2}b\cdot b \\[1ex]
                          -\ell b_{\nu} &-\tfrac{\ell^{2}}{2}b\cdot b
                          &1+\tfrac{\ell^{2}}{2}b\cdot b
                        \end{pmatrix}.\label{eq:2.69}
\end{align}
Multiplying these matrices on a point $X^{a}\in\overline{D}$ induces
the composition of conformal transformations \eqref{eq:2.13},
\eqref{eq:2.9} and \eqref{eq:2.11} and results in the diamond flow
\eqref{eq:2.67a} and \eqref{eq:2.67b}. It is not difficult to show
that \eqref{eq:2.67b} can also be obtained as a solution to the
following flow equation:
\begin{align}
  \frac{\partial x^{\mu}}{\partial\epsilon_{d0}}=k^{\mu d0}(x),\label{eq:2.70}
\end{align}
where $k^{\mu d0}(x)=-k^{\mu 0d}(x)$ is the conformal Killing vector
given in \eqref{eq:2.20b} and ${\epsilon^{d}}_{0}=\theta$. Namely, the
diamond flow is nothing but the conformal Killing vector flow
associated with $k^{\mu d0}$. Notice that the denominator
$1-2(b\cdot x)+(b\cdot b)(x\cdot x)$ does not vanish provided
$x^{\mu}=(x^{0},\boldsymbol{x})$ satisfies the condition
$|x^{0}|+|\boldsymbol{x}|<\ell$. Note also that
$(z(\theta),x^{\mu}(\theta))\to(0,\pm\ell,\boldsymbol{0})$ as
$\theta\to\pm\infty$.

Now we wish to give a physical interpretation of the diamond flow. As
discussed in \cite{Martinetti:2002sz}, the diamond flow coincides with
the worldline for a uniformly accelerating observer with finite
lifetime. To see this, let us first choose the initial point as
$x^{\mu}(0)=(x^{0},\boldsymbol{x})=(0,\ell\tfrac{a\ell\boldsymbol{\Omega}}{1+\sqrt{1+a^{2}\ell^{2}}})\in
D$, where $\boldsymbol{\Omega}$ is a constant unit vector that
satisfies $\boldsymbol{\Omega}\cdot\boldsymbol{\Omega}=1$ and
$a\in(-\infty,\infty)$ is another dimensionful parameter that plays
the role of constant proper acceleration. Substituting these into
\eqref{eq:2.67b}, we see that the diamond flow takes the following
simple forms:
\begin{align}
  x^{0}(\theta)=\ell\frac{\sinh\theta}{\cosh\theta+\sqrt{1+a^{2}\ell^{2}}}, \quad
  \boldsymbol{x}(\theta)=\ell\frac{a\ell\boldsymbol{\Omega}}{\cosh\theta+\sqrt{1+a^{2}\ell^{2}}}.\label{eq:2.71}
\end{align}
An important observation here is that \eqref{eq:2.71} satisfies the
following equation:
\begin{align}
  -\left(x^{0}(\theta)\right)^{2}+\left(\boldsymbol{x}(\theta)-\tfrac{1}{a}\sqrt{1+a^{2}\ell^{2}}\boldsymbol{\Omega}\right)^{2}
  =\frac{1}{a^{2}},\label{eq:2.72}
\end{align}
which implies that $x^{0}(\theta)$ and $\boldsymbol{x}(\theta)$ can be
put into the following alternative equivalent forms:
\begin{align}
  x^{0}(\tau)=\frac{1}{a}\sinh(a\tau), \quad
  \boldsymbol{x}(\tau)=\frac{1}{a}\left(\sqrt{1+a^{2}\ell^{2}}-\cosh(a\tau)\right)\boldsymbol{\Omega}.\label{eq:2.73}
\end{align}
This is nothing but the worldline for a uniformly accelerating
observer moving on the $(x^{0},\boldsymbol{\Omega})$-plane, where the
proper time $\tau$ and group parameter $\theta$ are related as
follows:
\begin{align}
  \tau
  =\frac{1}{a}\log
  \left[\frac{\cosh\left(\frac{\theta+a\tau_{0}}{2}\right)}{\cosh\left(\frac{\theta-a\tau_{0}}{2}\right)}\right]
  \in(-\tau_{0},\tau_{0}),\label{eq:2.74}
\end{align}
with $\tau_{0}=(1/a)\mathrm{arcsinh}(a\ell)>0$ being half the proper
lifetime of uniformly accelerating observer. Hence, up to this
parameterization difference, the diamond flow coincides with the
worldline for a uniformly accelerating observer with finite lifetime
$2\tau_{0}$. The conformal Killing vector flow associated with the
one-parameter subgroup \eqref{eq:2.66} is depicted in figure
\ref{figure:1c}.

Let us next move on to construct a coordinate patch for the diamond
$D$ where the temporal coordinate is identifies with the $SO(1,1)$
group parameter.\footnote{Again, if we work in a coordinate system in
  which the temporal coordinate is identified with the proper time, we
  have to deal with time-dependent temperature
  \cite{Martinetti:2002sz}.} Such coordinate system is turned out to
be of the form:
\begin{align}
  X^{a}=\lambda(Y)Y^{a}\in D,\label{eq:2.75}
\end{align}
where $Y^{a}=(Y^{0},\cdots,Y^{d+1})$ is a parameterization for the
foliation $\mathbb{H}^{1}\times\mathbb{H}^{d-1}\subset\overline{D}$
given by
\begin{align}
  Y^{a}=(\ell\sinh(t/\ell),\ell H^{i},\ell\cosh(t/\ell),\ell H^{d+1})\in\overline{D},\label{eq:2.76}
\end{align}
and $\lambda(Y)$ is a coordinate-dependent scaling factor that pulls
back a point $Y^{a}$ on the wedge $\overline{D}$ to a point $X^{a}$ on
the diamond $D$ given as follows:
\begin{align}
  \lambda(Y)=\frac{\ell}{Y^{d}+Y^{d+1}}=\frac{1}{\cosh(t/\ell)+H^{d+1}}>0.\label{eq:2.77}
\end{align}
Here $H^{a}=(H^{1},\cdots,H^{d-1},H^{d+1})$ parameterizes the
$(d-1)$-dimensional hyperbolic space $\mathbb{H}^{d-1}$ and is subject
to the following conditions:
\begin{align}
  H\cdot H\equiv(H^{1})^{2}+\cdots+(H^{d-1})^{2}-(H^{d+1})^{2}=-1
  \quad\text{and}\quad
  H^{d+1}\geq1.\label{eq:2.78}
\end{align}
The induced metric on the diamond takes the following form:
\begin{align}
  ds_{D}^{2}
  =\frac{-dt^{2}+\ell^{2}dH\cdot dH}{(\cosh(t/\ell)+H^{d+1})^{2}},\label{eq:2.79}
\end{align}
which is again conformal to the hyperbolic spacetime
$\mathbb{H}^{1}\times\mathbb{H}^{d-1}$. The two-point function
projected out onto the diamond then takes the following form:
\begin{align}
  \left.G_{\Delta}(X,X^{\prime})\right|_{X,X^{\prime}\in D}
  =\lambda^{-\Delta}(Y)\lambda^{-\Delta}(Y^{\prime})
  \left[
  \frac{1/(2\ell^{2})}{-\cosh\left(\frac{t-t^{\prime}}{\ell}\right)-H\cdot H^{\prime}}
  \right]^{\Delta}.\label{eq:2.80}
\end{align}
The overall factors can be removed by a conformal transformation,
after which \eqref{eq:2.80} describes the thermal two-point function
on the hyperbolic spacetime
$\mathbb{H}^{1}\times\mathbb{H}^{d-1}$. The remaining discussions are
exactly the same as those presented in section \ref{section:2.2} such
that we will stop here.

\subsection{A few remarks on the one-dimensional case}
\label{section:2.5}
Let us close our construction of thermal CFTs in $d$ dimensions by
making a few remarks on the case $d=1$.

\begin{figure}[t]
  \centering \subfigure[Light-cones =
  half-lines.]{
\xdefinecolor{rgb_000000}{rgb}{0,0,0}%
\xdefinecolor{rgb_0000ff}{rgb}{0,0,1}%
\xdefinecolor{rgb_ff0000}{rgb}{1,0,0}%
\setlength{\unitlength}{1cm}%
\begin{picture}(4.5,4.5)(0,0)%
\path(2.25,4.08971)(2.25,4.10294)
\path(2.20167,3.95796)(2.20191,3.95796)
\path(2.29809,3.95796)(2.29833,3.95796)
\path(2.20607,3.97114)(2.20653,3.97114)
\path(2.29347,3.97114)(2.29393,3.97114)
\path(2.21046,3.98432)(2.21115,3.98432)
\path(2.28885,3.98432)(2.28954,3.98432)
\path(2.21485,3.9975)(2.21578,3.9975)
\path(2.28422,3.9975)(2.28515,3.9975)
\path(2.21925,4.01068)(2.2204,4.01068)
\path(2.2796,4.01068)(2.28075,4.01068)
\path(2.22364,4.02386)(2.22503,4.02386)
\path(2.27497,4.02386)(2.27636,4.02386)
\path(2.22803,4.03704)(2.22965,4.03704)
\path(2.27035,4.03704)(2.27197,4.03704)
\path(2.23243,4.05022)(2.23428,4.05022)
\path(2.26572,4.05022)(2.26757,4.05022)
\path(2.23682,4.0634)(2.2389,4.0634)
\path(2.2611,4.0634)(2.26318,4.0634)
\path(2.24121,4.07658)(2.24353,4.07658)
\path(2.25647,4.07658)(2.25879,4.07658)
\path(2.24561,4.08976)(2.24815,4.08976)
\path(2.25185,4.08976)(2.25439,4.08976)
\path(2.28954,3.98235)(2.28954,3.98432)
\path(2.27636,4.01991)(2.27636,4.02386)
\path(2.26318,4.05747)(2.26318,4.0634)
\path(2.25,4.09503)(2.25,4.10294)
\path(2.23682,4.05747)(2.23682,4.0634)
\path(2.22364,4.01991)(2.22364,4.02386)
\path(2.21046,3.98235)(2.21046,3.98432)
\path(2.25,4.09503)(2.30272,3.94478)(2.25,4.10294)(2.19728,3.94478)(2.25,4.09503)
\allinethickness{0.8pt}%
\color{rgb_0000ff}%
\path(2.25,4.08971)(2.25,3.11029)
\allinethickness{0.0140584cm}%
\path(2.28766,3.22326)(2.28034,3.24209)
\path(2.28138,3.20444)(2.26674,3.24209)
\path(2.2751,3.18561)(2.25314,3.24209)
\path(2.26883,3.16678)(2.23955,3.24209)
\path(2.26255,3.14795)(2.22595,3.24209)
\path(2.25628,3.12912)(2.21235,3.24209)
\path(2.24436,3.24209)(2.21046,3.22891)
\path(2.28265,3.24209)(2.21485,3.21573)
\path(2.28991,3.23003)(2.21925,3.20255)
\path(2.28421,3.21292)(2.22364,3.18937)
\path(2.27851,3.19582)(2.22803,3.17619)
\path(2.27281,3.17871)(2.23243,3.16301)
\path(2.2671,3.16161)(2.23682,3.14983)
\path(2.2614,3.1445)(2.24121,3.13665)
\path(2.2557,3.1274)(2.24561,3.12347)
\allinethickness{0.8pt}%
\path(2.25,3.24209)(2.20607,3.24209)(2.25,3.11029)(2.29393,3.24209)(2.25,3.24209)
\path(2.25,3.11029)(2.25,2.11765)
\color{rgb_ff0000}%
\path(2.25,0.132353)(2.25,1.125)
\allinethickness{0.0140584cm}%
\path(2.25439,1.11182)(2.24174,1.10023)
\path(2.25879,1.09864)(2.23349,1.07547)
\path(2.26318,1.08546)(2.22523,1.0507)
\path(2.26757,1.07228)(2.21698,1.02593)
\path(2.27197,1.0591)(2.20872,1.00116)
\path(2.27636,1.04592)(2.21881,0.993203)
\path(2.28075,1.03274)(2.23759,0.993203)
\path(2.28515,1.01956)(2.25637,0.993203)
\path(2.28954,1.00638)(2.27515,0.993203)
\path(2.21095,1.00785)(2.22436,0.993203)
\path(2.21583,1.02249)(2.24266,0.993203)
\path(2.22071,1.03714)(2.26096,0.993203)
\path(2.22559,1.05178)(2.27925,0.993203)
\path(2.23047,1.06642)(2.29186,0.999411)
\path(2.23536,1.08107)(2.2814,1.03081)
\path(2.24024,1.09571)(2.27093,1.06221)
\path(2.24512,1.11036)(2.26047,1.0936)
\allinethickness{0.8pt}%
\path(2.25,0.993203)(2.29393,0.993203)(2.25,1.125)
  (2.20607,0.993203)(2.25,0.993203)
\path(2.25,1.125)(2.25,2.11765)
\put(2.25,4.20838){\makebox(0,0)[b]{\hbox{\color{rgb_000000}\scriptsize $x^{0}$}}}
\put(2.17971,2.01221){\makebox(0,0)[r]{\hbox{\color{rgb_000000}\scriptsize $0$}}}
\put(2.14456,3.11029){\makebox(0,0)[r]{\hbox{\color{rgb_000000}\scriptsize \color{blue}future light-cone$\left\{\parbox[c][1.8cm][c]{1pt}{}\right.$}}}
\put(2.35544,1.125){\makebox(0,0)[l]{\hbox{\color{rgb_000000}\scriptsize \color{red}$\left.\parbox[c][1.8cm][c]{1pt}{}\right\}$past light-cone}}}
\put(2.35544,3.18059){\makebox(0,0)[l]{\hbox{\color{rgb_000000}\scriptsize \color{blue}$t$}}}
\put(2.14456,1.05471){\makebox(0,0)[r]{\hbox{\color{rgb_000000}\scriptsize \color{red}$t$}}}
\end{picture}%
} \subfigure[Diamond =
  interval.]{
\xdefinecolor{rgb_000000}{rgb}{0,0,0}%
\xdefinecolor{rgb_ff0000}{rgb}{1,0,0}%
\setlength{\unitlength}{1cm}%
\begin{picture}(4.5,4.5)(0,0)%
\path(2.25,3.44118)(2.25,4.10294)
\path(2.20167,3.95796)(2.20191,3.95796)
\path(2.29809,3.95796)(2.29833,3.95796)
\path(2.20607,3.97114)(2.20653,3.97114)
\path(2.29347,3.97114)(2.29393,3.97114)
\path(2.21046,3.98432)(2.21115,3.98432)
\path(2.28885,3.98432)(2.28954,3.98432)
\path(2.21485,3.9975)(2.21578,3.9975)
\path(2.28422,3.9975)(2.28515,3.9975)
\path(2.21925,4.01068)(2.2204,4.01068)
\path(2.2796,4.01068)(2.28075,4.01068)
\path(2.22364,4.02386)(2.22503,4.02386)
\path(2.27497,4.02386)(2.27636,4.02386)
\path(2.22803,4.03704)(2.22965,4.03704)
\path(2.27035,4.03704)(2.27197,4.03704)
\path(2.23243,4.05022)(2.23428,4.05022)
\path(2.26572,4.05022)(2.26757,4.05022)
\path(2.23682,4.0634)(2.2389,4.0634)
\path(2.2611,4.0634)(2.26318,4.0634)
\path(2.24121,4.07658)(2.24353,4.07658)
\path(2.25647,4.07658)(2.25879,4.07658)
\path(2.24561,4.08976)(2.24815,4.08976)
\path(2.25185,4.08976)(2.25439,4.08976)
\path(2.28954,3.98235)(2.28954,3.98432)
\path(2.27636,4.01991)(2.27636,4.02386)
\path(2.26318,4.05747)(2.26318,4.0634)
\path(2.25,4.09503)(2.25,4.10294)
\path(2.23682,4.05747)(2.23682,4.0634)
\path(2.22364,4.01991)(2.22364,4.02386)
\path(2.21046,3.98235)(2.21046,3.98432)
\path(2.25,4.09503)(2.30272,3.94478)(2.25,4.10294)(2.19728,3.94478)(2.25,4.09503)
\path(2.25,0.132353)(2.25,0.794118)
\allinethickness{0.8pt}%
\color{rgb_ff0000}%
\path(2.25,0.794118)(2.25,1.45588)
\allinethickness{0.0140584cm}%
\path(2.28765,1.32408)(2.24372,1.43705)
\path(2.27405,1.32408)(2.23745,1.41823)
\path(2.26045,1.32408)(2.23117,1.3994)
\path(2.24686,1.32408)(2.2249,1.38057)
\path(2.23326,1.32408)(2.21862,1.36174)
\path(2.21966,1.32408)(2.21234,1.34291)
\path(2.25439,1.4427)(2.2443,1.43878)
\path(2.25879,1.42952)(2.2386,1.42167)
\path(2.26318,1.41634)(2.2329,1.40457)
\path(2.26757,1.40316)(2.22719,1.38746)
\path(2.27197,1.38998)(2.22149,1.37036)
\path(2.27636,1.3768)(2.21579,1.35325)
\path(2.28075,1.36362)(2.21009,1.33615)
\path(2.28515,1.35044)(2.21735,1.32408)
\path(2.28954,1.33726)(2.25564,1.32408)
\allinethickness{0.8pt}%
\path(2.25,1.32408)(2.29393,1.32408)(2.25,1.45588)(2.20607,1.32408)(2.25,1.32408)
\path(2.25,1.45588)(2.25,2.77941)
\allinethickness{0.0140584cm}%
\path(2.25439,2.76623)(2.24174,2.75464)
\path(2.25879,2.75305)(2.23349,2.72988)
\path(2.26318,2.73987)(2.22523,2.70511)
\path(2.26757,2.72669)(2.21698,2.68034)
\path(2.27197,2.71351)(2.20872,2.65558)
\path(2.27636,2.70033)(2.21881,2.64761)
\path(2.28075,2.68715)(2.23759,2.64761)
\path(2.28515,2.67397)(2.25637,2.64761)
\path(2.28954,2.66079)(2.27515,2.64761)
\path(2.21095,2.66226)(2.22436,2.64761)
\path(2.21583,2.6769)(2.24266,2.64761)
\path(2.22071,2.69155)(2.26096,2.64761)
\path(2.22559,2.70619)(2.27925,2.64761)
\path(2.23047,2.72084)(2.29186,2.65382)
\path(2.23536,2.73548)(2.2814,2.68522)
\path(2.24024,2.75012)(2.27093,2.71662)
\path(2.24512,2.76477)(2.26047,2.74801)
\allinethickness{0.8pt}%
\path(2.25,2.64761)(2.29393,2.64761)(2.25,2.77941)(2.20607,2.64761)(2.25,2.64761)
\path(2.25,2.77941)(2.25,3.44118)
\put(2.25,4.20838){\makebox(0,0)[b]{\hbox{\color{rgb_000000}\scriptsize $x^{0}$}}}
\put(2.14456,2.11765){\makebox(0,0)[r]{\hbox{\color{rgb_000000}\scriptsize $0$}}}
\put(2.14456,3.44118){\makebox(0,0)[r]{\hbox{\color{rgb_000000}\scriptsize $\ell$}}}
\put(2.14456,0.794118){\makebox(0,0)[r]{\hbox{\color{rgb_000000}\scriptsize $-\ell$}}}
\put(2.35544,2.11765){\makebox(0,0)[l]{\hbox{\color{rgb_000000}\scriptsize \color{red}$\left.\parbox[c][2.6cm][c]{1pt}{}\right\}$diamond}}}
\put(2.14456,2.70912){\makebox(0,0)[r]{\hbox{\color{rgb_000000}\scriptsize \color{red}$t$}}}
\end{picture}%
}
  \caption{Light-cones and diamond in $(0+1)$-dimensional
    spacetime. $x^{0}$ and the $SO(1,1)$ group parameter
    $t\in(-\infty,\infty)$ are related as (a)
    $x^{0}=\pm\ell\mathrm{e}^{-t/\ell}$ and (b)
    $x^{0}=\ell\tanh(t/(2\ell))$.}
  \label{figure:3}
\end{figure}
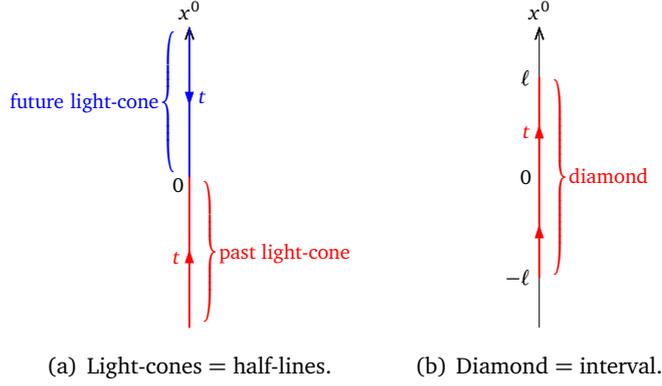

In $(0+1)$-dimensional spacetime, there is no immediate analogue of
Rindler wedge because of the absence of spatial directions. However,
we can still introduce the light-cones and diamond as half-lines and
interval of the time axis, and CFT restricted on these regions become
thermal as well under the identification of the temporal coordinates
with the $SO(1,1)\subset SO(2,1)$ group parameters; see figure
\ref{figure:3}. The embedding space formalism for one-dimensional CFT
can be developed equally well and the resultant thermal CFT turns out
to reside on the $(0+1)$-dimensional hyperbolic space
$\mathbb{H}^{1}$. It is a straightforward exercise to show that the
thermal two-point function for a scalar primary operator on
$\mathbb{H}^{1}$ takes the following form:
\begin{align}
  \left[\frac{\pi^{2}T^{2}}{-\sinh^{2}(\pi T(t-t^{\prime}))}\right]^{\Delta},\label{eq:2.81}
\end{align}
which of course satisfies the KMS condition (under the appropriate
$i\epsilon$-prescription). Notice that \eqref{eq:2.81} exactly
coincides with the result of conformal quantum mechanics
\cite{Nakayama:2011qh}.

\section{Intertwining Operator in \texorpdfstring{CFT$_{d}$}{CFTd}}
\label{section:3}
So far we have constructed thermal CFTs in $d$ dimensions by using the
embedding space formalism. Let us now turn to the
representation-theoretic aspect of conformal two-point functions: the
intertwining operators. Though it had been studied in the 1970s
\cite{Koller:1974ut,Dobrev:1977,Fradkin:1978pp,Todorov:1978rf} that
two-point functions of CFT are nothing but the integral kernels of
intertwining operators for two equivalent representations of conformal
algebra $\mathfrak{so}(2,d)$, this fact is hardly ever acknowledged
except to experts. We will see in the rest of the paper that the
intertwining operator and the resulting intertwining relations provide
powerful constraints on the structure of momentum-space thermal
two-point functions. Before embarking on this, however, in this
section we first review the basics of intertwining operators in
$d$-dimensional CFT and prepare necessary ingredients for our
analysis. Then we show as an application that, for the case of
zero-temperature CFT, the intertwining relations reduce to the
well-known conformal Ward--Takahashi identities for momentum-space
two-point functions.

We emphasize that the presentation of this section is meant to be
rather sketchy, yet enough for the purpose of this paper, leaving
mathematical rigors to the existing literature. For more details of
intertwining operators in CFT, we refer to
\cite{Koller:1974ut,Dobrev:1977,Fradkin:1978pp,Todorov:1978rf} (see
also \cite{Fradkin:1996} for more recent expositions.) We also note
that in this section we will use the standard $d$-dimensional approach
to $d$-dimensional CFT: we will not return to the embedding space
formalism in the remainder of the paper, though it may be possible to
reformulate everything in the subsequent sections in the
$(d+2)$-dimensional language.

Throughout this section we just focus on scalar two-point functions
for simplicity.

\subsection{Intertwining kernel = two-point function}
\label{section:3.1}
To begin with, let $x\mapsto x_{g}$ be a generic conformal
transformation associated with a group element $g\in SO(2,d)$. Let
$\mathcal{O}_{\Delta}(x)$ be a scalar primary operator of scaling
dimension $\Delta$ that satisfies the following transformation law:
\begin{align}
  U(g)\mathcal{O}_{\Delta}(x)U^{-1}(g)
  =\left|\frac{\partial x_{g}}{\partial x}\right|^{\Delta/d}
  \mathcal{O}_{\Delta}(x_{g}), \label{eq:3.1}
\end{align}
where $U(g)$ is a unitary representation of $g\in SO(2,d)$ and
$|\partial x_{g}/\partial x|$ stands for the Jacobian of conformal
transformation $x\mapsto x_{g}=x_{g}(x)$. Let $|\Omega\rangle$ be a
conformally invariant Minkowski vacuum satisfying
$U(g)|\Omega\rangle=|\Omega\rangle$ for any $g\in SO(2,d)$. Then the
positive-frequency two-point Wightman function $G_{\Delta}(x,y)$, for
example, is given by
\begin{align}
  G_{\Delta}(x,y)
  =\langle\Omega|\mathcal{O}_{\Delta}(x)\mathcal{O}_{\Delta}(y)|\Omega\rangle, \label{eq:3.2}
\end{align}
which satisfies the following identity (conformal Ward--Takahashi
identity):
\begin{align}
  G_{\Delta}(x,y)
  =\left|\frac{\partial x_{g}}{\partial x}\right|^{\Delta/d}
  \left|\frac{\partial y_{g}}{\partial y}\right|^{\Delta/d}
  G_{\Delta}(x_{g},y_{g}). \label{eq:3.3}
\end{align}
We note that, though \eqref{eq:3.2} is the simplest example, for the
following discussion $G_{\Delta}(x,y)$ is not necessarily to be the
positive-frequency two-point Wightman function. The following
discussion can be applied to any other two-point functions that
satisfy the conformal Ward--Takahashi identity.

Now, let us next introduce an intertwining operator $G_{\Delta}$,
which is defined through the following integral transform whose kernel
is the two-point function:
\begin{align}
  G_{\Delta}: \mathcal{O}_{d-\Delta}(x)\mapsto
  (G_{\Delta}\mathcal{O}_{d-\Delta})(x)
  :=\int\!d^{d}y\,G_{\Delta}(x,y)\mathcal{O}_{d-\Delta}(y). \label{eq:3.4}
\end{align}
This operator $G_{\Delta}$ maps a primary operator of scaling
dimension $d-\Delta$, which is called the ``shadow operator''
\cite{Ferrara:1972uq}, to another primary operator of scaling
dimension $\Delta$. Indeed, as easily checked, the operator
$(G_{\Delta}\mathcal{O}_{d-\Delta})(x)$ satisfies the following
transformation law:
\begin{align}
  U(g)(G_{\Delta}\mathcal{O}_{d-\Delta})(x)U^{-1}(g)
  =\left|\frac{\partial x_{g}}{\partial x}\right|^{\Delta/d}
  (G_{\Delta}\mathcal{O}_{d-\Delta})(x_{g}). \label{eq:3.5}
\end{align}
Hence $(G_{\Delta}\mathcal{O}_{d-\Delta})(x)$ is indeed a primary
operator of scaling dimension $\Delta$. The proof of \eqref{eq:3.5} is
straightforward and goes as follows:
\begin{align}
  U(g)(G_{\Delta}\mathcal{O}_{d-\Delta})(x)U^{-1}(g)
  &=U(g)\left(\int\!d^{d}y\,G_{\Delta}(x,y)\mathcal{O}_{d-\Delta}(y)\right)U^{-1}(g) \nonumber\\
  &=\int\!d^{d}y\,G_{\Delta}(x,y)U(g)\mathcal{O}_{d-\Delta}(y)U^{-1}(g) \nonumber\\
  &=\int\!d^{d}y\,
    \left|\frac{\partial x_{g}}{\partial x}\right|^{\Delta/d}
    \left|\frac{\partial y_{g}}{\partial y}\right|^{\Delta/d}
    G_{\Delta}(x_{g},y_{g})
    \left|\frac{\partial y_{g}}{\partial y}\right|^{(d-\Delta)/d}
    \mathcal{O}_{\Delta}(y_{g}) \nonumber\\
  &=\left|\frac{\partial x_{g}}{\partial x}\right|^{\Delta/d}
    \int\!d^{d}y\,
    \left|\frac{\partial y_{g}}{\partial y}\right|
    G_{\Delta}(x_{g},y_{g})
    \mathcal{O}_{\Delta}(y_{g}) \nonumber\\
  &=\left|\frac{\partial x_{g}}{\partial x}\right|^{\Delta/d}
    \int\!d^{d}y_{g}\,
    G_{\Delta}(x_{g},y_{g})
    \mathcal{O}_{\Delta}(y_{g}) \nonumber\\
  &=\left|\frac{\partial x_{g}}{\partial x}\right|^{\Delta/d}
    (G_{\Delta}\mathcal{O}_{d-\Delta})(x_{g}), \label{eq:3.6}
\end{align}
where the third equality follows from \eqref{eq:3.1} and
\eqref{eq:3.3}. The fifth equality follows from the change of
integration variable,
$\int\!d^{d}y|\partial y_{g}/\partial y|=\int\!d^{d}y_{g}$. As is
evident from this proof,
$U(g)(G_{\Delta}\mathcal{O}_{d-\Delta})(x)U^{-1}(g)$ and
$(G_{\Delta}U(g)\mathcal{O}_{d-\Delta}U^{-1}(g))(x)$ give the same
result. Hence the intertwining operator $G_{\Delta}$ should satisfy
the following operator identity (intertwining relation):
\begin{align}
  U(g)G_{\Delta}=G_{\Delta}U(g), \quad \forall g\in SO(2,d). \label{eq:3.7}
\end{align}
For practical calculations, however, it is more convenient to
introduce the infinitesimal form of \eqref{eq:3.7}. To this end, let
us next consider an infinitesimal conformal transformation associated
with a group element
${g^{a}}_{b}={\delta^{a}}_{b}+{\epsilon^{a}}_{b}$, where
$\epsilon_{ab}=-\epsilon_{ba}$ are infinitesimal parameters. The
conformal transformation $x\mapsto x_{1+\epsilon}=x_{1+\epsilon}(x)$
associated with such a group element $g=1+\epsilon$ can be
Taylor-expanded as follows:
\begin{align}
  x_{1+\epsilon}^{\mu}
  =x^{\mu}+\frac{1}{2}\epsilon_{ab}k^{\mu ab}(x)
  \quad\text{with}\quad
  k^{\mu ab}(x)
  :=\left.
  \frac{\partial x_{1+\epsilon}^{\mu}}{\partial\epsilon_{ab}}
  \right|_{\epsilon_{ab}=0}, \label{eq:3.8}
\end{align}
where $k^{\mu ab}(x)=-k^{\mu ba}(x)$ are conformal Killing vectors
which of course coincide with those given in
\eqref{eq:2.20a}--\eqref{eq:2.20d}. Similarly, the unitary operator
$U(g)$ associated with $g=1+\epsilon$ can also be Taylor-expanded:
\begin{align}
  U(1+\epsilon)=1+\frac{i}{2}\epsilon_{ab}J^{ab}, \label{eq:3.9}
\end{align}
where $J^{ab}=-J^{ba}$ are hermitian generators of the Lie algebra
$\mathfrak{so}(2,d)$ satisfying the following commutation relations:
\begin{align}
  [J^{ab},J^{cd}]=i(\eta^{ac}J^{bd}-\eta^{ad}J^{bc}-\eta^{bc}J^{ad}+\eta^{bd}J^{ac}). \label{eq:3.10}
\end{align}
Substituting \eqref{eq:3.8} and \eqref{eq:3.9} into \eqref{eq:3.1} we
get
\begin{align}
  \left(1+\frac{i}{2}\epsilon_{ab}J^{ab}\right)\mathcal{O}_{\Delta}(x)
  \left(1-\frac{i}{2}\epsilon_{ab}J^{ab}\right)
  =\left(1+\frac{\Delta}{2d}\epsilon_{ab}\partial_{\mu}k^{\mu ab}\right)
  \mathcal{O}_{\Delta}(x+\tfrac{1}{2}\epsilon k), \label{eq:3.11}
\end{align}
which, at the linear order of $\epsilon_{ab}$, reduces to the
following form:
\begin{align} [J^{ab}, \mathcal{O}_{\Delta}(x)]
  =-J_{\Delta}^{ab}\mathcal{O}_{\Delta}(x), \label{eq:3.12}
\end{align}
where $J_{\Delta}^{ab}=-J_{\Delta}^{ba}$ are coordinate realizations
of the $SO(2,d)$ generators given by
\begin{align}
  J_{\Delta}^{ab}
  =i\left(k^{\mu ab}\partial_{\mu}+\frac{\Delta}{d}\partial_{\mu}k^{\mu ab}\right). \label{eq:3.13}
\end{align}
Similarly, for the infinitesimal conformal transformations
\eqref{eq:3.8} and \eqref{eq:3.9} the conformal Ward--Takahashi
identity \eqref{eq:3.3} becomes
\begin{align}
  G_{\Delta}(x,y)
  =\left(1+\frac{\Delta}{2d}\epsilon_{ab}\partial_{\mu}k^{\mu ab}(x)\right)
  \left(1+\frac{\Delta}{2d}\epsilon_{ab}\partial_{\mu}k^{\mu ab}(y)\right)
  G_{\Delta}(x+\tfrac{1}{2}\epsilon k(x),y+\tfrac{1}{2}\epsilon k(y)), \label{eq:3.14}
\end{align}
which, at the linear order of $\epsilon_{ab}$, reduces to the
following linear differential equations (infinitesimal conformal
Ward--Takahashi identities):
\begin{align}
  \left(J_{\Delta}^{ab}(x,\partial_{x})+J_{\Delta}^{ab}(y,\partial_{y})\right)G_{\Delta}(x,y)=0, \label{eq:3.15}
\end{align}
where $J_{\Delta}^{ab}(x,\partial_{x})$ and
$J_{\Delta}^{ab}(y,\partial_{y})$ are the notations to emphasize these
are the differential operators acting on $x$ and $y$, respectively.

Now let us move on to the infinitesimal form of intertwining relation
\eqref{eq:3.7}. At the linear order of $\epsilon_{ab}$, the
intertwining relation \eqref{eq:3.7} becomes
$J^{ab}G_{\Delta}=G_{\Delta}J^{ab}$, or, equivalently,
\begin{align}
  (J_{\Delta}^{ab}G_{\Delta}\mathcal{O}_{d-\Delta})(x)
  =(G_{\Delta}J_{d-\Delta}^{ab}\mathcal{O}_{d-\Delta})(x). \label{eq:3.16}
\end{align}
Indeed, a straightforward calculation gives
\begin{align}
  (J_{\Delta}^{ab}G_{\Delta}\mathcal{O}_{d-\Delta})(x)
  &=\int\!d^{d}y\,
    \left(J_{\Delta}^{ab}(x,\partial_{x})G_{\Delta}(x,y)\right)
    \mathcal{O}_{d-\Delta}(y) \nonumber\\
  &=\int\!d^{d}y\,
    \left(-J_{\Delta}^{ab}(y,\partial_{y})G_{\Delta}(x,y)\right)
    \mathcal{O}_{d-\Delta}(y) \nonumber\\
  &=\int\!d^{d}y\,
    \left[
    -i\left(k^{\mu ab}(y)\frac{\partial}{\partial y^{\mu}}
    +\frac{\Delta}{d}\frac{\partial k^{\mu ab}(y)}{\partial y^{\mu}}\right)
    G_{\Delta}(x,y)
    \right]
    \mathcal{O}_{d-\Delta}(y) \nonumber\\
  &=\int\!d^{d}y\,
    G_{\Delta}(x,y)
    \left[
    i\left(k^{\mu ab}(y)\frac{\partial}{\partial y^{\mu}}
    +\frac{d-\Delta}{d}\frac{\partial k^{\mu ab}(y)}{\partial y^{\mu}}\right)
    \mathcal{O}_{d-\Delta}(y)
    \right] \nonumber\\
  &=\int\!d^{d}y\,
    G_{\Delta}(x,y)
    \left(J_{d-\Delta}^{ab}(y,\partial_{y})
    \mathcal{O}_{d-\Delta}(y)\right) \nonumber\\
  &=(G_{\Delta}J_{d-\Delta}^{ab}\mathcal{O}_{d-\Delta})(x), \label{eq:3.17}
\end{align}
where in the second line we have used the infinitesimal conformal
Ward--Takahashi identity \eqref{eq:3.15} and in the third line we have
used the explicit expression \eqref{eq:3.13} for the generator. The
fourth equality follows from the integration by parts, where we have
ignored the surface term which is of no significance for the following
discussions. It is obvious that one can also prove the identity
$(J_{d-\Delta}^{ab}G_{d-\Delta}\mathcal{O}_{\Delta})(x)=(G_{d-\Delta}J_{\Delta}^{ab}\mathcal{O}_{\Delta})(x)$
in exactly the same way. It is also obvious that, since
$\mathcal{O}_{d-\Delta}(x)$ and $\mathcal{O}_{\Delta}(x)$ are
arbitrary test primary operators in the above discussions, the
following operator identities must hold:
\begin{align}
  J_{\alpha}^{ab}G_{\alpha}=G_{\alpha}J_{d-\alpha}^{ab},
  \quad
  \alpha\in\{\Delta,d-\Delta\}. \label{eq:3.18}
\end{align}
It is these intertwining relations that we will use in the rest of the
paper.

\begin{figure}[t]
  \centering
\xdefinecolor{rgb_000000}{rgb}{0,0,0}%
\xdefinecolor{rgb_ff0000}{rgb}{1,0,0}%
\setlength{\unitlength}{1cm}%
\begin{picture}(4.8,3.2)(0,0)%
\allinethickness{0.8pt}%
\color{rgb_ff0000}%
\path(0.8,2.3)(0.8,1)
\allinethickness{0.0140584cm}%
\path(0.795607,1.01318)(0.804393,1.01318)
\path(0.791214,1.02636)(0.808786,1.02636)
\path(0.78682,1.03954)(0.81318,1.03954)
\path(0.782427,1.05272)(0.817573,1.05272)
\path(0.778034,1.0659)(0.821966,1.0659)
\path(0.773641,1.07908)(0.826359,1.07908)
\path(0.769247,1.09226)(0.830753,1.09226)
\path(0.764854,1.10544)(0.835146,1.10544)
\path(0.760461,1.11862)(0.839539,1.11862)
\path(0.756068,1.1318)(0.843932,1.1318)
\path(0.751674,1.14498)(0.848326,1.14498)
\path(0.747281,1.15816)(0.852719,1.15816)
\path(0.839539,1.11862)(0.839539,1.15816)
\path(0.826359,1.07908)(0.826359,1.15816)
\path(0.81318,1.03954)(0.81318,1.15816)
\path(0.8,1)(0.8,1.15816)
\path(0.78682,1.03954)(0.78682,1.15816)
\path(0.773641,1.07908)(0.773641,1.15816)
\path(0.760461,1.11862)(0.760461,1.15816)
\allinethickness{0.8pt}%
\path(0.8,1.15816)(0.747281,1.15816)(0.8,1)(0.852719,1.15816)(0.8,1.15816)
\path(1.5,2.6)(3.3,2.6)
\allinethickness{0.0140584cm}%
\path(3.28682,2.59561)(3.2829,2.6057)
\path(3.27364,2.59121)(3.26579,2.6114)
\path(3.26046,2.58682)(3.24869,2.6171)
\path(3.24728,2.58243)(3.23158,2.62281)
\path(3.2341,2.57803)(3.21448,2.62851)
\path(3.22092,2.57364)(3.19737,2.63421)
\path(3.20774,2.56925)(3.18027,2.63991)
\path(3.19456,2.56485)(3.16316,2.64561)
\path(3.18138,2.56046)(3.14606,2.65131)
\path(3.1682,2.55607)(3.14184,2.62386)
\path(3.15502,2.55167)(3.14184,2.58557)
\path(3.16161,2.64613)(3.14184,2.63844)
\path(3.18138,2.63954)(3.14184,2.62417)
\path(3.20115,2.63295)(3.14184,2.60989)
\path(3.22092,2.62636)(3.14184,2.59561)
\path(3.24069,2.61977)(3.14184,2.58134)
\path(3.26046,2.61318)(3.14184,2.56706)
\path(3.28023,2.60659)(3.14184,2.55278)
\allinethickness{0.8pt}%
\path(3.14184,2.6)(3.14184,2.54728)(3.3,2.6)(3.14184,2.65272)(3.14184,2.6)
\path(4,2.3)(4,1)
\allinethickness{0.0140584cm}%
\path(3.96777,1.15816)(3.95207,1.14378)
\path(3.98826,1.15816)(3.95687,1.1294)
\path(4.00874,1.15816)(3.96166,1.11502)
\path(4.02923,1.15816)(3.96645,1.10065)
\path(4.04972,1.15816)(3.97124,1.08627)
\path(4.04503,1.13509)(3.97604,1.07189)
\path(4.03603,1.10808)(3.98083,1.05751)
\path(4.02702,1.08106)(3.98562,1.04313)
\path(4.01801,1.05404)(3.99041,1.02876)
\path(4.00901,1.02702)(3.99521,1.01438)
\path(3.98972,1.03083)(4.00479,1.01438)
\path(3.97945,1.06165)(4.00959,1.02876)
\path(3.96917,1.09248)(4.01438,1.04313)
\path(3.9589,1.12331)(4.01917,1.05751)
\path(3.94862,1.15413)(4.02396,1.07189)
\path(3.9629,1.15816)(4.02876,1.08627)
\path(3.98086,1.15816)(4.03355,1.10065)
\path(3.99883,1.15816)(4.03834,1.11502)
\path(4.01679,1.15816)(4.04313,1.1294)
\path(4.03475,1.15816)(4.04793,1.14378)
\allinethickness{0.8pt}%
\path(4,1.15816)(3.94728,1.15816)(4,1)(4.05272,1.15816)(4,1.15816)
\path(1.3,0.6)(3.5,0.6)
\allinethickness{0.0140584cm}%
\path(3.34184,0.561809)(3.35942,0.553139)
\path(3.34184,0.576337)(3.37699,0.558996)
\path(3.34184,0.590865)(3.39456,0.564854)
\path(3.34184,0.605394)(3.41214,0.570712)
\path(3.34184,0.619922)(3.42971,0.576569)
\path(3.34184,0.63445)(3.44728,0.582427)
\path(3.34184,0.648978)(3.46485,0.588285)
\path(3.40924,0.630255)(3.48243,0.594142)
\path(3.48163,0.593877)(3.48682,0.604393)
\path(3.46326,0.587754)(3.47364,0.608786)
\path(3.4449,0.581632)(3.46046,0.61318)
\path(3.42653,0.575509)(3.44728,0.617573)
\path(3.40816,0.569386)(3.4341,0.621966)
\path(3.38979,0.563263)(3.42092,0.626359)
\path(3.37142,0.557141)(3.40774,0.630753)
\path(3.35305,0.551018)(3.39456,0.635146)
\path(3.34184,0.559403)(3.38138,0.639539)
\path(3.34184,0.590508)(3.3682,0.643932)
\path(3.34184,0.621614)(3.35502,0.648326)
\allinethickness{0.8pt}%
\path(3.34184,0.6)(3.34184,0.547281)(3.5,0.6)(3.34184,0.652719)(3.34184,0.6)
\path(1,1)(1,2.3)
\allinethickness{0.0140584cm}%
\path(1.03954,2.14184)(1.04229,2.17314)
\path(1.02636,2.14184)(1.03185,2.20444)
\path(1.01318,2.14184)(1.02142,2.23574)
\path(1,2.14184)(1.01099,2.26704)
\path(0.98682,2.14184)(1.00055,2.29834)
\path(0.973641,2.14184)(0.983061,2.24918)
\path(0.960461,2.14184)(0.965171,2.19551)
\path(1.00466,2.28603)(0.995607,2.28682)
\path(1.00932,2.27205)(0.991214,2.27364)
\path(1.01397,2.25808)(0.98682,2.26046)
\path(1.01863,2.2441)(0.982427,2.24728)
\path(1.02329,2.23013)(0.978034,2.2341)
\path(1.02795,2.21616)(0.973641,2.22092)
\path(1.03261,2.20218)(0.969247,2.20774)
\path(1.03726,2.18821)(0.964854,2.19456)
\path(1.04192,2.17423)(0.960461,2.18138)
\path(1.04658,2.16026)(0.956068,2.1682)
\path(1.05124,2.14628)(0.951674,2.15502)
\allinethickness{0.8pt}%
\path(1,2.14184)(1.05272,2.14184)(1,2.3)(0.947281,2.14184)(1,2.14184)
\path(3.8,1)(3.8,2.3)
\allinethickness{0.0140584cm}%
\path(3.80439,2.28682)(3.79466,2.28399)
\path(3.80879,2.27364)(3.78933,2.26798)
\path(3.81318,2.26046)(3.78399,2.25196)
\path(3.81757,2.24728)(3.77865,2.23595)
\path(3.82197,2.2341)(3.77331,2.21994)
\path(3.82636,2.22092)(3.76798,2.20393)
\path(3.83075,2.20774)(3.76264,2.18791)
\path(3.83515,2.19456)(3.7573,2.1719)
\path(3.83954,2.18138)(3.75196,2.15589)
\path(3.84393,2.1682)(3.75338,2.14184)
\path(3.84833,2.15502)(3.80305,2.14184)
\path(3.75432,2.16295)(3.76046,2.14184)
\path(3.76135,2.18406)(3.77364,2.14184)
\path(3.76839,2.20516)(3.78682,2.14184)
\path(3.77542,2.22627)(3.8,2.14184)
\path(3.78246,2.24738)(3.81318,2.14184)
\path(3.78949,2.26848)(3.82636,2.14184)
\path(3.79653,2.28959)(3.83954,2.14184)
\allinethickness{0.8pt}%
\path(3.8,2.14184)(3.85272,2.14184)(3.8,2.3)(3.74728,2.14184)(3.8,2.14184)
\put(0.9,2.6){\makebox(0,0)[c]{\hbox{\color{rgb_000000}$\mathcal{V}_{d-\Delta}$}}}
\put(3.9,2.6){\makebox(0,0)[c]{\hbox{\color{rgb_000000}$\mathcal{V}_{d-\Delta}$}}}
\put(0.9,0.6){\makebox(0,0)[c]{\hbox{\color{rgb_000000}$\mathcal{V}_{\Delta}$}}}
\put(3.9,0.6){\makebox(0,0)[c]{\hbox{\color{rgb_000000}$\mathcal{V}_{\Delta}$}}}
\put(0.659416,1.6){\makebox(0,0)[r]{\hbox{\color{rgb_000000}$G_{\Delta}$}}}
\put(4.14058,1.6){\makebox(0,0)[l]{\hbox{\color{rgb_000000}$G_{\Delta}$}}}
\put(1.14058,1.6){\makebox(0,0)[l]{\hbox{\color{rgb_000000}$G_{d-\Delta}$}}}
\put(3.65942,1.6){\makebox(0,0)[r]{\hbox{\color{rgb_000000}$G_{d-\Delta}$}}}
\put(2.4,2.70544){\makebox(0,0)[b]{\hbox{\color{rgb_000000}$J_{d-\Delta}^{ab}$}}}
\put(2.4,0.494562){\makebox(0,0)[t]{\hbox{\color{rgb_000000}$J_{\Delta}^{ab}$}}}
\end{picture}%
  \caption{Commutative diagram of intertwining operators.}
  \label{figure:4}
\end{figure}
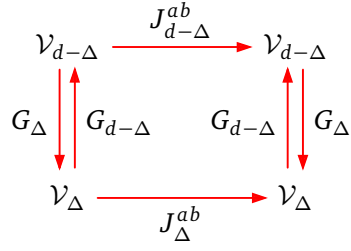

Several comments are in order.
\begin{itemize}
\item \emph{Primary states.} The intertwining operator $G_{\alpha}$
  also maps a primary state of scaling dimension $d-\alpha$ to another
  primary state of scaling dimension $\alpha$, where $\alpha$ is
  either $\Delta$ or $d-\Delta$. In fact, by defining the
  coordinate-dependent primary state as
  \begin{align}
    |\mathcal{O}_{\alpha}(x)\rangle:=\mathcal{O}_{\alpha}(x)|\Omega\rangle,
    \quad
    \alpha\in\{\Delta,d-\Delta\}, \label{eq:3.19}
  \end{align}
  which satisfies the transformation law
  $U(g)|\mathcal{O}_{\alpha}(x)\rangle=|\partial x_{g}/\partial
  x|^{\alpha/d}|\mathcal{O}_{\alpha}(x)\rangle$, one can easily show
  that the state
  $G_{\alpha}|\mathcal{O}_{d-\alpha}(x)\rangle:=(G_{\alpha}\mathcal{O}_{d-\alpha})(x)|\Omega\rangle$
  also satisfies the transformation law
  $U(g)G_{\alpha}|\mathcal{O}_{d-\alpha}(x)\rangle=|\partial
  x_{g}/\partial
  x|^{\alpha/d}G_{\alpha}|\mathcal{O}_{d-\alpha}(x)\rangle$ and hence
  is the coordinate-dependent primary state of scaling dimension
  $\alpha$. $|\mathcal{O}_{\alpha}(x)\rangle$ is an element of the
  representation space $\mathcal{V}_{\alpha}$ of conformal algebra
  $\mathfrak{so}(2,d)$ in which the quadratic Casimir operator
  $C_{2}[\mathfrak{so}(2,d)]$ takes the value $\Delta(d-\Delta)$.  The
  intertwining relations between $\mathcal{V}_{d-\Delta}$ and
  $\mathcal{V}_{\Delta}$ are schematically summarized in the
  commutative diagram in figure \ref{figure:4}.
  
\item \emph{Normalization.} The intertwining operators $G_{\Delta}$
  and $G_{d-\Delta}$ are basically inverse to each other. Namely,
  under appropriate normalizations they can be chosen to satisfy the
  following identity:\footnote{There is, however, a subtle problem in
    the identity \eqref{eq:3.20}. A careful analysis shows that the
    identity \eqref{eq:3.21} (or
    $\Tilde{G}_{\Delta}(p)\Tilde{G}_{d-\Delta}(p)=1$) does not hold
    when $\Delta=d/2+n$ ($n\in\mathbb{Z}_{\geq0}$), which may well be
    related to the conformal anomaly; see, e.g.,
    \cite{Bzowski:2015pba}. We will not address this issue in the
    present paper.}
  \begin{align}
    G_{\alpha}G_{d-\alpha}=1_{\mathcal{V}_{\alpha}},
    \quad
    \alpha\in\{\Delta,d-\Delta\}, \label{eq:3.20}
  \end{align}
  or, equivalently,
  \begin{align}
    \int\!d^{d}z\,G_{\alpha}(x,z)G_{d-\alpha}(z,y)=\delta^{(d)}(x-y), \label{eq:3.21}
  \end{align}
  where $1_{\mathcal{V}_{\alpha}}$ stands for the identity operator in
  the representation space $\mathcal{V}_{\alpha}$. Notice that
  eq.~\eqref{eq:3.21} is most simply expressed in the momentum space
  as $\Tilde{G}_{\Delta}(p)\Tilde{G}_{d-\Delta}(p)=1$, where
  $\Tilde{G}_{\alpha}(p)$ stands for the Fourier transform of
  two-point function $G_{\alpha}(x,y)$.

\item \emph{Unitarity bound.} It is well-known that in $d$-dimensional
  spacetime the unitarity bound for scaling dimension $\Delta$ of
  scalar primary operator is $\Delta>(d-2)/2$, where the lower bound
  is the scaling dimension of free scalar field. In general,
  $d-\Delta$ violates this bound if $\Delta$ is big enough. However,
  there exists a range of $\Delta$ in which both the scaling
  dimensions of $\mathcal{O}_{\Delta}(x)$ and its shadow
  $\mathcal{O}_{d-\Delta}(x)$ fulfill the unitarity bound:
  \begin{align}
    \frac{d-2}{2}<\Delta<\frac{d+2}{2}. \label{eq:3.22}
  \end{align}
  In the AdS/CFT correspondence this range of $\Delta$ corresponds to
  the well-known range of scalar field mass-squared
  $-d^{2}/4<(mR)^{2}<-d^{2}/4+1$ ($R$: AdS radius) in which there
  exist two different quantizations \cite{Klebanov:1999tb}.

\item \emph{Conformal algebra.} It is worth to write down here the
  $d$-dimensional conformal algebra because the relation between the
  conformal generators $\{D,P^{\mu},K^{\mu},M^{\mu\nu}\}$ and
  $SO(2,d)$ generators $J^{ab}$ often differs in the literature. In
  our convention the dilatation generator $D$, momentum generators
  $P^{\mu}$, special conformal transformation generators $K^{\mu}$,
  and Lorentz transformation generators $M^{\mu\nu}$ that are
  consistent with \eqref{eq:2.8}, \eqref{eq:2.10}, \eqref{eq:2.12} and
  \eqref{eq:2.14} are given by the following linear combinations:
  \begin{align}
    D=J^{d,d+1},\quad
    P^{\mu}=\frac{1}{\ell}(J^{d+1,\mu}+J^{d\mu}),\quad
    K^{\mu}=\ell(J^{d+1,\mu}-J^{d\mu}),\quad
    M^{\mu\nu}=J^{\mu\nu}, \label{eq:3.23}
  \end{align}
  which satisfy the commutation relations of $d$-dimensional conformal
  algebra:
  \begin{align}
    \begin{split}
      &[D,P^{\mu}]=iP^{\mu},\quad [D,K^{\mu}]=-iK^{\mu},\quad
      [P^{\mu},K^{\nu}]=2i(\eta^{\mu\nu}D-J^{\mu\nu}),\\
      &[M^{\mu\nu},P^{\rho}]=i(\eta^{\mu\rho}P^{\nu}-\eta^{\nu\rho}P^{\mu}),\quad
      [M^{\mu\nu},K^{\rho}]=i(\eta^{\mu\rho}K^{\nu}-\eta^{\nu\rho}K^{\mu}),\\
      &[M^{\mu\nu},M^{\rho\sigma}]=i(\eta^{\mu\rho}M^{\nu\sigma}-\eta^{\mu\sigma}M^{\nu\rho}-\eta^{\nu\rho}M^{\mu\sigma}+\eta^{\nu\sigma}M^{\mu\rho}). \label{eq:3.24}
    \end{split}
  \end{align}
  With this definition the infinitesimal conformal transformation
  $U(1+\epsilon)=1+\tfrac{i}{2}\epsilon_{ab}J^{ab}$ is expressed as
  follows:
  \begin{align}
    U(1+\epsilon)=1+i\varphi D-ia_{\mu}P^{\mu}+ib_{\mu}K^{\mu}+\frac{i}{2}\omega_{\mu\nu}M^{\mu\nu}, \label{eq:3.25}
  \end{align}
  where the infinitesimal parameters ${\epsilon^{a}}_{b}$ and
  $(\varphi,a^{\mu},b^{\mu},{\omega^{\mu}}_{\nu})$ are identified
  under the rule \eqref{eq:2.16}. Finite conformal transformations are
  obtained by exponentiating the last four terms in
  \eqref{eq:3.25}. For example, one can easily check that, by using
  the $(d+2)\times(d+2)$-matrix representation of conformal generators
  \begin{align}
    \begin{split}
      &{(D)^{a}}_{b}=-i
      \begin{pmatrix}
        \boldsymbol{0}_{d}
        &&\\
        &0
        &1\\
        &1 &0
      \end{pmatrix},\quad {(M^{\mu\nu})^{a}}_{b}=-i
      \begin{pmatrix}
        \eta^{\mu a}{\delta^{\nu}}_{b}-\eta^{\nu a}{\delta^{\mu}}_{b}
        &&\\
        &0
        &0\\
        &0 &0
      \end{pmatrix},\\
      &{(P^{\mu})^{a}}_{b}=\frac{i}{\ell}
      \begin{pmatrix}
        \boldsymbol{0}_{d} &\eta^{\mu a}
        &\eta^{\mu a}\\
        -{\delta^{\mu}}_{b} &0
        &0\\
        +{\delta^{\mu}}_{b} &0 &0
      \end{pmatrix},\quad {(K^{\mu})^{a}}_{b}=-i\ell
      \begin{pmatrix}
        \boldsymbol{0}_{d} &\eta^{\mu a}
        &-\eta^{\mu a}\\
        -{\delta^{\mu}}_{b} &0
        &0\\
        -{\delta^{\mu}}_{b} &0 &0
      \end{pmatrix}, \label{eq:3.26}
    \end{split}
  \end{align}
  which follow from the matrix representation of $SO(2,d)$ generators
  $i{(J^{ab})^{c}}_{d}=\eta^{ac}{\delta^{b}}_{d}-\eta^{bc}{\delta^{a}}_{d}$,
  the $(d+2)\times(d+2)$ matrices $\exp(i\varphi D)\in SO(1,1)$,
  $\exp(-ia_{\mu}P^{\mu})\in E(1)^{d}$ and
  $\exp(ib_{\mu}K^{\mu})\in E(1)^{d}$ exactly coincides with
  \eqref{eq:2.8}, \eqref{eq:2.10} and \eqref{eq:2.12}. (These matrix
  representations are not unitary representations of course.) For our
  purpose, however, the coordinate realization of conformal generators
  is more important. Substituting the conformal Killing vectors into
  \eqref{eq:3.13}, one can easily check that the coordinate
  realization
  $\{D_{\alpha},P^{\mu}_{\alpha},K^{\mu}_{\alpha},M^{\mu\nu}_{\alpha}\}$
  that acts on the representation space $\mathcal{V}_{\alpha}$ of
  scaling dimension $\alpha\in\{\Delta,d-\Delta\}$ is given as
  follows:
  \begin{subequations}
    \begin{align}
      D_{\alpha}&=-i(x\cdot\partial+\alpha),\label{eq:3.27a}\\
      P^{\mu}_{\alpha}&=-i\partial^{\mu},\label{eq:3.27b}\\
      K^{\mu}_{\alpha}&=-i\left[(x\cdot x)\partial^{\mu}-2x^{\mu}(x\cdot\partial+\alpha)\right],\label{eq:3.27c}\\
      M^{\mu\nu}_{\alpha}&=-i(x^{\mu}\partial^{\nu}-x^{\nu}\partial^{\mu}).\label{eq:3.27d}
    \end{align}
  \end{subequations}
  We shall use these differential operators in the next section.
\end{itemize}

\subsection{Intertwining relations in the
  \texorpdfstring{$E(1)$}{E(1)} basis}
\label{section:3.2}
The intertwining relations are the operator identities in the
representation space $\mathcal{V}_{\alpha}$ of fixed scaling dimension
$\alpha\in\{\Delta,d-\Delta\}$. In the last section we have seen that
the operator identities \eqref{eq:3.18} hold when they act on primary
operators or primary states. However, the intertwining relations still
hold when they are applied to the basis of representation space
$\mathcal{V}_{\alpha}$. In this section we study the intertwining
relations in the basis where the $E(1)$ transformations
\eqref{eq:2.10} become diagonal. In other words, we apply the identity
\eqref{eq:3.18} to the simultaneous eigenfunctions of $E(1)$
generators \eqref{eq:3.27b} which provide a complete orthonormal basis
of $\mathcal{V}_{\alpha}$. We will see that in this case the
intertwining relations just reduce to the well-known conformal
Ward--Takahashi identities for momentum-space two-point function of
zero-temperature CFT.

To begin with, let $\{f_{p}(x)\}$ be a complete orthonormal basis that
satisfies the following orthonormality and completeness relations:
\begin{subequations}
  \begin{align}
    \int\!d^{d}x\,f_{p}^{\ast}(x)f_{q}(x)&=(2\pi)^{d}\delta^{(d)}(p-q), \label{eq:3.28a}\\
    \int\!\!\frac{d^{d}p}{(2\pi)^{d}}\,f_{p}(x)f_{p}^{\ast}(y)&=\delta^{(d)}(x-y). \label{eq:3.28b}
  \end{align}
\end{subequations}
In the flat Minkowski spacetime such complete orthonormal basis is of
course given by the plane wave, $f_{p}(x)=\mathrm{e}^{ip\cdot x}$,
which is nothing but the simultaneous eigenfunction of $E(1)$
generators $P_{\alpha}^{\mu}=-i\partial^{\mu}$ with eigenvalue
$p^{\mu}$. The primary operator and two-point function can then be
expanded in terms of this complete orthonormal basis:
\begin{subequations}
  \begin{align}
    \mathcal{O}_{\alpha}(x)
    &=\int\!\!\frac{d^{d}p}{(2\pi)^{d}}\,
      \Tilde{\mathcal{O}}_{\alpha}(p)f_{p}(x), \label{eq:3.29a}\\
    G_{\alpha}(x,y)
    &=\int\!\!\frac{d^{d}p}{(2\pi)^{d}}\,
      \Tilde{G}_{\alpha}(p)f_{p}(x)f_{p}^{\ast}(y), \quad \alpha\in\{\Delta,d-\Delta\}, \label{eq:3.29b}
  \end{align}
\end{subequations}
which are of course just the Fourier transforms. Plugging these into
the left- and right-hand sides of the identity
$(J_{\Delta}^{ab}G_{\Delta}\mathcal{O}_{d-\Delta})(x)=(G_{\Delta}J_{d-\Delta}^{ab}\mathcal{O}_{d-\Delta})(x)$,
we get
\begin{subequations}
  \begin{align}
    (J_{\Delta}^{ab}G_{\Delta}\mathcal{O}_{d-\Delta})(x)
    &=\int\!\!\frac{d^{d}p}{(2\pi)^{d}}\int\!\!\frac{d^{d}q}{(2\pi)^{d}}\,
      \langle p|J_{\Delta}^{ab}|q\rangle
      \Tilde{G}_{\Delta}(q)
      \Tilde{\mathcal{O}}_{d-\Delta}(q)
      f_{p}(x), \label{eq:3.30a}\\
    (G_{\Delta}J_{d-\Delta}^{ab}\mathcal{O}_{d-\Delta})(x)
    &=\int\!\!\frac{d^{d}p}{(2\pi)^{d}}\int\!\!\frac{d^{d}q}{(2\pi)^{d}}\,
      \Tilde{G}_{\Delta}(p)
      \langle p|J_{d-\Delta}^{ab}|q\rangle
      \Tilde{\mathcal{O}}_{d-\Delta}(q)
      f_{p}(x), \label{eq:3.30b}
  \end{align}
\end{subequations}
where $\langle p|J_{\alpha}^{ab}|q\rangle$ are ``matrix elements''
given by
\begin{align}
  \langle p|J_{\alpha}^{ab}|q\rangle
  :=\int\!d^{d}x\,f_{p}^{\ast}(x)J_{\alpha}^{ab}f_{q}(x),
  \quad
  \alpha\in\{\Delta,d-\Delta\}. \label{eq:3.31}
\end{align}
Comparing the expressions \eqref{eq:3.30a} and \eqref{eq:3.30b} and
noting $\Tilde{\mathcal{O}}_{d-\Delta}(q)$ is an arbitrary primary
operator, we arrive at the following identities for the momentum-space
two-point function:
\begin{align}
  \langle p|J_{\Delta}^{ab}|q\rangle\Tilde{G}_{\Delta}(q)
  =\Tilde{G}_{\Delta}(p)\langle p|J_{d-\Delta}^{ab}|q\rangle. \label{eq:3.32}
\end{align}
Though it is less obvious, the intertwining relation \eqref{eq:3.32}
is just another form of the conformal Ward--Takahashi identity. To see
this, let us first compute the ``matrix elements''
$\langle p|J_{\alpha}^{ab}|q\rangle$ by using the coordinate
realizations \eqref{eq:3.13}. Substituting
\eqref{eq:3.27a}--\eqref{eq:3.27d} into \eqref{eq:3.31} we get the
following ``matrix elements'':
\begin{subequations}
  \begin{align}
    \langle p|D_{\alpha}|q\rangle
    &=-i(q\cdot\partial_{q}+\alpha)
      \delta^{(d)}(p-q). \label{eq:3.33a}\\
    \langle p|P_{\alpha}^{\mu}|q\rangle
    &=q^{\mu}\delta^{(d)}(p-q), \label{eq:3.33b}\\
    \langle p|K_{\alpha}^{\mu}|q\rangle
    &=\left[
      -q^{\mu}\partial_{q}\cdot\partial_{q}
      +2(q\cdot\partial_{q}+\alpha)\partial_{q}^{\mu}
      \right]
      \delta^{(d)}(p-q), \label{eq:3.33c}\\
    \langle p|M_{\alpha}^{\mu\nu}|q\rangle
    &=i(q^{\mu}\partial_{q}^{\nu}-q^{\nu}\partial_{q}^{\mu})
      \delta^{(d)}(p-q). \label{eq:3.33d}
  \end{align}
\end{subequations}
Now it is a straightforward exercise to show that the intertwining
relation \eqref{eq:3.32} boils down to the following differential
equations for the momentum-space two-point function:
\begin{subequations}
  \begin{align}
    \left(p^{\mu}\partial^{\nu}-p^{\nu}\partial^{\mu}\right)
    \Tilde{G}_{\Delta}(p)
    &=0, \label{eq:3.34a}\\
    \left[p^{\mu}\partial\cdot\partial-2(p\cdot\partial-\Delta+d)\partial^{\mu}\right]
    \Tilde{G}_{\Delta}(p)
    &=0, \label{eq:3.34b}\\
    \left(p\cdot\partial-2\Delta+d\right)
    \Tilde{G}_{\Delta}(p)
    &=0, \label{eq:3.34c}
  \end{align}
\end{subequations}
which are nothing but the well-known conformal Ward--Takahashi
identities in the momentum space. These differential equations are
solved as follows. First, eq.~\eqref{eq:3.34a} implies
$\Tilde{G}_{\Delta}$ must be a function of the Lorentz invariant
$p\cdot p$. And, for such a function, eq.~\eqref{eq:3.34c} reduces to
$[(p\cdot p)\frac{d}{d(p\cdot
  p)}-(\Delta-\frac{d}{2})]\Tilde{G}_{\Delta}=0$, which is easily
solved with the result
\begin{align}
  \Tilde{G}_{\Delta}(p)\propto(p\cdot p)^{\Delta-d/2}. \label{eq:3.35}
\end{align}
Notice that the solution \eqref{eq:3.35} automatically satisfies
\eqref{eq:3.34b}.

Before closing this section it is wise to reexamine here the
intertwining relations \eqref{eq:3.32} in a little bit more abstract
way. To this end, we first define the action of intertwining operator
$G_{\alpha}$ on the basis function $f_{p}(x)$ as follows:
\begin{align}
  (G_{\alpha}f_{p})(x):=\int\!d^{d}y\,G_{\alpha}(x,y)f_{p}(y), \label{eq:3.36}
\end{align}
which, after substituting the expansion \eqref{eq:3.29b}, can be
reduced to the following:
\begin{align}
  (G_{\alpha}f_{p})(x)=\Tilde{G}_{\alpha}(p)f_{p}(x). \label{eq:3.37}
\end{align}
This is best expressed in the following bra-ket notation:
\begin{align}
  G_{\alpha}|p\rangle=\Tilde{G}_{\alpha}(p)|p\rangle. \label{eq:3.38}
\end{align}
That is to say, the momentum-space two-point function
$\Tilde{G}_{\alpha}(p)$ is the eigenvalue of intertwining operator
$G_{\alpha}$. Now it is obvious that the intertwining relations
\eqref{eq:3.32} are equivalent to
$\langle p|J_{\Delta}^{ab}G_{\Delta}|q\rangle=\langle
p|G_{\Delta}J_{d-\Delta}^{ab}|q\rangle$, which is nothing but the
``matrix elements'' of the operator identities \eqref{eq:3.18}.

To summarize, in zero-temperature CFT, the intertwining relations are
just the conformal Ward--Takahashi identities in disguise. At finite
temperature, however, we shall see that the intertwining relations
lead to novel constraints on momentum-space two-point functions.

\section{Thermal Correlator Recursions}
\label{section:4}
In the last section we have studied the intertwining relations in the
$E(1)$ basis in which the $E(1)$ generators $P_{\alpha}^{\mu}$ become
diagonal. For the case of finite-temperature CFT, however, we need to
work in the basis in which the $SO(1,1)$ generator becomes diagonal
because, as we have seen in section \ref{section:2}, the
time-translation generator generates the one-parameter subgroup
$SO(1,1)$ in geometrically thermalized CFTs. In this section we shall
first develop the representation of conformal algebra
$\mathfrak{so}(2,d)$ in the $SO(1,1)$ basis and derive the basis
function explicitly. We then show that the intertwining relations in
the $SO(1,1)$ basis result in the recurrence relations for
momentum-space thermal two-point functions.

For the sake of notational brevity, throughout this section we will
work in the units $\ell=1$ (i.e., $2\pi T=1$). The Unruh temperature
$T$ is easily restored by dimensional analysis.

\subsection{Representation of the Lie algebra
  \texorpdfstring{$\mathfrak{so}(2,d)$}{so(2,d)} in the
  \texorpdfstring{$SO(1,1)$}{SO(1,1)} basis}
\label{section:4.1}
Let us first construct the $SO(1,1)$ basis for the representation
space of the Lie algebra $\mathfrak{so}(2,d)$. As we have seen in
sections \ref{section:2.2}--\ref{section:2.4}, in thermal CFTs on the
Rindler wedge, light-cone, and diamond the time translations are given
by the $SO(1,1)$ hyperbolic rotations on the $(X^{0},X^{1})$-,
$(X^{d},X^{d+1})$-, and $(X^{0},X^{d})$-planes. This motivates us to
introduce the following linear combinations of the $SO(2,d)$
generators:
\begin{subequations}
  \begin{alignat}{4}
    \text{Rindler wedge:}&\quad &&H=J^{10},\quad &&E^{\pm a}=J^{0a}\pm
    J^{1a},\quad
    &&M^{ab}=J^{ab},\label{eq:4.1a}\\
    \text{Light-cone:}&\quad &&H=J^{d,d+1},\quad &&E^{\pm
      a}=J^{d+1,a}\pm J^{da},\quad
    &&M^{ab}=J^{ab},\label{eq:4.1b}\\
    \text{Diamond:}&\quad &&H=J^{d0},\quad &&E^{\pm a}=J^{0a}\pm
    J^{da},\quad &&M^{ab}=J^{ab},\label{eq:4.1c}
  \end{alignat}
\end{subequations}
where the indices $a$ and $b$ run through $\{2,\cdots,d+1\}$,
$\{0,\cdots,d-1\}$, and $\{1,\cdots,d-1,d+1\}$ for the Rindler wedge,
light-cone, and diamond, respectively. Irrespective of these linear
combinations, it follows from \eqref{eq:3.10} that the set of
operators $\{H, E^{\pm a}, M^{ab}\}$ satisfy the following same
commutation relations:
\begin{subequations}
  \begin{align}
    [H,E^{\pm a}]
    &=\pm iE^{\pm a}, \label{eq:4.2a}\\
    [E^{+a},E^{-b}]
    &=2i(\eta^{ab}H-M^{ab}), \label{eq:4.2b}\\
    [M^{ab},E^{\pm c}]
    &=i(\eta^{ac}E^{\pm b}-\eta^{bc}E^{\pm a}), \label{eq:4.2c}\\
    [M^{ab},M^{cd}]
    &=i(\eta^{ac}M^{bd}
      -\eta^{ad}M^{bc}
      -\eta^{bc}M^{ad}
      +\eta^{bd}M^{ac}), \label{eq:4.2d}
  \end{align}
\end{subequations}
with other commutators vanishing. The quadratic Casimir operator of
the Lie algebra $\mathfrak{so}(2,d)$ is then expressed as follows:
\begin{align}
  C_{2}[\mathfrak{so}(2,d)]
  &=\frac{1}{2}J^{ab}J_{ab} \nonumber\\
  &=-H(H\pm id)
    -\eta_{ab}E^{\mp a}E^{\pm b}
    +C_{2}[\mathfrak{so}(1,d-1)], \label{eq:4.3}
\end{align}
where $C_{2}[\mathfrak{so}(1,d-1)]$ is the quadratic Casimir operator
of the subalgebra $\mathfrak{so}(1,d-1)$ given by
\begin{align}
  C_{2}[\mathfrak{so}(1,d-1)]=\frac{1}{2}M^{ab}M_{ab}. \label{eq:4.4}
\end{align}
Notice that $C_{2}[\mathfrak{so}(1,d-1)]$ commutes with $H$ and
$M^{ab}$, whereas it does not commute with $E^{\pm a}$. The quadratic
Casimir operator of the whole algebra $C_{2}[\mathfrak{so}(2,d)]$, on
the other hand, commutes with all the generators of course.

Now, we wish to find a complete orthonormal basis for the
representation space $\mathcal{V}_{\alpha}$ of scaling dimension
$\alpha\in\{\Delta,d-\Delta\}$ where the time-translation generator
$H$ becomes diagonal. This can be done as follows. Since $H$ commutes
with the subalgebra $\mathfrak{so}(1,d-1)$, there exists a
simultaneous eigenstate of $C_{2}[\mathfrak{so}(2,d)]$, $H$, and
$C_{2}[\mathfrak{so}(1,d-1)]$ that satisfies the following eigenvalue
equations:
\begin{subequations}
  \begin{align}
    C_{2}[\mathfrak{so}(2,d)]|\alpha,\omega,j;\sigma\rangle
    &=\Delta(\Delta-d)|\alpha,\omega,j;\sigma\rangle, \label{eq:4.5a}\\
    H|\alpha,\omega,j;\sigma\rangle
    &=\omega|\alpha,\omega,j;\sigma\rangle, \label{eq:4.5b}\\
    C_{2}[\mathfrak{so}(1,d-1)]|\alpha,\omega,j;\sigma\rangle
    &=j(j-d+2)|\alpha,\omega,j;\sigma\rangle, \label{eq:4.5c}
  \end{align}
\end{subequations}
where the eigenvalues $\Delta(\Delta-d)$, $\omega$, and $j(j-d+2)$ are
all real. The label $\sigma$ denotes the set of eigenvalues for all
other operators that commute with $H$ and
$C_{2}[\mathfrak{so}(1,d-1)]$, which, as we will see shortly for the
case of Rindler wedge, can be chosen to the eigenvalues of momentum
operators perpendicular to the $(x^{0},x^{1})$-plane. Before going to
construct the eigenstate $|\alpha,\omega,j;\sigma\rangle$, it is
worthwhile to point out here the meaning of the operators $E^{\pm
  a}$. It just follows from the commutation relations
$[H,E^{\pm a}]=\pm iE^{\pm a}$ that the states
$E^{\pm a}|\alpha,\omega,j;\sigma\rangle$ satisfy the following
eigenvalue equations:
\begin{align}
  HE^{\pm a}|\alpha,\omega,j;\sigma\rangle
  =(\omega\pm i)E^{\pm a}|\alpha,\omega,j;\sigma\rangle, \label{eq:4.6}
\end{align}
which implies the operators $E^{\pm a}$ raise and lower the eigenvalue
$\omega$ by $\pm i$. Thus one may write
\begin{align}
  E^{\pm a}|\alpha,\omega,j;\sigma\rangle
  =\sum_{j^{\prime}\sigma^{\prime}}c_{j\sigma;j^{\prime}\sigma^{\prime}}^{\pm a}(\alpha,\omega)
  |\alpha,\omega\pm i,j^{\prime};\sigma^{\prime}\rangle. \label{eq:4.7}
\end{align}
We shall see that it is these coefficients
$c_{j\sigma;j^{\prime}\sigma^{\prime}}^{\pm a}(\alpha,\omega)$ that
determines the momentum-space two-point function of thermal CFT on the
hyperbolic spacetime $\mathbb{H}^{1}\times\mathbb{H}^{d-1}$. In order
to find these coefficients, we first need to solve the eigenvalue
equations \eqref{eq:4.5a}--\eqref{eq:4.5c} and construct the
eigenstate $|\alpha,\omega,j;\sigma\rangle$. To this end, from now on
we shall focus on the linear combinations \eqref{eq:4.1a} and utilize
the coordinate realizations of the generators in the Rindler
coordinate system $(t,x,\boldsymbol{x}_{\perp})$. A straightforward
(yet a little bit tedious) calculation shows that, in the Rindler
coordinate system given in \eqref{eq:2.48}, the coordinate
realizations of the operators
$C_{2}[\mathfrak{so}(2,d)]_{\alpha}=\frac{1}{2}\sum_{a,b=0}^{d+1}J_{\alpha}^{ab}J_{\alpha,ab}$,
$H_{\alpha}=J_{\alpha}^{10}$, and
$C_{2}[\mathfrak{so}(1,d-1)]_{\alpha}=\frac{1}{2}\sum_{a,b=2}^{d+1}J_{\alpha}^{ab}J_{\alpha,ab}$
that act on the representation space $\mathcal{V}_{\alpha}$ of scaling
dimension $\alpha\in\{\Delta,d-\Delta\}$ take the following
forms:\footnote{To derive these, substitute the following into
  \eqref{eq:3.13} with \eqref{eq:2.20a}--\eqref{eq:2.20d}:
  \begin{align}
    x^{0}=\pm x\sinh t,\quad
    x^{1}=\pm x\cosh t,\quad
    \partial_{0}=\pm\frac{\cosh t}{x}\partial_{t}\mp\sinh t\partial_{x},\quad
    \partial_{1}=\mp\frac{\sinh t}{x}\partial_{t}\pm\cosh t\partial_{x}. \nonumber
  \end{align}}
\begin{subequations}
  \begin{align}
    C_{2}[\mathfrak{so}(2,d)]_{\alpha}
    &=\Delta(\Delta-d), \label{eq:4.8a}\\
    H_{\alpha}
    &=i\partial_{t}, \label{eq:4.8b}\\
    C_{2}[\mathfrak{so}(1,d-1)]_{\alpha}
    &=x^{2}\left(\bigtriangleup_{\mathbb{R}^{d-2}}
      +\partial_{x}^{2}
      +\frac{2\alpha-d+3}{x}\partial_{x}
      +\frac{\alpha(\alpha-d+2)}{x^{2}}\right), \label{eq:4.8c}
  \end{align}
\end{subequations}
where
$\bigtriangleup_{\mathbb{R}^{d-2}}=\sum_{i=2}^{d-1}\frac{\partial}{\partial
  x_{\perp}^{i}}\frac{\partial}{\partial x_{\perp}^{i}}$ is the
Laplacian of $(d-2)$-dimensional space $\mathbb{R}^{d-2}$ that is
perpendicular to the $(x^{0},x^{1})$-plane. Notice that the quadratic
Casimir operator \eqref{eq:4.8a} becomes constant and hence the
eigenvalue equation \eqref{eq:4.5a} does not give us any
information. What we need to solve is therefore the following
differential equations:
\begin{subequations}
  \begin{align}
    i\partial_{t}f
    &=\omega f, \label{eq:4.9a}\\
    x^{2}\left(\bigtriangleup_{\mathbb{R}^{d-2}}
    +\partial_{x}^{2}
    +\frac{2\alpha-d+3}{x}\partial_{x}
    +\frac{\alpha(\alpha-d+2)}{x^{2}}\right)f
    &=j(j-d+2)f. \label{eq:4.9b}
  \end{align}
\end{subequations}
The first equation implies that the $t$-dependence of $f$ is just the
plane wave $\mathrm{e}^{-i\omega t}$. In order to solve the second
equation, we first note that eq.~\eqref{eq:4.9b} can be cast into the
following Schr\"{o}dinger-like equation:
\begin{align}
  \left[-\partial_{x}^{2}+\frac{(j-\tfrac{d-2}{2})^{2}-\tfrac{1}{4}}{x^{2}}\right]\Bar{f}
  =-\boldsymbol{p}_{\perp}^{2}\Bar{f}, \label{eq:4.10}
\end{align}
where $\Bar{f}:=x^{\alpha-(d-3)/2}f$ and $\boldsymbol{p}_{\perp}^{2}$
is the eigenvalue of $-\bigtriangleup_{\mathbb{R}^{d-2}}$ whose
eigenfunction is the plane wave
$\mathrm{e}^{i\boldsymbol{p}_{\perp}\cdot\boldsymbol{x}_{\perp}}$.
Notice that eq.~\eqref{eq:4.10} is the bound-state problem of
inverse-square potential, which has no solution unless
$(j-(d-2)/2)^{2}$ becomes negative. Hence $j$ must be of the form
\begin{align}
  j=\frac{d-2}{2}+ik, \quad k\in(0,\infty). \label{eq:4.11}
\end{align}
In other words, the representation of the subalgebra
$\mathfrak{so}(1,d-1)$ must be the principal series representation,
which is one of the continuous series representations of the
indefinite orthogonal group. It should be emphasized that, though $j$
is complex, the combination $j(j-d+2)=-k^{2}-(d-2)^{2}/4$ is real. The
solution to the bound-state problem \eqref{eq:4.10} that converges as
$x\to\infty$ is given by $x^{1/2}K_{ik}(|\boldsymbol{p}_{\perp}|x)$,
where $K_{\nu}(z)$ is the modified Bessel function of the second
kind. Collecting the above pieces, we arrive at the following
eigenfunction:
\begin{align}
  f_{\alpha,\omega,k;\boldsymbol{p}_{\perp}}(t,x,\boldsymbol{x}_{\perp})
  =\sqrt{\frac{4k\sinh(\pi k)}{\pi}}x^{-\alpha+(d-2)/2}K_{ik}(|\boldsymbol{p}_{\perp}|x)
  \mathrm{e}^{-i\omega t+i\boldsymbol{p}_{\perp}\cdot\boldsymbol{x}_{\perp}}, \label{eq:4.12}
\end{align}
where the normalization factor is chosen for later convenience. Notice
that $\boldsymbol{p}_{\perp}$ corresponds to $\sigma$ in
eqs.~\eqref{eq:4.5a}--\eqref{eq:4.5c}. It can be shown that the
eigenfunction \eqref{eq:4.12} provides a complete orthonormal basis on
the Rindler wedge; that is, it satisfies the
orthonormality\footnote{Eqs.~\eqref{eq:4.13a} and \eqref{eq:4.13b}
  follow from the following identities for the modified Bessel
  function of the second kind (also known as the Macdonald function)
  with imaginary index (see, e.g.,
  \cite{Yakubovich:2006,Passian:2009,Szmytkowski:2010}):
  \begin{alignat}{4}
    \frac{4k\sinh(\pi k)}{\pi}\int_{0}^{\infty}\!\frac{dx}{x}\,
    K_{ik}(x)K_{ik^{\prime}}(x) &=2\pi\delta(k-k^{\prime})
    &\quad\text{for}\quad&&k,k^{\prime}&>0; \nonumber\\
    \int_{0}^{\infty}\!\frac{dk}{2\pi}\,\frac{4k\sinh(\pi k)}{\pi}
    K_{ik}(x)K_{ik}(x^{\prime}) &=x\delta(x-x^{\prime})
    &\quad\text{for}\quad&&x,x^{\prime}&>0. \nonumber
  \end{alignat}
  We note that these identities are mutually related through the
  Kontorovich--Lebedev transform.}
\begin{subequations}
  \begin{align}
    \begin{split}
      \int_{t,x,\boldsymbol{x}_{\perp}}\!\!\!
      f_{\alpha,\omega,k,\boldsymbol{p}_{\perp}}^{\ast}(t,x,\boldsymbol{x}_{\perp})
      f_{d-\alpha,\omega^{\prime},k^{\prime},\boldsymbol{p}_{\perp}^{\prime}}(t,x,\boldsymbol{x}_{\perp}) \\
      =(2\pi)^{d}\delta(\omega-\omega^{\prime})
      \delta(k-k^{\prime})\delta^{(d-2)}(\boldsymbol{p}_{\perp}-\boldsymbol{p}_{\perp}^{\prime}), \label{eq:4.13a}
    \end{split}
  \end{align}
  and the completeness
  \begin{align}
    \begin{split}
      \int_{\omega,k,\boldsymbol{p}_{\perp}}\!\!\!
      f_{d-\alpha,\omega,k,\boldsymbol{p}_{\perp}}(t,x,\boldsymbol{x}_{\perp})
      f_{\alpha,\omega,k,\boldsymbol{p}_{\perp}}^{\ast}(t^{\prime},x^{\prime},\boldsymbol{x}_{\perp}^{\prime}) \\
      =\frac{1}{x}\delta(t-t^{\prime})
      \delta(x-x^{\prime})\delta^{(d-2)}(\boldsymbol{x}_{\perp}-\boldsymbol{x}_{\perp}^{\prime}), \label{eq:4.13b}
    \end{split}
  \end{align}
\end{subequations}
where
\begin{subequations}
  \begin{align}
    \int_{t,x,\boldsymbol{x}_{\perp}}
    &:=\int_{-\infty}^{\infty}\!\!dt
      \int_{0}^{\infty}\!\!xdx
      \int\!d^{d-2}\boldsymbol{x}_{\perp}, \label{eq:4.14a}\\
    \int_{\omega,k,\boldsymbol{p}_{\perp}}
    &:=\int_{-\infty}^{\infty}\!\frac{d\omega}{2\pi}
      \int_{0}^{\infty}\!\frac{dk}{2\pi}
      \int\!\frac{d^{d-2}\boldsymbol{p}_{\perp}}{(2\pi)^{d-2}}. \label{eq:4.14b}
  \end{align}
\end{subequations}
The two-point function on the Rindler wedge can then be expanded as
follows:
\begin{align}
  G_{\alpha}(t,x,\boldsymbol{x}_{\perp};t^{\prime},x^{\prime},\boldsymbol{x}_{\perp}^{\prime})
  &=\int_{\omega,k,\boldsymbol{p}_{\perp}}\!\!\!
    \Tilde{G}_{\alpha}(\omega,k)
    f_{\alpha,\omega,k,\boldsymbol{p}_{\perp}}(t,x,\boldsymbol{x}_{\perp})
    f_{\alpha,\omega,k,\boldsymbol{p}_{\perp}}^{\ast}(t^{\prime},x^{\prime},\boldsymbol{x}_{\perp}^{\prime}). \label{eq:4.15}
\end{align}
It should be noted that, because $G_{\alpha}$ is an
$SO(1,1)\times SO(1,d-1)$ scalar, its Fourier transform
$\Tilde{G}_{\Delta}$ only depends on the $SO(1,1)$ eigenvalue $\omega$
and the $SO(1,d-1)$ invariant $k$: it does not depend on the
eigenvalues $\boldsymbol{p}_{\perp}$ of the generators of the
subalgebra $\mathfrak{so}(1,d-1)$.\footnote{This would not be the case
  for two-point functions of generic primary tensors.} Note also that,
just as $G_{\Delta}$ maps a primary operator of scaling dimension
$d-\Delta$ to another primary operator of scaling dimension $\Delta$,
the intertwining operator also maps a basis function
$f_{d-\Delta,\omega,k;\boldsymbol{p}_{\perp}}(t,x,\boldsymbol{x}_{\perp})$
of the representation space $\mathcal{V}_{d-\Delta}$ to another basis
function
$f_{\Delta,\omega,k;\boldsymbol{p}_{\perp}}(t,x,\boldsymbol{x}_{\perp})$
of the representation space $\mathcal{V}_{\Delta}$ and satisfies the
following equation:
\begin{align}
  (G_{\Delta}f_{d-\Delta,\omega,k;\boldsymbol{p}_{\perp}})(t,x,\boldsymbol{x}_{\perp})
  &:=\int_{t^{\prime},x^{\prime},\boldsymbol{x}_{\perp}^{\prime}}\!\!\!
    G_{\Delta}(t,x,\boldsymbol{x}_{\perp};t^{\prime},x^{\prime},\boldsymbol{x}_{\perp}^{\prime})
    f_{d-\Delta,\omega,k;\boldsymbol{p}_{\perp}}(t^{\prime},x^{\prime},\boldsymbol{x}_{\perp}^{\prime}) \nonumber\\
  &\phantom{:}=\Tilde{G}_{\Delta}(\omega,k)
    f_{\Delta,\omega,k;\boldsymbol{p}_{\perp}}(t,x,\boldsymbol{x}_{\perp}), \label{eq:4.16}
\end{align}
where the last equality follows from eqs.~\eqref{eq:4.15} and
\eqref{eq:4.13a}. Eq.~\eqref{eq:4.16} is most simply expressed by the
bra-ket notation:
\begin{align}
  G_{\Delta}|d-\Delta,\omega,k;\boldsymbol{p}_{\perp}\rangle
  =\Tilde{G}_{\Delta}(\omega,k)|\Delta,\omega,k;\boldsymbol{p}_{\perp}\rangle. \label{eq:4.17}
\end{align}
This equation, together with the ladder equations \eqref{eq:4.7},
enables us to extract powerful constraints on
$\Tilde{G}_{\Delta}(\omega,k)$ from the intertwining relations. To see
this, we need to explicitly compute the coefficients
$c_{j\sigma;j^{\prime}\sigma^{\prime}}^{\pm a}(\alpha,\omega)$ in
\eqref{eq:4.7}. We shall do this in the next section and show that the
intertwining relations result in certain linear recurrence relations
for $\Tilde{G}_{\Delta}(\omega,k)$ in the complex momentum space.

\subsection{Intertwining relations in the
  \texorpdfstring{$SO(1,1)$}{SO(1,1)} basis}
\label{section:4.2}
In general, it is quite complicated to compute all the coefficients of
the ladder equations \eqref{eq:4.7}. For our purpose, however, there
is no need to compute all of those coefficients. It turns out that it
is sufficient to consider the ladder equations for the linear
combinations $E_{\alpha}^{\pm d}+E_{\alpha}^{\pm(d+1)}$, which take
the following simple coordinate realizations in the Rindler coordinate
system:\footnote{These are the coordinate realizations in $W_{R}$. In
  $W_{L}$, just replace $x$ and $\partial_{x}$ to $-x$ and
  $-\partial_{x}$.}
\begin{align}
  E_{\alpha}^{\pm d}+E_{\alpha}^{\pm(d+1)}
  =i\left(-\partial_{0}\pm\partial_{1}\right)
  =\mathrm{e}^{\pm t}\left(-\frac{i\partial_{t}}{x}\pm i\partial_{x}\right). \label{eq:4.18}
\end{align}
By applying this to the eigenfunction \eqref{eq:4.12} and using the
recurrence relations of the modified Bessel function
\begin{align}
  \frac{1}{z}K_{\nu}(z)=-\frac{K_{\nu-1}(z)-K_{\nu+1}(z)}{2\nu}
  \quad\text{and}\quad
  \frac{d}{dz}K_{\nu}(z)=-\frac{K_{\nu-1}(z)+K_{\nu+1}(z)}{2}, \label{eq:4.19}
\end{align}
one immediately arrives, in the bra-ket notation, at the following
ladder equations:
\begin{align}
  \left(E_{\alpha}^{\pm d}+E_{\alpha}^{\pm(d+1)}\right)
  |\alpha,\omega,k;\boldsymbol{p}_{\perp}\rangle
  &=A^{\pm}
    \left[\alpha-\frac{d-2}{2}\mp i(\omega\pm k)\right]
    |\alpha,\omega\pm i,k+i;\boldsymbol{p}_{\perp}\rangle \nonumber\\
  &\quad
    +B^{\pm}
    \left[\alpha-\frac{d-2}{2}\mp i(\omega\mp k)\right]
    |\alpha,\omega\pm i,k-i;\boldsymbol{p}_{\perp}\rangle, \label{eq:4.20}
\end{align}
where $A^{\pm}$ and $B^{\pm}$ are $\alpha$-independent irrelevant
factors.

Now we use the intertwining relations:
\begin{align}
  \left(E_{\Delta}^{\pm d}+E_{\Delta}^{\pm(d+1)}\right)G_{\Delta}
  =G_{\Delta}\left(E_{d-\Delta}^{\pm d}+E_{d-\Delta}^{\pm(d+1)}\right). \label{eq:4.21}
\end{align}
Applying this to the basis
$|d-\Delta,\omega,k;\boldsymbol{p}_{\perp}\rangle$, one immediately
sees that the intertwining relations \eqref{eq:4.21} result in the
following identities:
\begin{subequations}
  \begin{align}
    \left[\Delta-\frac{d-2}{2}\mp i(\omega\mp k)\right]\Tilde{G}_{\Delta}(\omega,k)
    &=\left[\Tilde{\Delta}-\frac{d-2}{2}\mp i(\omega\mp k)\right]
      \Tilde{G}_{\Delta}(\omega\pm i,k-i), \label{eq:4.22a}\\
    \left[\Delta-\frac{d-2}{2}\mp i(\omega\pm k)\right]\Tilde{G}_{\Delta}(\omega,k)
    &=\left[\Tilde{\Delta}-\frac{d-2}{2}\mp i(\omega\pm k)\right]
      \Tilde{G}_{\Delta}(\omega\pm i,k+i), \label{eq:4.22b}
  \end{align}
\end{subequations}
or, equivalently, to the following recurrence relations in the complex
momentum space:
\begin{subequations}
  \begin{align}
    \Tilde{G}_{\Delta}(\omega\pm i,k\pm i)
    &=\frac{\Delta-(d-2)/2\mp i(\omega+k)}{\Tilde{\Delta}-(d-2)/2\mp i(\omega+k)}
      \Tilde{G}_{\Delta}(\omega,k), \label{eq:4.23a}\\
    \Tilde{G}_{\Delta}(\omega\pm i,k\mp i)
    &=\frac{\Delta-(d-2)/2\mp i(\omega-k)}{\Tilde{\Delta}-(d-2)/2\mp i(\omega-k)}
      \Tilde{G}_{\Delta}(\omega,k), \label{eq:4.23b}
  \end{align}
\end{subequations}
where $\Tilde{\Delta}=d-\Delta$ is the scaling dimension of the
``shadow operator''.

Now, the problem is to reduce to a problem to solve these recurrence
relations exactly. In fact, the solution is not unique: there are a
number of nontrivial solutions that satisfy \eqref{eq:4.23a} and
\eqref{eq:4.23b}. Among them are the following ``minimal'' solutions:
\begin{subequations}
  \begin{align}
    \Tilde{G}^{A/R}_{\Delta}(\omega,k)
    &=\frac{\Gamma\left(\tfrac{\Delta-(d-2)/2\pm i(\omega+k)}{2}\right)}
      {\Gamma\left(\tfrac{\Tilde{\Delta}-(d-2)/2\pm i(\omega+k)}{2}\right)}
      \frac{\Gamma\left(\tfrac{\Delta-(d-2)/2\pm i(\omega-k)}{2}\right)}
      {\Gamma\left(\tfrac{\Tilde{\Delta}-(d-2)/2\pm i(\omega-k)}{2}\right)}, \label{eq:4.24a}\\
    \Tilde{G}^{\pm}_{\Delta}(\omega,k)
    &=\mathrm{e}^{\pm\pi\omega}
      \left|\Gamma\left(\tfrac{\Delta-(d-2)/2+i(\omega+k)}{2}\right)\right|^{2}
      \left|\Gamma\left(\tfrac{\Delta-(d-2)/2+i(\omega-k)}{2}\right)\right|^{2}, \label{eq:4.24b}
  \end{align}
\end{subequations}
where for simplicity we have set the overall normalization factors to
be unity. Note that $|\Gamma(x+iy)|^{2}=\Gamma(x+iy)\Gamma(x-iy)$ for
$x,y\in\mathbb{R}$. These solutions are the ``minimal'' solutions in
the sense that they do not contain a factor that is invariant under
the shifts $(\omega,k)\to(\omega\pm i,k\pm i)$ and
$(\omega,k)\to(\omega\pm i,k\mp i)$. A typical example of such factor
is $g=g(\mathrm{e}^{2\pi\omega},\mathrm{e}^{2\pi k})$, where $g$ is an
arbitrary function. One can easily check that \eqref{eq:4.24a} and
\eqref{eq:4.24b} indeed satisfy \eqref{eq:4.23a} and \eqref{eq:4.23b}
and hence give the solutions to the recurrence relations.

Finally, let us give physical interpretations of the solutions
\eqref{eq:4.24a} and
\eqref{eq:4.24b}. $\Tilde{G}^{A/R}_{\Delta}(\omega,k)$ are interpreted
as the advanced and retarded two-point functions. Indeed, they have
desired analytic structures in the complex $\omega$-plane:
$\Tilde{G}^{R}_{\Delta}(\omega,k)$ has simple poles at
$\omega=\pm k-i(\Delta-(d-2)/2+2n)$ ($n\in\mathbb{Z}_{\geq0}$) in the
lower half complex $\omega$-plane, whereas
$\Tilde{G}^{A}_{\Delta}(\omega,k)$ has simple poles at
$\omega=\pm k+i(\Delta-(d-2)/2+2n)$ ($n\in\mathbb{Z}_{\geq0}$) in the
upper half complex $\omega$-plane, both of which are consistent with
the causal structures of retarded and advanced two-point
functions.\footnote{Here the causal structures mean that the
  position-space representations
  $G_{\Delta}^{R/A}(t,x,\boldsymbol{x}_{\perp};
  t^{\prime},x^{\prime},\boldsymbol{x}_{\perp}^{\prime})$ given
  through \eqref{eq:4.15} vanish for $t-t^{\prime}<0$
  ($t-t^{\prime}>0$). (Note that the integral \eqref{eq:4.15} contains
  the factor $\mathrm{e}^{-i\omega(t-t^{\prime})}$. And thanks to this
  factor, the integration contour can be deformed to the infinite
  semicircle in the upper (lower) half $\omega$-plane for
  $t-t^{\prime}<0$ ($t-t^{\prime}>0$), which results in
  $G_{\Delta}^{R/A}(t,x,\boldsymbol{x}_{\perp};
  t^{\prime},x^{\prime},\boldsymbol{x}_{\perp}^{\prime})=0$ for
  $t-t^{\prime}<0$ ($t-t^{\prime}>0$) if there is no simple poles in
  the upper (lower) half $\omega$-plane.)} (Note that we have assumed
the scaling dimension satisfies the unitarity bound $\Delta>(d-2)/2$.)
$\Tilde{G}^{\pm}_{\Delta}(\omega,k)$, on the other hand, are
interpreted as the positive- and negative-frequency two-point Wightman
functions. Indeed, they satisfy the KMS condition in the momentum
space (see eq.~\eqref{eq:2.44}):
\begin{align}
  \Tilde{G}_{\Delta}^{+}(\omega,k)=\mathrm{e}^{2\pi\omega}\Tilde{G}_{\Delta}^{-}(\omega,k). \label{eq:4.25}
\end{align}
In appendix \ref{section:A.2} we check that the Fourier transforms of
positive- and negative-frequency two-point Wightman functions indeed
coincide with the solutions \eqref{eq:4.24b}. It should be noted that,
since the recurrence relations are linear, any linear combination of
the solutions also satisfies \eqref{eq:4.24a} and \eqref{eq:4.24b} and
hence gives another solution to the recurrence relations.

\section{Conclusions}
\label{section:5}
It is well-known that $d$-dimensional CFT can be easily thermalized by
just putting it on the Rindler wedge, light-cone, or diamond under the
identifications of temporal coordinates with $SO(1,1)\subset SO(2,d)$
group parameters. Though this is essentially the Unruh effect that has
been vastly studied over the last four decades, correlation functions
of thus obtained thermal CFT have not been well-explored before. In
this paper we studied thermal two-point functions for scalar primary
operators by using the intertwining operator. Though it has long been
known that two-point functions of CFT are nothing but the integral
kernels of intertwining operators, its implications to thermal CFT
have not been well-explored either. Much inspired by the Kerimov's
intertwining operator approach to exact S-matrix
\cite{Kerimov:1998zz}, in this work we showed that, by using the
unconventional $SO(1,1)$ continuous basis for the representation space
of conformal algebra $\mathfrak{so}(2,d)$, the intertwining relations
result in the recurrence relations for thermal two-point functions in
the complex momentum space. By solving these recurrence relations, we
obtained the advanced/retarded two-point functions as well as the
positive/negative-frequency two-point Wightman functions in the
momentum space.

There are several remaining issues that deserve to be addressed in the
future. In this paper we just focused on scalar correlation functions
for simplicity. For practical applications, however, it is important
to understand the structure of momentum-space correlation functions
for generic primary tensors. It is also interesting to generalize our
approach to the AdS/CFT correspondence. Since thermal CFT on
$\mathbb{H}^{1}\times\mathbb{H}^{d-1}$ can also be obtained as a
boundary theory of certain AdS wedge regions, it must be possible to
develop a similar algebraic approach to momentum-space two-point
functions by utilizing the intertwining operators developed in
\cite{Dobrev:1998md,Aizawa:2014yqa}. We would like to address this
issue elsewhere.

\subsection*{Acknowledgments}
I would like to thank Naruhiko Aizawa, Vladimir Dobrev, and Mihail
Mintchev for discussions. I am indebted to Satoru Odake for reminding
me the ambiguity of solutions to the recurrence relations. This work
is supported in part by JSPS Grant-in-Aid for Research Activity
Startup \#15H06641.

\titleformat{\section}[block]{\filright\bfseries\boldmath}{\appendixname~\thesection.}{0.5em}{}[]
\titleformat{\subsection}[block]{\filright\bfseries\boldmath}{\appendixname~\thesubsection.}{0.5em}{}
\appendix
\section{Fourier Transform}
\label{section:A}
Thermal correlation functions are notorious for being complicated to
Fourier-transform even in the well-studied two-dimensional CFT. In
this appendix we present computational details for the Fourier
transforms of position-space two-point Wightman functions of thermal
CFT living on the hyperbolic spacetime
$\mathbb{H}^{1}\times\mathbb{H}^{d-1}$. To do this, we first need to
find a complete orthonormal basis on
$\mathbb{H}^{1}\times\mathbb{H}^{d-1}$ in an appropriate coordinate
system. This is done in appendix \ref{section:A.1}. In appendix
\ref{section:A.2} we compute the Fourier transforms and show that the
resulting momentum-space two-point functions exactly coincide with
\eqref{eq:4.24b}.

Throughout this appendix we will work in the units $\ell=1$ (i.e.,
$2\pi T=1$).

\subsection{Harmonic analysis on the hyperbolic spacetime}
\label{section:A.1}
The $d$-dimensional hyperbolic spacetime
$\mathbb{H}^{1}\times\mathbb{H}^{d-1}$ is a foliation of the
$(d+1)$-dimensional cone \eqref{eq:2.1} and can be embedded into the
$(d+2)$-dimensional space
$\mathbb{R}^{2,d}\ni Y^{a}=(Y^{0},\cdots,Y^{d+1})$ as follows:
\begin{align}
  (Y^{0})^{2}-(Y^{1})^{2}=-1=(Y^{2})^{2}+\cdots+(Y^{d})^{2}-(Y^{d+1})^{2},\quad
  Y^{1}\geq1,\quad
  Y^{d+1}\geq1. \label{eq:A.1}
\end{align}
In order to find a complete orthonormal basis on this spacetime, we
first need to introduce an appropriate coordinate patch. The most
convenient coordinate system is turned out to be of the form:
\begin{align}
  Y^{0}=\sinh t, \quad
  Y^{1}=\cosh t, \quad
  Y^{i}=\frac{x_{\perp}^{i}}{x}, \quad
  Y^{d}=\frac{1-x^{2}-\boldsymbol{x}_{\perp}^{2}}{2x}, \quad
  Y^{d+1}=\frac{1+x^{2}+\boldsymbol{x}_{\perp}^{2}}{2x}, \label{eq:A.2}
\end{align}
where $i\in\{2,\cdots,d-1\}$, $t\in(-\infty,\infty)$,
$x\in(0,\infty)$, and $\boldsymbol{x}_{\perp}\in\mathbb{R}^{d-2}$. The
induced metric on the hyperbolic spacetime
$\mathbb{H}^{1}\times\mathbb{H}^{d-1}$ then takes the following form:
\begin{align}
  ds_{\mathbb{H}^{1}\times\mathbb{H}^{d-1}}^{2}
  &=\left.
    -(dY^{0})^{2}+(dY^{1})^{2}+\cdots+(dY^{d})^{2}-(dY^{d+1})^{2}
    \right|_{Y\in\mathbb{H}^{1}\times\mathbb{H}^{d-1}} \nonumber\\
  &=-dt^{2}+\frac{dx^{2}+d\boldsymbol{x}_{\perp}^{2}}{x^{2}}. \label{eq:A.3}
\end{align}
The Laplace--Beltrami operator on
$\mathbb{H}^{1}\times\mathbb{H}^{d-1}$ is therefore given by
\begin{align}
  \bigtriangleup_{\mathbb{H}^{1}\times\mathbb{H}^{d-1}}
  &=\frac{1}{\sqrt{|g|}}\partial_{\mu}\sqrt{|g|}g^{\mu\nu}\partial_{\nu}
    =-\partial_{t}^{2}+\bigtriangleup_{\mathbb{H}^{d-1}}, \label{eq:A.4}
\end{align}
where
\begin{align}
  \bigtriangleup_{\mathbb{H}^{d-1}}
  =x^{2}\left(x^{d-3}\partial_{x}\frac{1}{x^{d-3}}\partial_{x}+\bigtriangleup_{\mathbb{R}^{d-2}}\right). \label{eq:A.5}
\end{align}
Here $\bigtriangleup_{\mathbb{R}^{d-2}}$ is the Laplacian on
$\mathbb{R}^{d-2}$. Notice that the metric \eqref{eq:A.3} is
time-independent, which means that in this coordinate system the time
translation is an isometry of
$\mathbb{H}^{1}\times\mathbb{H}^{d-1}$. In order to find the complete
orthonormal basis on the hyperbolic spacetime, we then need to solve
the following eigenvalue equations:
\begin{subequations}
  \begin{align}
    i\partial_{t}f&=\omega f, \label{eq:A.6a}\\
    \bigtriangleup_{\mathbb{H}^{d-1}}f&=j(j-d+2)f, \label{eq:A.6b}\\
    -\bigtriangleup_{\mathbb{R}^{d-2}}f&=\boldsymbol{p}_{\perp}^{2}f. \label{eq:A.6c}
  \end{align}
\end{subequations}
The solutions to the first and third equations are just the plane
waves $\mathrm{e}^{-i\omega t}$ and
$\mathrm{e}^{i\boldsymbol{p}_{\perp}\cdot\boldsymbol{x}_{\perp}}$, so
in what follows we shall focus on the second equation
\eqref{eq:A.6b}. Substituting \eqref{eq:A.5} into \eqref{eq:A.6b}, we
see that the eigenvalue equation \eqref{eq:A.6b} reduces to the
following Schr\"{o}dinger-like equation:
\begin{align}
  \left[-\partial_{x}^{2}+\frac{(j-\frac{d-2}{2})^{2}-\frac{1}{4}}{x^{2}}\right]\Bar{f}
  =-\boldsymbol{p}_{\perp}^{2}\Bar{f}, \label{eq:A.7}
\end{align}
where
\begin{align}
  \Bar{f}=x^{-(d-3)/2}f. \label{eq:A.8}
\end{align}
The Schr\"{o}dinger-like equation \eqref{eq:A.7} does not admit any
normalizable solutions unless $(j-(d-2)/2)^{2}$ becomes
negative. Hence $j$ must be of the form
\begin{align}
  j=\frac{d-2}{2}+ik, \quad k\in(0,\infty). \label{eq:A.9}
\end{align}
Notice that, though $j$ is complex, the eigenvalue of the
Laplace--Beltrami operator $\bigtriangleup_{\mathbb{H}^{d-1}}$ itself
is real, $j(j-d+2)=-k^{2}-(d-2)^{2}/4$. The solution to the equation
\eqref{eq:A.7} that converges as $x\to\infty$ is
$x^{1/2}K_{ik}(|\boldsymbol{p}_{\perp}|x)$, where $K_{\nu}(z)$ is the
modified Bessel function of the second kind. Collecting the above
pieces, we find the following complete orthonormal basis on the
hyperbolic spacetime $\mathbb{H}^{1}\times\mathbb{H}^{d-1}$:
\begin{align}
  f_{\omega,k,\boldsymbol{p}_{\perp}}(t,x,\boldsymbol{x}_{\perp})
  =\sqrt{\frac{4k\sinh(\pi k)}{\pi}}x^{(d-2)/2}K_{ik}(|\boldsymbol{p}_{\perp}|x)
  \mathrm{e}^{-i\omega t+i\boldsymbol{p}_{\perp}\cdot\boldsymbol{x}_{\perp}}, \label{eq:A.10}
\end{align}
which satisfies the orthonormality
\begin{subequations}
  \begin{align}
    \begin{split}
      \int_{-\infty}^{\infty}\!\!dt\int_{0}^{\infty}\!\!\frac{dx}{x^{d-1}}\int\!\!d^{d-2}\boldsymbol{x}_{\perp}\,
      f_{\omega,k,\boldsymbol{p}_{\perp}}^{\ast}(t,x,\boldsymbol{x}_{\perp})
      f_{\omega^{\prime},k^{\prime},\boldsymbol{p}_{\perp}^{\prime}}(t,x,\boldsymbol{x}_{\perp}) \\
      =(2\pi)^{d}\delta(\omega-\omega^{\prime})\delta(k-k^{\prime})\delta^{(d-2)}(\boldsymbol{p}_{\perp}-\boldsymbol{p}_{\perp}^{\prime}), \label{eq:A.11a}
    \end{split}
  \end{align}
  and the completeness
  \begin{align}
    \begin{split}
      \int_{-\infty}^{\infty}\!\!\frac{d\omega}{2\pi}\int_{0}^{\infty}\!\!\frac{dk}{2\pi}\int\!\!\frac{d^{d-2}\boldsymbol{p}_{\perp}}{(2\pi)^{d-2}}\,
      f_{\omega,k,\boldsymbol{p}_{\perp}}(t,x,\boldsymbol{x}_{\perp})
      f_{\omega,k,\boldsymbol{p}_{\perp}}^{\ast}(t^{\prime},x^{\prime},\boldsymbol{x}_{\perp}^{\prime}) \\
      =x^{d-1}\delta(t-t^{\prime})\delta(x-x^{\prime})\delta^{(d-2)}(\boldsymbol{x}_{\perp}-\boldsymbol{x}_{\perp}^{\prime}). \label{eq:A.11b}
    \end{split}
  \end{align}
\end{subequations}

\subsection{Two-point Wightman function in the momentum space}
\label{section:A.2}
Let us next turn to the problem of calculating the Fourier transform
of two-point functions. As noted in section \ref{section:2.2}, in the
coordinate system \eqref{eq:A.2} the positive- and negative-frequency
two-point Wightman functions are given by
\begin{align}
  G_{\Delta}^{\pm}(Y,Y^{\prime})
  =\frac{1}{(-2Y\cdot Y^{\prime})^{\Delta}}
  =\left[
  \frac{1/2}{-\cosh(t-t^{\prime}\mp i\epsilon)+\frac{x^{2}+x^{\prime2}+|\boldsymbol{x}_{\perp}-\boldsymbol{x}_{\perp}^{\prime}|^{2}}{2xx^{\prime}}}
  \right]^{\Delta}, \label{eq:A.12}
\end{align}
where $\epsilon$ is a positive infinitesimal. In order to compute the
momentum-space representations of \eqref{eq:A.12}, we first note that
the two-point functions \eqref{eq:A.12} can be expanded as follows:
\begin{align}
  G_{\Delta}^{\pm}(Y,Y^{\prime})
  =\int_{-\infty}^{\infty}\!\!\frac{d\omega}{2\pi}
  \int_{0}^{\infty}\!\!\frac{dk}{2\pi}
  \int\!\!\frac{d^{d-2}\boldsymbol{p}_{\perp}}{(2\pi)^{d-2}}\,
  \Tilde{G}_{\Delta}^{\pm}(\omega,k)
  f_{\omega,k,\boldsymbol{p}_{\perp}}(Y)
  f_{\omega,k,\boldsymbol{p}_{\perp}}^{\ast}(Y^{\prime}), \label{eq:A.13}
\end{align}
where
$f_{\omega,k,\boldsymbol{p}_{\perp}}(Y)\equiv
f_{\omega,k,\boldsymbol{p}_{\perp}}(t,x,\boldsymbol{x}_{\perp})$.  It
then follows from the orthonormality \eqref{eq:A.11a} that the
following equality holds:
\begin{align}
  (G_{\Delta}^{\pm}f_{\omega,k,\boldsymbol{p}_{\perp}})(Y)
  &:=\int_{\mathbb{H}^{1}\times\mathbb{H}^{d-1}}\!\!dY^{\prime}\,
    G_{\Delta}^{\pm}(Y,Y^{\prime})
    f_{\omega,k,\boldsymbol{p}_{\perp}}(Y^{\prime}) \nonumber\\
  &\phantom{:}=\Tilde{G}_{\Delta}^{\pm}(\omega,k)
    f_{\omega,k,\boldsymbol{p}_{\perp}}(Y), \label{eq:A.14}
\end{align}
where $\int_{\mathbb{H}^{1}\times\mathbb{H}^{d-1}}\!dY^{\prime}$ is a
shorthand for
$\int_{-\infty}^{\infty}\!dt^{\prime}\int_{0}^{\infty}\!dx^{\prime}{x^{\prime}}^{-d+1}\int\!d^{d-2}\boldsymbol{x}_{\perp}^{\prime}$. The
momentum-space two-point Wightman functions
$\Tilde{G}_{\Delta}^{\pm}(\omega,k)$ can be obtained from this
identity. In other words, $\Tilde{G}_{\Delta}^{\pm}(\omega,k)$ are
obtained by calculating the following $d$-dimensional integral:
\begin{align}
  (G_{\Delta}^{\pm}f_{\omega,k,\boldsymbol{p}_{\perp}})(Y)
  &=\sqrt{\frac{4k\sinh(\pi k)}{\pi}}
    \int_{-\infty}^{\infty}\!\!dt^{\prime}
    \int_{0}^{\infty}\!\!\frac{dx^{\prime}}{x^{\prime d-1}}
    \int\!\!d^{d-2}\boldsymbol{x}_{\perp}^{\prime} \nonumber\\
  &\quad\times
    \left[
    \frac{1/2}{-\cosh(t-t^{\prime}\mp i\epsilon)+\frac{x^{2}+x^{\prime2}+|\boldsymbol{x}_{\perp}-\boldsymbol{x}_{\perp}^{\prime}|^{2}}{2xx^{\prime}}}
    \right]^{\Delta}
    {x^{\prime}}^{\frac{d-2}{2}}K_{ik}(|\boldsymbol{p}_{\perp}|x^{\prime})
    \mathrm{e}^{-i\omega t^{\prime}+i\boldsymbol{p}_{\perp}\cdot\boldsymbol{x}_{\perp}^{\prime}}. \label{eq:A.15}
\end{align}
In order to compute this integral, we first change the integration
variables as $t^{\prime}\to t^{\prime}+t$ and
$\boldsymbol{x}_{\perp}^{\prime}\to\boldsymbol{x}_{\perp}^{\prime}+\boldsymbol{x}_{\perp}$,
and then shift the $t^{\prime}$-integration contour slightly
upward/downward by $\pm i\pi$ for $G_{\Delta}^{\pm}$. The resultant
integral takes the following form:
\begin{align}
  (G_{\Delta}^{\pm}f_{\omega,k,\boldsymbol{p}_{\perp}})(Y)
  &=\sqrt{\frac{4k\sinh(\pi k)}{\pi}}
    \mathrm{e}^{-i\omega t+i\boldsymbol{p}_{\perp}\cdot\boldsymbol{x}_{\perp}}
    \int_{-\infty\pm i\pi}^{\infty\pm i\pi}\!\!dt^{\prime}
    \int_{0}^{\infty}\!\!dx^{\prime}{x^{\prime}}^{-d/2}
    \int\!\!d^{d-2}\boldsymbol{x}_{\perp}^{\prime} \nonumber\\
  &\quad\times
    \left[
    \frac{1/2}{-\cosh t^{\prime}+\frac{x^{2}+x^{\prime2}+\boldsymbol{x}_{\perp}^{\prime2}}{2xx^{\prime}}}
    \right]^{\Delta}
    K_{ik}(|\boldsymbol{p}_{\perp}|x^{\prime})
    \mathrm{e}^{-i\omega t^{\prime}+i\boldsymbol{p}_{\perp}\cdot\boldsymbol{x}_{\perp}^{\prime}} \nonumber\\
  &=\mathrm{e}^{\pm\pi\omega}
    \sqrt{\frac{4k\sinh(\pi k)}{\pi}}\mathrm{e}^{-i\omega t+i\boldsymbol{p}_{\perp}\cdot\boldsymbol{x}_{\perp}}
    \int_{-\infty}^{\infty}\!\!dt^{\prime}
    \int_{0}^{\infty}\!\!dx^{\prime}{x^{\prime}}^{-d/2}
    \int\!\!d^{d-2}\boldsymbol{x}_{\perp}^{\prime} \nonumber\\
  &\quad\times
    \left[
    \frac{1/2}{\cosh t^{\prime}+\frac{x^{2}+x^{\prime2}+\boldsymbol{x}_{\perp}^{\prime2}}{2xx^{\prime}}}
    \right]^{\Delta}
    K_{ik}(|\boldsymbol{p}_{\perp}|x^{\prime})
    \mathrm{e}^{-i\omega t^{\prime}+i\boldsymbol{p}_{\perp}\cdot\boldsymbol{x}_{\perp}^{\prime}}, \label{eq:A.16}
\end{align}
where in the second equality we have changed the integration variable
as $t^{\prime}\to t^{\prime}\pm i\pi$. Next we use the following
identity:
\begin{align}
  \left[\frac{1/2}{\cosh t^{\prime}+\frac{x^{2}+x^{\prime2}+\boldsymbol{x}_{\perp}^{\prime2}}{2xx^{\prime}}}\right]^{\Delta}
  =\frac{1}{2^{\Delta}\Gamma(\Delta)}
  \int_{0}^{\infty}\!\!dz\,z^{\Delta-1}
  \mathrm{e}^{-z\left(\cosh t+\frac{x^{2}+x^{\prime2}+\boldsymbol{x}_{\perp}^{\prime2}}{2xx^{\prime}}\right)}, \label{eq:A.17}
\end{align}
which just follows from the integral representation of the Gamma
function. Substituting this into \eqref{eq:A.16} we get
\begin{align}
  (G_{\Delta}^{\pm}f_{\omega,k,\boldsymbol{p}_{\perp}})(Y)
  &=\frac{\mathrm{e}^{\pm\pi\omega}}{2^{\Delta}\Gamma(\Delta)}
    \sqrt{\frac{4k\sinh(\pi k)}{\pi}}
    \mathrm{e}^{-i\omega t+i\boldsymbol{p}_{\perp}\cdot\boldsymbol{x}_{\perp}}
    \int_{0}^{\infty}\!\!dz\,z^{\Delta-1}
    \int_{-\infty}^{\infty}\!\!dt^{\prime}\,
    \mathrm{e}^{-i\omega t^{\prime}-z\cosh t^{\prime}} \nonumber\\
  &\quad\times
    \int_{0}^{\infty}\!\!dx^{\prime}\,{x^{\prime}}^{-d/2}K_{ik}(|\boldsymbol{p}_{\perp}|x^{\prime})
    \mathrm{e}^{-\frac{z}{2xx^{\prime}}(x^{2}+x^{\prime2})}
    \int\!\!d^{d-2}\boldsymbol{x}_{\perp}^{\prime}\,
    \mathrm{e}^{i\boldsymbol{p}_{\perp}\cdot\boldsymbol{x}_{\perp}^{\prime}-\frac{z}{2xx^{\prime}}\boldsymbol{x}_{\perp}^{\prime2}}. \label{eq:A.18}
\end{align}
The $t^{\prime}$- and $\boldsymbol{x}_{\perp}^{\prime}$-integrals are
calculated as follows:
\begin{subequations}
  \begin{align}
    \int_{-\infty}^{\infty}\!\!dt^{\prime}\,
    \mathrm{e}^{-i\omega t^{\prime}-z\cosh t^{\prime}}
    &=K_{i\omega}(z), \label{eq:A.19a}\\
    \int\!\!d^{d-2}\boldsymbol{x}_{\perp}^{\prime}\,
    \mathrm{e}^{i\boldsymbol{p}_{\perp}\cdot\boldsymbol{x}_{\perp}^{\prime}-\frac{z}{2xx^{\prime}}\boldsymbol{x}_{\perp}^{\prime2}}
    &=\left(\frac{2\pi xx^{\prime}}{z}\right)^{(d-2)/2}
      \mathrm{e}^{-\frac{xx^{\prime}}{2z}\boldsymbol{p}_{\perp}^{2}}, \label{eq:A.19b}
  \end{align}
\end{subequations}
from which we get
\begin{align}
  (G_{\Delta}^{\pm}f_{\omega,k,\boldsymbol{p}_{\perp}})(Y)
  &=\frac{(2\pi)^{\frac{d-2}{2}}\mathrm{e}^{\pm\pi\omega}}{2^{\Delta}\Gamma(\Delta)}
    \sqrt{\frac{4k\sinh(\pi k)}{\pi}}x^{(d-2)/2}
    \mathrm{e}^{-i\omega t+i\boldsymbol{p}_{\perp}\cdot\boldsymbol{x}_{\perp}}
    \int_{0}^{\infty}\!\!dz\,z^{\Delta-d/2}K_{i\omega}(z) \nonumber\\
  &\quad\times
    \int_{0}^{\infty}\!\!\frac{dx^{\prime}}{x^{\prime}}\,K_{ik}(|\boldsymbol{p}_{\perp}|x^{\prime})
    \mathrm{e}^{-\frac{z}{2xx^{\prime}}(x^{2}+x^{\prime2})-\frac{xx^{\prime}}{2z}\boldsymbol{p}_{\perp}^{2}}. \label{eq:A.20}
\end{align}
Making use of the following identity for the modified Bessel function
\begin{align}
  \int_{0}^{\infty}\!\!\frac{dx^{\prime}}{x^{\prime}}\,K_{ik}(|\boldsymbol{p}_{\perp}|x^{\prime})
  \mathrm{e}^{-\frac{z}{2xx^{\prime}}(x^{2}+x^{\prime2})-\frac{xx^{\prime}}{2z}\boldsymbol{p}_{\perp}^{2}}
  =2K_{ik}(z)K_{ik}(|\boldsymbol{p}_{\perp}|x), \label{eq:A.21}
\end{align}
we finally get
\begin{align}
  (G_{\Delta}^{\pm}f_{\omega,k,\boldsymbol{p}_{\perp}})(Y)
  =\Tilde{G}_{\Delta}^{\pm}(\omega,k)f_{\omega,k,\boldsymbol{p}_{\perp}}(Y), \label{eq:A.22}
\end{align}
where
\begin{align}
  \Tilde{G}_{\Delta}^{\pm}(\omega,k)
  =\frac{2(2\pi)^{\frac{d-2}{2}}}{2^{\Delta}\Gamma(\Delta)}\mathrm{e}^{\pm\pi\omega}
  \int_{0}^{\infty}\!\!dz\,z^{\Delta-d/2}K_{i\omega}(z)K_{ik}(z). \label{eq:A.23}
\end{align}
This integral is exactly calculable. To see this, we first use the
following identities:
\begin{align}
  K_{i\omega}(z)=\int_{-\infty}^{\infty}\!\!dt\,\mathrm{e}^{i\omega t-z\cosh z}
  \quad\text{and}\quad
  K_{ik}(z)=\int_{-\infty}^{\infty}\!\!dx\,\mathrm{e}^{ikx-z\cosh z}. \label{eq:A.24}
\end{align}
Substituting these into eq.~\eqref{eq:A.23} we get
\begin{align}
  \Tilde{G}_{\Delta}^{\pm}(\omega,k)
  &=\frac{2(2\pi)^{\frac{d-2}{2}}}{2^{\Delta}\Gamma(\Delta)}\mathrm{e}^{\pm\pi\omega}
    \int_{-\infty}^{\infty}\!\!dt
    \int_{-\infty}^{\infty}\!\!dx\,\mathrm{e}^{i\omega t+ikx}
    \int_{0}^{\infty}\!\!dz\,z^{\Delta-d/2}\mathrm{e}^{-z(\cosh t+\cosh x)} \nonumber\\
  &=\frac{2\pi^{\frac{d-2}{2}}\Gamma(\Delta-\tfrac{d-2}{2})}{\Gamma(\Delta)}
    \mathrm{e}^{\pm\pi\omega}
    \int_{-\infty}^{\infty}\!\!dt
    \int_{-\infty}^{\infty}\!\!dx\,
    \frac{\mathrm{e}^{i\omega t+ikx}}{[4\cosh(\tfrac{t+x}{2})\cosh(\tfrac{t-x}{2})]^{\Delta-(d-2)/2}}, \label{eq:A.25}
\end{align}
where the second equality follows from the identity
\begin{align}
  \frac{1}{2^{\Delta-\frac{d-2}{2}}}\int_{0}^{\infty}\!\!dz\,z^{\Delta-d/2}\mathrm{e}^{-z(\cosh t+\cosh x)}
  =\frac{\Gamma(\Delta-\tfrac{d-2}{2})}{[4\cosh(\frac{t+x}{2})\cosh(\frac{t-x}{2})]^{\Delta-(d-2)/2}}. \label{eq:A.26}
\end{align}
Let us next introduce the light-cone coordinates $x^{\pm}=t\pm x$, in
which the integration measure becomes $dtdx=(1/2)dx^{+}dx^{-}$. The
integral \eqref{eq:A.25} is then evaluated as follows:
\begin{align}
  \Tilde{G}_{\Delta}^{\pm}(\omega,k)
  &=\frac{\pi^{\frac{d-2}{2}}\Gamma(\Delta-\tfrac{d-2}{2})}{\Gamma(\Delta)}
    \mathrm{e}^{\pm\pi\omega}
    \int_{-\infty}^{\infty}\!\!dx^{+}\,
    \frac{\mathrm{e}^{i\frac{\omega+k}{2}x^{+}}}{[2\cosh(\tfrac{x^{+}}{2})]^{\Delta-\frac{d-2}{2}}}
    \int_{-\infty}^{\infty}\!\!dx^{-}\,
    \frac{\mathrm{e}^{i\frac{\omega-k}{2}x^{-}}}{[2\cosh(\tfrac{x^{-}}{2})]^{\Delta-\frac{d-2}{2}}} \nonumber\\
  &=\frac{\pi^{\frac{d-2}{2}}\Gamma(\Delta-\tfrac{d-2}{2})}{\Gamma(\Delta)}
    \mathrm{e}^{\pm\pi\omega}
    \int_{0}^{\infty}\!\!du\,
    \frac{u^{\frac{\Delta-(d-2)/2+i(\omega+k)}{2}-1}}{(1+u)^{\Delta-\frac{d-2}{2}}}
    \int_{0}^{\infty}\!\!dv\,
    \frac{v^{\frac{\Delta-(d-2)/2+i(\omega-k)}{2}-1}}{(1+v)^{\Delta-\frac{d-2}{2}}} \nonumber\\
  &=\frac{\pi^{\frac{d-2}{2}}\Gamma(\Delta-\tfrac{d-2}{2})}{\Gamma(\Delta)}
    \mathrm{e}^{\pm\pi\omega}
    \frac{\Gamma(\frac{\Delta-(d-2)/2+i(\omega+k)}{2})\Gamma(\frac{\Delta-(d-2)/2-i(\omega+k)}{2})}{\Gamma(\Delta-\frac{d-2}{2})} \nonumber\\
  &\quad\times
    \frac{\Gamma(\frac{\Delta-(d-2)/2+i(\omega-k)}{2})\Gamma(\frac{\Delta-(d-2)/2-i(\omega-k)}{2})}{\Gamma(\Delta-\frac{d-2}{2})}, \label{eq:A.27}
\end{align}
where in the second equality we have changed the integration variables
as $u=\mathrm{e}^{x^{+}}$ and $v=\mathrm{e}^{x^{-}}$ and in the last
equality we have used the integration formula for the beta function
\begin{align}
  B(p,q)=\int_{0}^{\infty}\!\!dx\,\frac{x^{p-1}}{(1+x)^{p+q}}=\frac{\Gamma(p)\Gamma(q)}{\Gamma(p+q)}
  \quad\text{for}\quad
  \mathrm{Re}\,p,~\mathrm{Re}\,q>0. \label{eq:A.28}
\end{align}
To summarize, we have found that the positive- and negative-frequency
two-point Wightman functions in the momentum space take the following
forms:
\begin{align}
  \Tilde{G}_{\Delta}^{\pm}(\omega,k)
  =\frac{\pi^{\frac{d-2}{2}}}{\Gamma(\Delta)\Gamma(\Delta-\tfrac{d-2}{2})}
  \mathrm{e}^{\pm\pi\omega}
  \left|\Gamma\left(\tfrac{\Delta-(d-2)/2+i(\omega+k)}{2}\right)\right|^{2}
  \left|\Gamma\left(\tfrac{\Delta-(d-2)/2+i(\omega-k)}{2}\right)\right|^{2}, \label{eq:A.29}
\end{align}
which, up to the normalization factor, exactly coincide with the
solutions \eqref{eq:4.24b}.

\bibliographystyle{utphys} \bibliography{bibliography}

\end{document}